\documentclass[aps,onecolumn,notitlepage,footinbib,superscriptaddress,showpacs,longbibliography]{revtex4-1}%
\let\revappendix\appendix
\usepackage[english]{babel}
\usepackage[T1]{fontenc}
\usepackage{endnotes}
\usepackage{amssymb,amsmath,amsfonts}
\usepackage{textcomp}
\usepackage{graphicx,color}
\usepackage[utf8x]{inputenc}
\usepackage{float}
\usepackage[normalem]{ulem}
\usepackage{siunitx}
\usepackage{xcolor}
\usepackage{tikz}
\usepackage{epstopdf}
\usepackage{booktabs}
\usepackage{hyperref}
\usepackage{graphicx}
\usepackage[export]{adjustbox}
\draft 
\graphicspath{{./figures/}}

\begin{document}

\newcommand{\lsolid}[1]{\raisebox{2pt}{\tikz{\draw[#1,solid,line width=1pt](0,0)--(8mm,0);}}}
\newcommand{\ldash}[1]{\raisebox{2pt}{\tikz{\draw[#1,dashed,line width=1pt](0,0)--(8mm,0);}}}
\newcommand{\ldott}[1]{\raisebox{2pt}{\tikz{\draw[#1,dotted,line width=1pt](0,0)--(8mm,0);}}}
\newcommand{\ldashdot}[1]{\raisebox{2pt}{\tikz{\draw[#1,dashdotted,line width=1pt](0,0)--(8mm,0);}}}
\newcommand{\linecircle}[1]{\raisebox{2pt}{\tikz{\draw[#1,solid,fill,scale=0.5]circle(1.0mm);\draw[#1,solid,line width = 1.0pt](4.2mm,0.1mm)-- (-4.5mm,0.1mm)}}}
\newcommand{\linetriangle}[1]{\raisebox{2pt}{\tikz{\draw[#1,solid,fill,scale=0.6](0,0) --(0.2cm,0) -- (0.1cm,0.2cm);\draw[#1,solid,line width = 1.0pt](4.1mm,0.5mm)-- (-2.8mm,0.5mm)}}}
\newcommand{\linesquare}[1]{\raisebox{2pt}{\tikz{\draw[#1,solid,fill,line width=1pt]rectangle(0.8mm,0.8mm);\draw[#1,solid,line width = 1.0pt](4.6mm,0.5mm) -- (-3.5mm,0.5mm)}}}
\definecolor{darkgreen}{rgb}{0.0, 0.5, 0.0}
\definecolor{oorange}{rgb}{1.0, 0.65, 0.0}

\title{Finite-rate chemistry effects in turbulent hypersonic boundary layers: a direct numerical simulation study}

\author{D. Passiatore}
\affiliation{DMMM, Politecnico di Bari, via Re David 200, 70125 Bari, Italy}
\email[Corresponding author: ]{donatella.passiatore@poliba.it}
\affiliation{Laboratoire DynFluid, Arts et M\'{e}tiers ParisTech, 151 Bd. de l'H\^{o}pital, 75013 Paris, France}

\author{L. Sciacovelli}
\affiliation{Laboratoire DynFluid, Arts et M\'{e}tiers ParisTech, 151 Bd. de l'H\^{o}pital, 75013 Paris, France}
\author{P. Cinnella}
\affiliation{Laboratoire DynFluid, Arts et M\'{e}tiers ParisTech, 151 Bd. de l'H\^{o}pital, 75013 Paris, France}
\author{G. Pascazio}
\affiliation{DMMM, Politecnico di Bari, via Re David 200, 70125 Bari, Italy}

\date{\today}

\begin{abstract}
The influence of high-enthalpy effects on hypersonic turbulent boundary layers is investigated by means of direct numerical simulations (DNS). A quasi-adiabatic flat-plate air flow at free-stream Mach number equal to 10 is simulated up to fully-developed turbulent conditions using  a five-species, chemically-reacting model. A companion DNS based on a frozen-chemistry assumption is also carried out, in order to isolate the effect of finite-rate chemical reactions and assess their influence on turbulent quantities. In order to reduce uncertainties associated with turbulence generation at the inlet of the computational domain, both simulations are initiated in the laminar flow region and the flow is let to evolve up to the fully turbulent regime. Modal forcing by means of localized suction and blowing is used to trigger laminar-to-turbulent transition.
The high temperatures reached in the near wall region including the viscous and buffer sublayers activate significant dissociation of both oxygen and nitrogen. This modifies in turn the thermodynamic and transport properties of the reacting mixture, affecting the first-order statistics of thermodynamic quantities. Due to the endothermic nature of the chemical reactions in the forward direction, temperature and density fluctuations in the reacting layer are smaller than in the frozen-chemistry flow.
However, the first- and second-order statistics of the velocity field are found to be little affected by the chemical reactions under a scaling that accounts for the modified fluid properties. We also observed that the Strong Reynolds Analogy (SRA) remains well respected despite the severe hypersonic conditions and that the computed skin friction coefficient distributions match well the results of the Renard-Deck decomposition extended to compressible flows.
\end{abstract}

\maketitle

\section{Introduction}
The accurate prediction of turbulent hypersonic flows is a major subject for the design of planetary atmosphere reentry bodies or hypersonic aircrafts. Recently, vehicles concepts involving hypersonic flight are driving renewed attention not only in the defense and military fields, but also in the areas of spatial tourism and trans-atmospheric flight (a recent review can be found in Ref. \onlinecite{leyva2017relentless}).
The massive amount of kinetic energy in a hypersonic free stream is converted into internal energy as the gas flows across the bow shock; the high temperatures reached in this regime can cause vibrational excitation and gas dissociation effects that strongly modify the forces and heat transfer acting on the surface. More generally, internal energy relaxation, chemical reactions and gas-surface interactions can occur in hypersonic flows at rates that are similar to the fluid motion ones, resulting in a nonequilibrium thermochemical state. These processes have a strong effect on aerodynamic performance of reentry objects and may vary global properties of the system \cite{colonna2019impact}; at the same time, heat transfer rates, ablation and instabilities growth (which lead to boundary layer transition and breakdown to turbulence) could be altered as well \citep{candler2019rate}. Such high-temperature effects not only modify the inviscid flow behavior, but also affect turbulence dynamics. Carrying out physical experiments in the working conditions of interest is generally a costly (sometimes infeasible) task \citep{bertin2006critical}; hence the necessity of performing high-fidelity numerical simulations to palliate the lack of experimental data. More specifically, Direct Numerical Simulations (DNS), which ensure the resolution of the whole active range of temporal and spatial scales, represents a powerful tool for a deeper understanding of out-of-equilibrium, high-speed flows and for the development of improved Reynolds--Averaged Navier--Stokes (RANS) models.
Turbulence models currently employed in hypersonic vehicle design
have been developed under perfect-gas assumptions, and their predictive performance for chemically reacting real gas flows is highly uncertain \citep{roy2006review}.

Significant research effort has been put into linear and non linear boundary layer stability analyses of hypersonic laminar boundary layers, with and without thermochemical non-equilibrium effects (see Refs. \onlinecite{malik1990stability,marxen2011high,zhong2012direct,marxen2013method,marxen2014direct,bitter2015stability,mortensen2016real,miro2019high}), since accurate predictions of  laminar-to-turbulent transition onset are of crucial importance for the evaluation of aerodynamic and heat transfer coefficients and for the design of the thermal protection system (TPS).
Such studies highlighted the existence of multiple modes in the supersonic regime, and the dominant role played by Mack's mode \citep{mack1969boundary}, commonly referred to as the second mode. This mode is strongly influenced (destabilized or stabilized) by the thermodynamic properties at the wall\cite{robinet2019instabilities}. A few studies have investigated the initial stages of transition: for instance, Franko \emph{et al.}\cite{franko2010effects} assessed the influence of different chemistry models in predicting the $2^\text{nd}$ mode growth rate and Marxen \textit{et al.} \cite{marxen2014direct} performed an analysis of the influence of primary and secondary amplitude perturbations, without encompassing transition.\\
 Regarding fully developed turbulence in boundary layer configurations, a handful of high-fidelity DNS studies have been carried out at low-enthalpy, non-reacting conditions. Duan \textit{et al.} \cite{duan2010direct2,duan2010direct} performed DNS of temporally-evolving, zero-pressure-gradient turbulent boundary layers in the high-Mach regime and varying wall temperatures. Mach numbers up to 20 were considered in the work of Lagha \textit{et al.} \cite{lagha2011numerical}. A comparative study between low- and high-enthalpy, reactive boundary layers can be found in the work of Duan \textit{et al.} \cite{duan2011direct4}, who carried out temporally-evolving boundary layer simulations, albeit a moderate temperature value was imposed at the wall ($\approx2400$ K) resulting in weak chemical activity. Recently, more attention has been paid to hypersonic, cold-wall boundary layers in spatially-evolving configurations \citep{zhang2018direct,huang2020simulation}, but thermodynamic conditions were such that a non-equilibrium thermochemical state is not present.
Chemical non-equilibrium in high-enthalpy turbulent boundary layer flows has been extensively studied by Duan and Mart\'in \cite{duan2009effectchem,duan2011assessment} at moderate Reynolds numbers, whereas thermal non-equilibrium has been recently investigated on isotropic decaying turbulence \citep{neville2014effect,khurshid2019decaying} and mixing layers \citep{neville2015thermal}.

%
The objective of the current paper is to investigate finite-rate chemistry effects in spatially-evolving, high-enthalpy boundary layers of hypersonic flows. For that purpose we focus on a configuration widely employed in the past for stability studies of boundary layers \cite{malik1991real, hudson1997linear, perraud1999studies, franko2010effects, marxen2014direct, miro2018diffusion}. Specifically, we consider a flat plate flow of air, modeled as a 5-species mixture, with a free-stream Mach number equal to 10 and a quasi-adiabatic wall conditions. The resulting wall temperature of approximately \SI{5300}{K} is such that the mixture components undergo strong dissociation and recombination reactions in the near-wall region. The calculation is started in the laminar region, where suction and blowing forcing is introduced to trigger boundary layer instability and transition. The computational domain is sufficiently extended in the streamwise direction to achieve a significant portion of flow characterized by fully turbulent regime. A companion DNS is run at the same free-stream conditions and wall temperature by assuming frozen chemical composition. This allows isolating the contribution of finite-rate chemistry effects by direct comparison of the two simulations.\\
The paper is structured as follows. The governing equations, the numerical set-up and the flow parameters are described in Section~\ref{sec:simulation_details}. Numerical results are presented in Section~\ref{sec:results}, encompassing the analysis of the transitional zone in the first place and of the turbulent integral and statistical properties afterwards. The validity of the classical and modified Reynolds analogies are discussed and verified, as well as the skin friction decomposition for boundary layer flows. Lastly, conclusions are drawn in Section~\ref{sec:conclusions}.
\section{Simulation details}
\label{sec:simulation_details}
\subsection{Governing equations}
We consider flows governed by the compressible Navier--Stokes equations for multicomponent, chemically-reacting gases:
\begin{align}
\frac{\partial \rho}{\partial t} + \frac{\partial \rho u_j }{\partial x_j} & = 0 \\
\label{eq:momentum}
 \frac{\partial \rho u_i}{\partial t} + \frac{\partial \left( \rho u_i u_j + p \delta_{ij} \right)}{\partial x_j} & = \frac{\partial \tau_{ij}}{\partial x_j} \\
\label{eq:energy}
 \frac{\partial \rho E}{\partial t} + \frac{\partial \left[\left(\rho E + p \right)u_j\right]}{\partial x_j} & = \frac{\partial (u_i \tau_{ij} - q_j)}{\partial x_j} -\frac{\partial}{\partial x_j}\left(\sum_{n=1}^\text{NS} \rho_n u_{n,j}^D h_n \right) \\
  \label{eq:species}
\frac{\partial \rho_n}{\partial t} + \frac{\partial \left( \rho_n u_j \right)}{\partial x_j} & = -\frac{\partial \rho_n u_{n,j}^D }{\partial x_j} + \dot{\omega}_n \qquad \qquad (n=1,\dots,\text{NS}-1)
\end{align}
In the preceding equations, $\rho_n$ is the density of the $n$-th species, $\rho=\sum_{n=1}^\text{NS} \rho_n$ is the mixture density, NS is the total number of species, $u_i$ ($i=1,2,3$) are the components of the velocity vector in the direction $x_i$, $p$ is the pressure, $E=e + \frac{1}{2}u_i u_i$ is the specific total energy of the mixture, $u_{n,j}^D$, $h_n$, and $\dot{\omega}_n$ are the $n$-th species diffusion velocity in the $j$-th direction, specific enthalpy and rate of production, respectively. $\tau_{ij}$ denotes the viscous stress tensor:
\begin{equation}
 \tau_{ij} = \mu\left(\frac{\partial u_i}{\partial x_j} + \frac{\partial u_j}{\partial x_i} \right) -\frac{2}{3}\mu\frac{\partial u_k}{\partial x_k}\delta_{ij},
 \end{equation}
being $\mu$ the mixture dynamic viscosity, $\delta_{ij}$ the Kronecker symbol and $q_j = -\lambda \frac{\partial T}{\partial x_j}$ represents the heat flux, with $\lambda$ the mixture thermal conductivity and $T$ the temperature.\\
Air is modeled as a five-species mixture of N$_2$, O$_2$, NO, O and N. Species' conservation equations are written for $\text{NS}-1$ species, the NS-th species' partial density being computed as $\rho_\text{NS} = \rho - \sum_{n=1}^{\text{NS}-1} \rho_n$. In the following simulations, the NS-th species is chosen to be Nitrogen due to its large mass fraction throughout the computational domain. Each species is assumed to behave as a perfect gas; based on Dalton's law, the mixture pressure equation of state writes:
\begin{equation}
   p = \sum_{n=1}^\text{NS} p_n = \mathcal{R} \rho T \sum_{n=1}^\text{NS} \frac{Y_n}{W_n} = T \sum_{n=1}^\text{NS} \rho_n R_n,
\end{equation}
with $Y_n$, $W_n$ and $R_n$ the mass fraction, molecular weight and gas constant of the $n$-th species, respectively, and $\mathcal{R} = 8.314$ J/mol~K the universal constant of gases.
The thermodynamic properties of high-$T$ air species are computed considering the contributions of translational, rotational and vibrational modes \cite{gnoffo1989conservation}; specifically, the internal energy reads:
\begin{equation}
e = \sum_{n=1}^\text{NS} Y_n h_n - \frac{p}{\rho},
\end{equation}
with $h_n = h^0_{f,n} + \int_{T_\text{ref}}^T (c^\text{tr}_{p,n}+c^\text{rot}_{p,n}) \text{ d}T' + e^\text{vib}_n$ the $n$-th species enthalpy. Here, $h^0_{f,n}$ is the $n$-th species enthalpy of formation at the reference temperature ($T_\text{ref} = \SI{298.15}{K}$), $c^\text{tr}_{p,n}$ and $c^\text{rot}_{p,n}$ the translational and rotational contributions to the isobaric heat capacity of the $n$-th species, computed as
\begin{equation}
   c^\text{tr}_{p,n} = \frac{5}{2} R_n, \qquad
   c^\text{rot}_{p,n} =
   \begin{cases}
    R_n & \text{for diatomic species} \\
    0 & \text{for monoatomic species}
   \end{cases}
 \end{equation}
and $e^\text{vib}_n$ the vibrational energy of species $n$, given by
\begin{equation}
    e^\text{vib}_n = \begin{cases}
    \frac{\theta_n R_n}{\exp{(\theta_n/T)} - 1}& \text{for diatomic species} \\
    0 & \text{for monoatomic species}
   \end{cases}
\end{equation}
with $\theta_n$ the characteristic vibrational temperature of each molecule (equal to \SI{3393}{K}, \SI{2273}{K} and \SI{2739}{K} for N$_2$, O$_2$ and NO, respectively \cite{park1989nonequilibrium}).
After the numerical integration of the conservation equations, an iterative Newton-Raphson method is implemented to compute the temperature from the conservative variables.\\
With regard to the transport coefficients, pure species' viscosity and thermal conductivity are computed using Blottner's model and Eucken's formula, respectively~\cite{blottner1971}; the corresponding mixture properties are evaluated by means of Wilke's mixing rule. In equation~\eqref{eq:species}, the mass diffusion phenomenon is governed by Fick's law
\begin{equation}\label{eq:fick}
 u^D_{n,j} \rho_n= -\rho D_{n} \frac{\partial Y_n}{\partial x_j} + \rho_n \sum_{n=1}^N D_n \frac{\partial Y_n}{\partial x_j},
 \end{equation}
where the first term on the r.h.s. represents the effective diffusion velocity and the second one is a mass corrector term, needed in order to satisfy the mass conservation equation when dealing with non-constant species diffusion coefficients \cite{poinsot2005theoretical,giovangigli2012multicomponent}. Specifically, $D_n$ is an equivalent diffusion coefficient of species $n$ into the mixture, computed following Hirschfelder's approximation \cite{hirschfelder1964molecular} as
\begin{equation}
 D_n= \frac{1-Y_n}{ \sum_{\substack{m = 1 \\ m \neq n}}^\text{NS} \frac{X_n}{D_{mn}} }
\end{equation}
 with $D_{mn}$ is the binary diffusion coefficient of species $m$ into species $n$:
\begin{equation}
 D_{mn} = \frac{1}{p}\exp{(A_{4,mn})} T^{\left[ A_{1,mn} (\ln T)^2 + A_{2,mn} \ln T + A_{3,mn} \right]},
\end{equation}
where $A_{1,mn}, ..., A_{4,mn}$ are curve-fitted coefficients computed as in Gupta \textit{et al.} \cite{gupta1990review}. The five species interact with each other through a 17-reaction mechanism:
\begin{alignat}{3}
 \notag \text{R1}: & \qquad \text{N}_2 + \text{M} && \Longleftrightarrow 2\text{N} + \text{M} \\
 \notag \text{R2}: & \qquad \text{O}_2 + \text{M} && \Longleftrightarrow 2\text{O} + \text{M} \\
        \text{R3}: & \qquad \text{NO}  + \text{M} && \Longleftrightarrow  \text{N} + \text{O} + \text{M} \\
 \notag \text{R4}: & \qquad \text{N}_2 + \text{O} && \Longleftrightarrow  \text{NO}+ \text{N} \\
 \notag \text{R5}: & \qquad \text{NO}  + \text{O} && \Longleftrightarrow  \text{N} + \text{O}_2
\end{alignat}
being M any of the five species considered. The mass rate of production of the \textit{n}-th species is governed by the law of mass action
\begin{widetext}
  \begin{equation}
 \dot{\omega }_n =  W_n \sum_{r=1}^\text{NR} \left( \nu_{nr}'' - \nu_{nr}' \right) \times \left\lbrace k_{f,r} \prod_{n=1}^\text{NS} \left(\frac{\rho Y_n}{W_n}\right)^{\nu_{nr}'} - k_{b,r} \prod_{n=1}^\text{NS} \left(\frac{\rho Y_n}{W_n}\right)^{\nu_{nr}''} \right\rbrace,
 \label{mass_action}
\end{equation}
\end{widetext}
where $\nu_{nr}'$ and  $\nu_{nr}''$ are the stoichiometric coefficients for reactants and products in the $r$-th reaction for $n$-th species, respectively, and NR is the total number of reactions. $k_{f,r}$ and $k_{b,r}$ are the forward and backward reaction rates of reaction $r$, modeled according to Arrhenius' law. Further details can be found in the work of Park \cite{park1993review}.

A frozen-chemistry model is also used for some of the following results. In this case, the source terms in equation~\eqref{eq:species} are simply set to zero and $\text{NS}=2$, the only species present being $\text{N}_2$ and $\text{O}_2$.


\subsection{Numerical methods}
The governing equations are approximated using high-order finite-difference schemes. The convective fluxes are discretized by means of the tenth-order central scheme using 11-points in each direction, whereas standard fourth-order schemes are used for viscous fluxes. A selective 10$^\text{th}$-order centered filter, with an amplitude equal to 0.1, is applied sequentially in each direction to damp grid-to-grid oscillations. The numerical discretization is supplemented with a shock-capturing term based on the Localized Artificial Diffusivity (LAD) approach, initially introduced by Cook \& Cabot\cite{cook2004} and later extended to multicomponent flows \cite{kawai2010assessment}. In the present implementation, only artificial bulk viscosity and thermal conductivity are introduced, with $C_\beta=0.5$ and $C_\lambda=0.005$ (further details can be found in Kawai \emph{et al.}\cite{kawai2010assessment}), while the artificial shear viscosity and mass diffusion coefficients are set to zero. Time integration is carried out by means of a third-order TVD Runge-Kutta scheme \cite{gottlieb1998total}.


\subsection{Computational setup and definitions}
We simulate two hypersonic, zero-pressure-gradient, spatially-evolving boundary layers over a flat plate, in chemical nonequilibrium (``CN'' case) and under a frozen-chemistry assumption (``FR'' case), respectively. The free-stream conditions are similar to those considered in previous stability studies \cite{malik1990stability,marxen2011high, marxen2013method, marxen2014direct,miro2018diffusion}, namely, an external Mach number of $M_\infty=10$, an external Reynolds number per unit length of $Re_u {=} \SI{6.0e6}{m^{-1}}$, and a free-stream pressure and temperature of $p_\infty = \SI{3596}{Pa}$ and $T_\infty=\SI{350}{K}$, respectively.
The calculations are initiated in the laminar region, where air with standard composition ($Y_{\text{O}_2}{=}0.233$ and $Y_{\text{N}_2}{=}0.767$) enters the computational domains.
Self-similar profiles of the flow properties, corresponding to a compressible Blasius boundary layer solution under frozen-chemistry assumptions, are prescribed at the computational domain inlet which is located to a distance $x_0$ from the plate leading edge (not included in the domain). Based on the initial displacement thickness $\delta^*_\text{in}$, we define a dimensionless streamwise coordinate as
\begin{equation}
\hat{x} = (x-x_0)/\delta^*_\text{in}
\end{equation}
that will be used in the following. At the inflow (i.e., $\hat{x}=0$), the $\delta^*_\text{in}$-based Reynolds number is set to $Re_{\delta^*}{=}5000$; a sponge layer is applied from $\hat{x} = 0$ to $\hat{x}=100$ to prevent distortions of the boundary layer similarity profiles. At the plate wall, the adiabatic wall temperature resulting from the similarity solution, $T_w{=}\SI{5323}{K}$, is prescribed along with no-slip, non-catalytic conditions. Characteristics-based boundary conditions \cite{poinsot1992boundary} are used at the inflow, outflow and free-stream boundaries, whereas periodicity is imposed in the spanwise direction.
The dimensions of the computational domain are $L_x \times L_y \times L_z = 5200 \delta^*_\text{in} \times 240 \delta^*_\text{in} \times 50\pi \delta_\text{in}^*$ in the streamwise ($x$), wall-normal ($y$) and spanwise ($z$) directions, respectively; the domain is discretized with $N_x \times N_y \times N_z = 5236 \times 256 \times 240$ points, corresponding to a total of approximately $ 3.2 \times 10^8$ grid points.
The grid spacing is uniform in the streamwise and spanwise directions, whereas a grid stretching is applied in the wall-normal direction to ensure fine resolutions near the wall; specifically, the stretching function is
\begin{equation}
\frac{y(j)}{L_y} = (1-\alpha)\left(\frac{j-1}{N_y-1}\right)^3 + \alpha \frac{j-1}{N_y-1}
\end{equation}
where $j\in [1,N_y]$ and $\alpha{=}0.25$. The same computational grid is used for both the frozen- and finite-rate chemistry DNS; resolutions are reported in table~\ref{tab:dns_param} for the latter case at several downstream locations in wall viscous units, the viscous length scale being $l_v = \overline{\mu}_w/(\overline{\rho}_w u_\tau)$. Here, $\overline{\mu}_w$ and $\overline{\rho}_w$ denote the time- and spanwise-averaged wall values of viscosity and density, and $u_\tau = \sqrt{ \overline{\tau}_w / \overline{\rho}_w }$ the friction velocity based on the averaged wall shear stress $\overline{\tau}_w$. The values obtained along the turbulent portion of the domain (with $\Delta x^+<2.6$, $\Delta y_w^+<0.6$ and $\Delta z^+<1.8$ everywhere) denote an excellent spatial resolution; note that similar values were obtained for the frozen-chemistry DNS. Additionally, in order to assess the adequacy of the domain size and grid refinement, two-point correlations and one-dimensional kinetic energy spectra are presented in Appendix~\ref{app:grid_study}.

\begin{table*}
\centering
\caption{Boundary layer properties for finite-rate chemistry DNS at five downstream stations. In the following, $Re_x{=} \frac{\rho_\infty u_\infty x}{\mu_\infty}$ is the Reynolds number based on the distance from the leading edge $x$; $Re_\theta{=}\frac{\rho_\infty u_\infty \theta}{\mu_\infty}$ is the Reynolds number based on the local momentum thickness; $Re_\theta^\text{inc}{=}\frac{\mu_\infty}{\overline{\mu}_w} Re_\theta$ is the momentum-thickness-based Reynolds number in the incompressible scaling and $Re_\tau{=}\frac{\rho_w u_\tau \delta_\text{99}}{\overline{\mu}_w}$ the friction Reynolds number. $\Delta x^+$,  $\Delta y^+_\text{w}$, $\Delta y^+_\delta$ and  $\Delta z^+$ denote the grid size in inner variables in the $x$-direction, $y$-direction at the wall and at the boundary layer edge, and in the $z$-direction, respectively. Finally, $Ma_\tau=\frac{u_\tau}{c_w}$ stands for the friction Mach number and $H=\frac{\delta^*}{\theta}$ is the boundary layer shape factor. The last station, $\hat{x}{=}5100$, will be considered in the following for data analysis. } \label{tab:dns_param}
\begin{ruledtabular}
\begin{tabular}{cccccc}
$\hat{x}$ & 1000 & 2000 & 3000 & 4000 & 5100\\
\hline
Re$_x$                       & $5.04\times 10^6$ & $10.03 \times 10^6$  & $15.04 \times 0^6$  & $20.04 \times 10^6$  & $25.54 \times 10^6$ \\
Re$_\theta$                  & 1741              & 3141                 & 4494                & 5784                 & 7149                \\
Re$_{\theta}^\text{inc}$     & 272               & 494                  & 705                 & 907                  & 1120                \\
Re$_\tau$                    & 35                & 89                   & 120                 & 154                  & 185                 \\
$\Delta x^+$                 & 2.04              & 2.78                 & 2.58                & 2.50                 & 2.45                \\
$\Delta y^+_\text{w}$        & 0.48              & 0.66                 & 0.61                & 0.59                 & 0.58                \\
$\Delta y^+_\delta$          & 0.73              & 1.47                 & 1.80                & 2.15                 & 2.42                \\
$\Delta z^+$                 & 1.35              & 1.83                 & 1.70                & 1.65                 & 1.61                \\
Ma$_\tau$                    & 0.14              & 0.17                 & 0.16                & 0.16                 & 0.15                \\
$H$                          & 35.6              & 33.8                 & 34.6                & 34.9                 & 34.4                \\
\end{tabular}
\end{ruledtabular}
\end{table*}

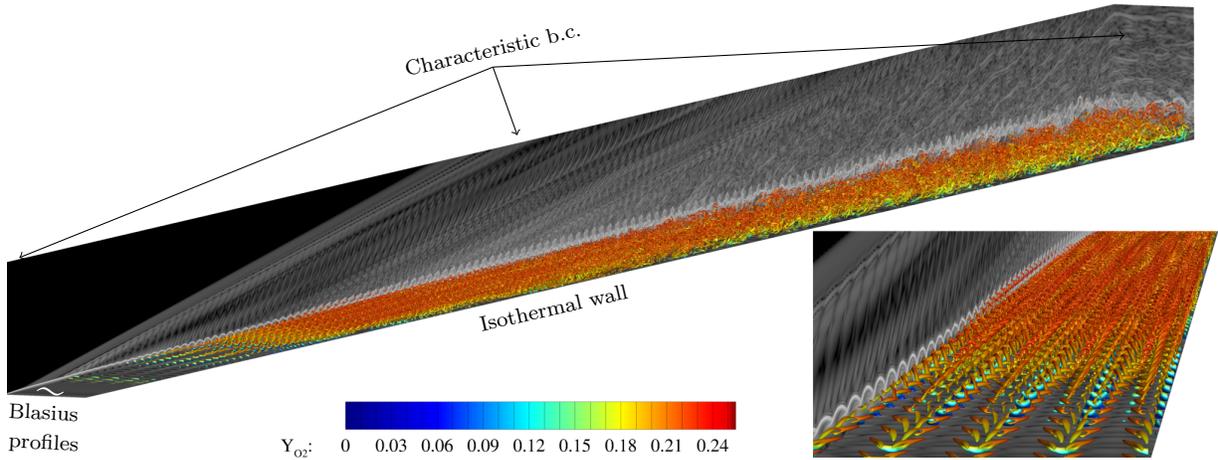
\begin{figure*}[tb]
   \begin{tikzpicture}
      \node[anchor=south west,inner sep=0] (a) at (0,0) {\includegraphics[width=0.9\textwidth, trim={5 5 5 5}, clip]{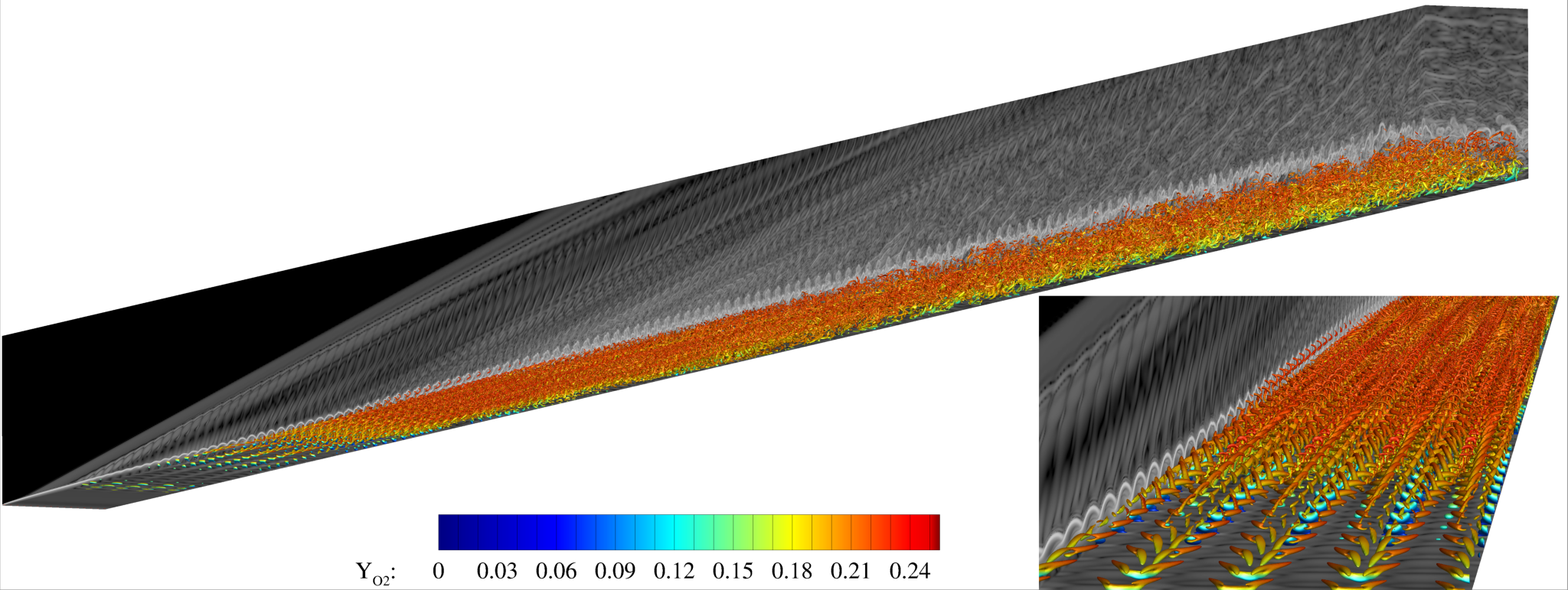}};
      \begin{scope}[x={(a.south east)},y={(a.north west)}]
        \draw [->] (0.4,0.86) -- (0.92,0.93);
        \draw [->] (0.4,0.86) -- (0.42,0.71);
        \draw [->] (0.4,0.86) -- (0.01,0.44);
        \node [align=center,rotate= 13] at (0.4,0.90) {\footnotesize Characteristic b.c.};
        \node [align=center,rotate= 13] at (0.45,0.33) {\footnotesize Isothermal wall};
        \node [align=center,rotate=-13] at (0.032,0.105) {\Huge{\color{white}\textasciitilde}};
        \node [align=center,rotate= -3] at (0.03,0.10) {\footnotesize Blasius};
        \node [align=center,rotate= -3] at (0.03,0.03) {\footnotesize profiles};
      \end{scope}
  \end{tikzpicture}
\vspace{-0.5cm}
\caption{Isosurfaces of $Q$-criterion, coloured with the local values of O$_2$ mass fraction for the CN case. The entire computational domain is displayed, along with a zoom on the laminar-to-turbulent transition region.}
\label{fig:cne1}
\end{figure*}

Transition to turbulence is induced by means of a suction-and-blowing forcing applied at the wall along a spanwise strip located close to the inflow.
In this region, a time-and-space-varying vertical velocity is prescribed, corresponding to two-dimensional waves with different amplitudes and phase angles, modulated in the $z$ direction by a cosine spanwise perturbation, used to speed-up flow tridimensionalisation and reduce the transition length:
\begin{widetext}
\begin{equation}
\frac{v_\text{wall}}{u_\infty} = f(x) A \left\{\sin(2 \pi \xi - \Omega t ) + \cos(4\chi z)\left[0.008\sin(2 \pi \xi - \Omega t+\frac{\pi}{4}) + 1 \right] \right\},
\end{equation}
\end{widetext}
where $\xi{=}(x-x_\text{forc})/L_\text{forc}$, $f(x) {=} \exp(-0.4 \xi^2)$ and $\chi{=}2 \pi/\lambda_z$. Here, $x_\text{forc}$ and $L_\text{forc}$ denote the centerline of the forcing strip and its streamwise extent, $\lambda_z$ the spanwise wavelength, $A$ the forcing amplitude and $\Omega$ the dimensional suction-and-blowing frequency.
The excitation frequencies and wavelengths are derived from the stability study of Marxen \textit{et al.} \cite{marxen2014direct}; namely,
 $\chi \delta^*_\text{in}{=}0.04$. In the present simulations, we set
$\Omega {=}1.70c_{\infty}/\delta^*_\text{in}$, $\chi {=}0.04\delta^*_\text{in}$ and $A{=}0.025$ (with $u_{\infty}$ and $c_\infty$ the free-stream velocity and speed of sound). Lastly, the forcing strip is located at $\hat{x}_\text{forc}=300$ and it extends over $L_\text{forc}=30\delta^*_\text{in}$.\\
A fully turbulent state is achieved at streamwise locations corresponding to a momentum Reynolds number, Re$_{\theta \text{,tr}}$, approximately equal to 2300 and 2600 for the FR and CN simulations, respectively. In both cases, the fully turbulent region extends over approximately the last third of the computational domain. A global view of the computational domain is given in figure~\ref{fig:cne1} showing the boundary conditions, the location of the forcing strip and the fully turbulent domain. A close-up view of the transitional zone is presented in the inset.

In the following, first- and second-order moments of various flow quantities will be presented and discussed; most of the analyses will focus on the turbulent region. For a given variable $f$, we denote with $\overline{f} = f - f'$ the standard time- and spanwise average, being $f'$ the corresponding fluctuation, whereas $\widetilde{f}= f - f''$ represents the density-weighted Favre averaging, with $f''$ the Favre fluctuation and $\widetilde{f}=  \overline{\rho f} / \overline{\rho}$.
Flow statistics are collected for more than three turnover times, after that the initial transient is evacuated and the flow has reached a statistically steady state. The sampling time interval is constant and equal to $\Delta t _\text{stats}^+= \Delta t_\text{stats} \frac{u_\tau}{l_v}= \num{5.4e-2}$, for a total of $\approx 70000$ samples.


\section{Results}
\label{sec:results}
\begin{figure}
\centering
\begin{tikzpicture}
      \node[anchor=south west,inner sep=0] (a) at (0,0) {\includegraphics[trim={0 0 0 0}, clip, width=0.90\columnwidth]{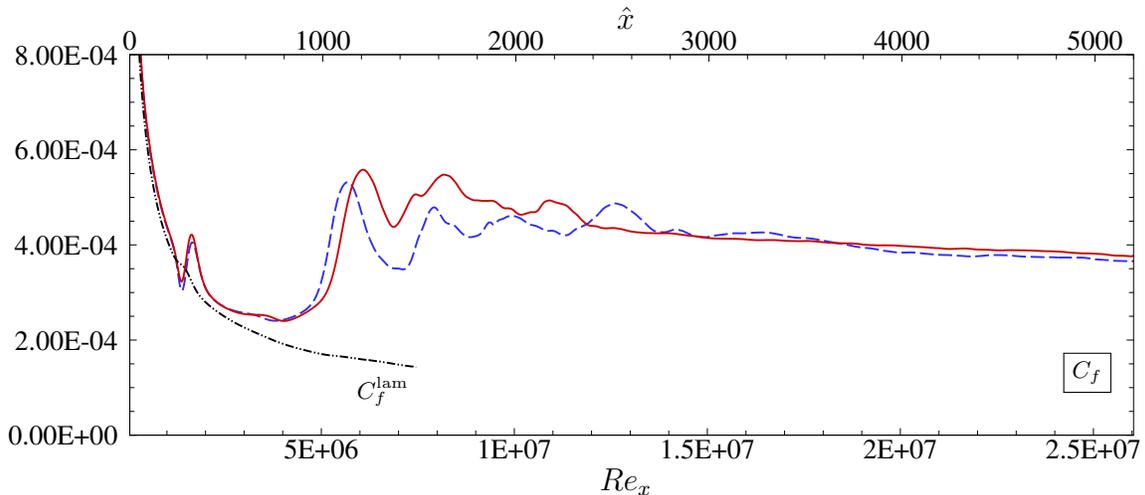}};
      \begin{scope}[x={(a.south east)},y={(a.north west)}]
        \node [align=center] at (0.53,0.02) {\large $Re_x$};
        \node [align=center] at (0.53,1.0) {\large $\hat{x}$};
        \node [align=center] at (0.91,0.25) {$\boxed{C_f}$};
        \node [align=center] at (0.33,0.21) {\small{$C_f^\text{lam}$}};
      \end{scope}
\end{tikzpicture}
\vspace{-0.5cm}
\caption{Wall distributions of the skin friction coefficient $C_f$ as a function of $Re_x$. (\protect\lsolid{red}), CN case; (\protect\ldash{blue}), FR case. The black dash-dotted lines denote the laminar correlation $C_f^\text{lam}$.}
\label{fig:cf}
\end{figure}

\begin{figure}
\centering
\begin{tikzpicture}
      \node[anchor=south west,inner sep=0] (a) at (0,0) {\includegraphics[width=0.50\columnwidth]
      {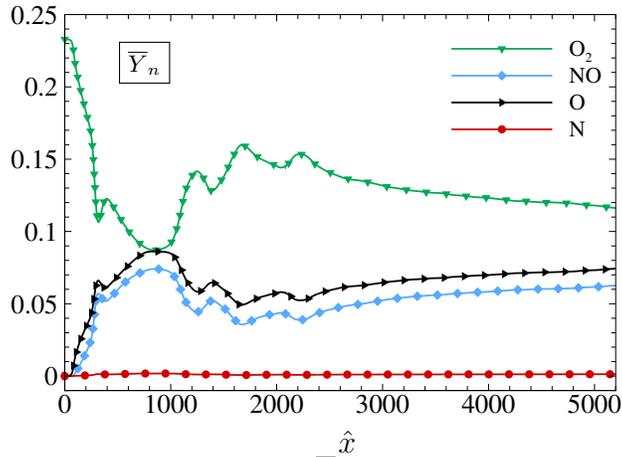}};
      \begin{scope}[x={(a.south east)},y={(a.north west)}]
        \node [align=center] at (0.55,0.00) {\large $\hat{x}$};
        \node [align=center] at (0.25,0.84) {{$\boxed{\overline{Y}_n}$}};
      \end{scope}
\end{tikzpicture}
\vspace{-0.5cm}
\caption{Streamwise evolution of the averaged mass fractions $\overline{Y}_n$ at the wall for species O$_2$, NO, O and N. $\overline{Y}_{\text{N}_2}$ is not shown being outside the $y$-axis bounds.}
\label{fig:species_stream}
\end{figure}

The streamwise evolution of selected flow properties along the wall is analyzed first. Figure~\ref{fig:cf} shows the distribution of the skin friction coefficient computed as $C_f=\frac{2\overline{\tau}_w}{\rho_\infty u_\infty^2}$. In the same figure, we also report Blasius' laminar correlation rescaled for compressible boundary layers, $C_{f}^\text{lam}=\frac{0.664}{\sqrt{Re_x}}\sqrt{\frac{\overline{\rho}_w \overline{\mu}_w}{\rho_\infty \mu_\infty}}$, which is in excellent agreement with the computed $C_f$ up to the suction-and-blowing forcing location.
The large temperatures close to the wall lead to a strong friction heating of the boundary layer, and consequently to rather small values of the scaled momentum-thickness Reynolds number $Re_\theta^\text{inc}=\frac{\mu_\infty}{\overline{\mu}_w}Re_{\theta}$, in the range $270\div 950$ for the fully turbulent region. Accordingly, the friction Reynolds number is $Re_{\tau}\approx 200$ at the rear end of the plate, as shown in Table~\ref{tab:dns_param}.
Despite the very high wall temperature, finite-rate chemistry little affects the skin friction distribution in the laminar and fully turbulent regions, where the chemical non-equilibrium and frozen flow models give very similar results. Significant quantitative discrepancies are observed only in the transition region, albeit the qualitative trends are close-by.\\
\begin{figure}
\centering
\begin{tikzpicture}
\node[anchor=south west,inner sep=0] (image) at (0,0) {
\includegraphics[width=0.96\textwidth,trim={2 2 2 2}, clip]{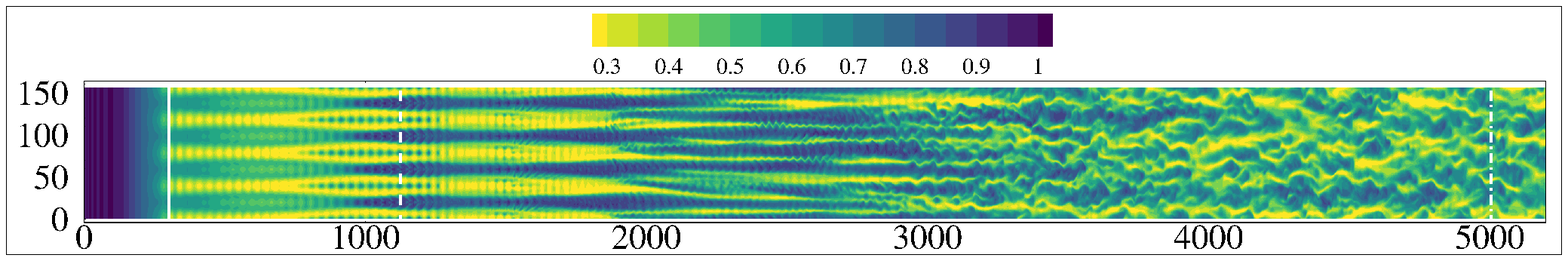}};
\begin{scope}[x={(image.south east)},y={(image.north west)}]
   \node [align=center]            at (0.32,0.88)  {\large {$\frac{u}{u_\infty}$}};
   \node [align=center, rotate=90] at (-0.02,0.45) {$z/\delta^*_\text{in}$};
\end{scope}
\end{tikzpicture}
\begin{tikzpicture}
  \node[anchor=south west,inner sep=0] (image) at (0,0) {
  \includegraphics[width=0.96\textwidth,trim={2 2 2 22}, clip]{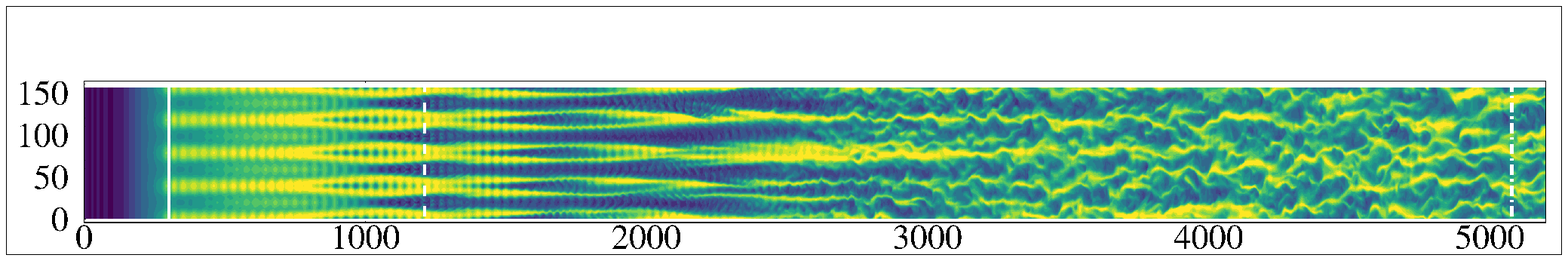}};
  \begin{scope}[x={(image.south east)},y={(image.north west)}]
    \node [align=center] at (0.54,-0.1)             {\large $\hat{x}$};
    \node [align=center, rotate=90] at (-0.02,0.55) {$z/\delta^*_\text{in}$};
  \end{scope}
\end{tikzpicture}
\vspace{-0.5cm}
\caption{Instantaneous visualizations of the normalized streamwise velocity in a $x-z$ plane for FR case (top) and CN case (bottom) at $y\approx 4.5\delta^*_\text{in}$ (corresponding to $y^+\approx 30$ at $\hat{x}=5100$). The spanwise white solid lines denote the blowing-and-suction forcing location; the dashed lines mark the position of the first peak of $C_f$ and the dash-dot lines indicate the streamwise position at which $Re_\tau=185$. The spanwise direction is stretched for better visualization.}
\label{fig:streaks}
\end{figure}
\begin{figure}
\centering
\begin{tikzpicture}
      \node[anchor=south west,inner sep=0] (a) at (0,-0.1) {\includegraphics[width=0.49\columnwidth]
      {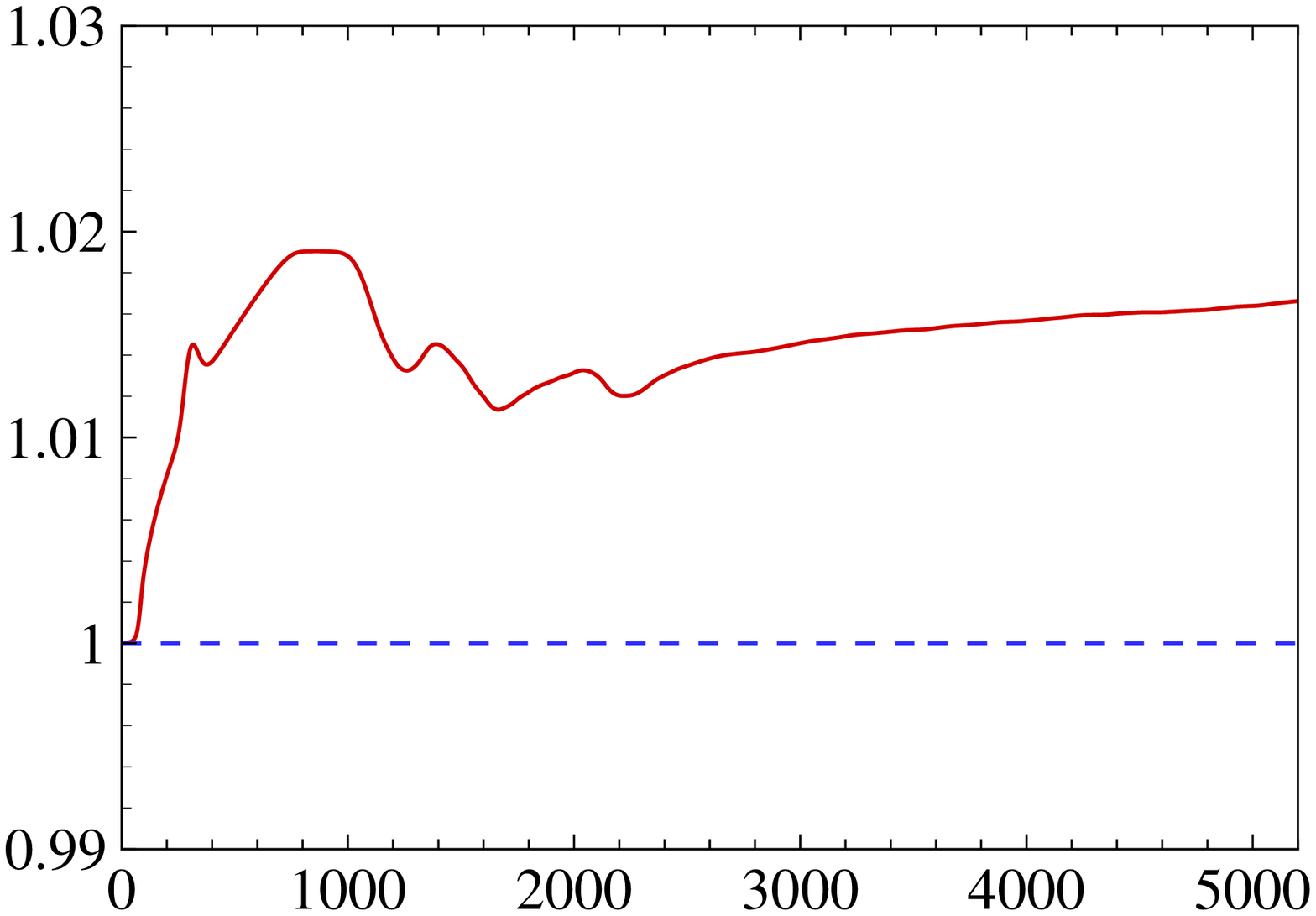}};
      \begin{scope}[x={(a.south east)},y={(a.north west)}]
        \node [align=center] at (0.02,0.95)  {(a)};
        \node [align=center] at (0.57,-0.01) {\large $\hat{x}$};
        \node [align=center] at (0.80,0.84)  {\small $\boxed{\overline{\mu}_w/\overline{\mu}_w^\text{FR}}$};
      \end{scope}
\end{tikzpicture}
\hspace{-0.5cm}
\centering
\begin{tikzpicture}
      \node[anchor=south west,inner sep=0] (a) at (0,-0.1) {\includegraphics[trim={0 0 0 0}, clip,width=0.49\columnwidth]
      {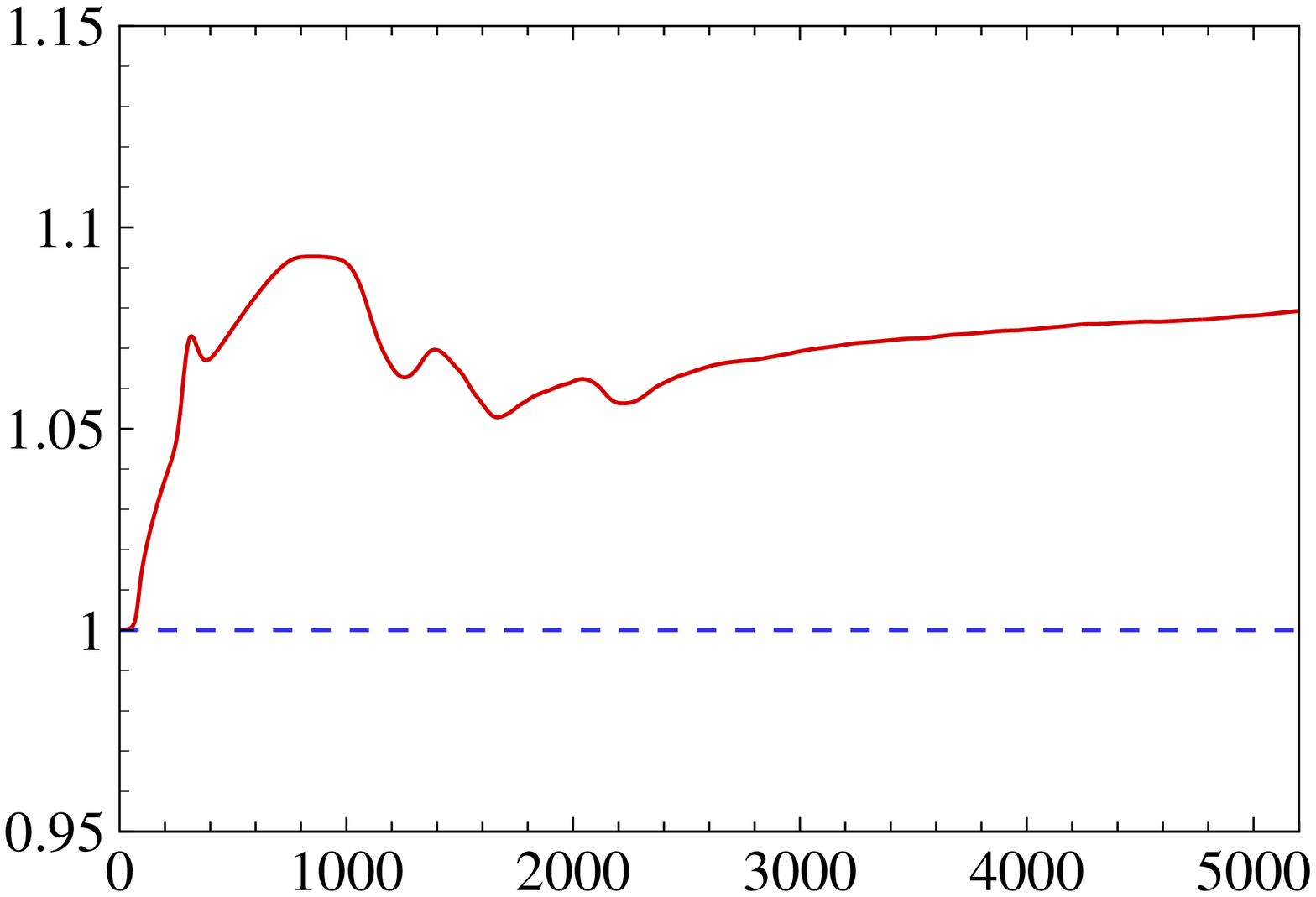}};
      \begin{scope}[x={(a.south east)},y={(a.north west)}]
        \node [align=center] at (0.02,0.95)  {(b)};
        \node [align=center] at (0.57,-0.01) {\large $\hat{x}$};
        \node [align=center] at (0.80,0.84)  {\small $\boxed{\overline{\lambda}_w/\overline{\lambda}_w^\text{FR}}$};
      \end{scope}
\end{tikzpicture}
\vspace{-0.5cm}
\caption{Streamwise evolution of averaged dynamic viscosity $\overline{\mu}$ (a) and thermal conductivity $\overline{\lambda}$ (b) at the wall, normalized with respect the constant values of the FR case. (\protect\lsolid{red}), CN case; (\protect\ldash{blue}), FR case.}
\label{fig:mulambda}
\end{figure}
Figure~\ref{fig:streaks} shows instantaneous visualizations of the streamwise velocity in a plane parallel to wall extracted at $y\approx 4.5\delta^*_\text{in}$. Due to the significant spanwise distortion introduced at the forcing point, streamwise vortical structures and streaks are generated immediately downstream, leading to a sudden increase of $C_f$ which deviates from the laminar correlation. Their complex interaction produces a sharp increase of the skin friction and leads to to instabilities growth in the region ranging from $\hat{x} \approx 1000$ to $\hat{x} \approx 2000$, characterized by sinuous streak motions and interactions, and by an oscillatory behavior of the skin friction. The flow finally bursts into turbulence at $\hat{x} \approx 3000$ and relaxes subsequently toward a fully turbulent state.
We observe that the initial overshoot is slightly delayed in the CN flow, and fine details of the transitional region are different, both in the instantaneous field and in the average quantities. \\
Finite-rate chemistry effects alter the mixture composition as the flow evolves along the plate and modify its thermo-physical properties, as shown in figure~\ref{fig:mulambda}(a)-(b) displaying the streamwise evolution of the averaged molecular viscosity and thermal conductivity along the wall. Unlike the FR case, where the transport properties are univocally fixed through the imposed wall temperature, for CN these vary according to the evolving local composition, with deviations of the order of 1.5\% for the viscosity and 7\% for the thermal conductivity. This leads in turn to slightly lower local Reynolds numbers for the chemically-reacting case. Both quantities rapidly depart from their inlet values, reaching a peak approximately at the same location where $C_f$ overshoots. The variation of the transport coefficients becomes more smooth in the fully turbulent region, where they gradually increase as a consequence of O$_2$ dissociation and atomic oxygen formation (the latter being characterized by larger diffusion coefficients).\\
Wall distributions of the averaged species mass fractions for O$_2$, NO, O and N are shown in Figure~\ref{fig:species_stream}. At the selected flow conditions, dissociation of O$_2$ (and, to a much smaller extent, of N$_2$) is quickly activated downstream of the inlet boundary, leading to sudden formation of atomic oxygen and nitric oxide. The amounts of molecular oxygen and nitrogen decrease (and, conversely, increase for the other species) until the location of maximum $C_f$ is reached. The trend is reverted from this point on, due to increased mixing with the external layers. The subsequent series of secondary peaks in the range $\hat{x} \in [1000,2300]$ are registered roughly in correspondence of those observed in the $C_f$ distribution. For $\hat{x}>2300$, the mass fractions tend to their chemical equilibrium composition values at the given $T_w$ and $\overline{p}_w$, albeit their slow variation indicates that the characteristic chemical time scales are much smaller than the local residence flow time. This is confirmed by the increasing downstream amounts of NO, an intermediate product generated by the Zel'dovich mechanism that would be subsequently consumed to produce atomic nitrogen, and by streamwise values of $Y_\text{N}$, which are almost negligible across the entire flat plate.

%

In the following, the analysis will focus on the fully turbulent region. Due to the relatively small range of Reynolds numbers covered in the turbulent portion of the computational domain, the nondimensional flow profiles do not vary substantially along the flat plate; therefore, a single streamwise station will be considered for data analysis. Specifically, we examine the station at which $Re_\tau = 185$ (see table~\ref{tab:dns_param}), corresponding to $\hat{x}=5100$ for the CN case and $\hat{x}=5045$ for the FR case. Unless specified otherwise, wall-normal profiles are plotted in inner scaling, i.e. against the wall coordinate $y^+$.

\subsection{First-order statistics}
\begin{figure}
\centering
\begin{tikzpicture}
      \node[anchor=south west,inner sep=0] (a) at (0,0) {\includegraphics[width=0.49\columnwidth]
      {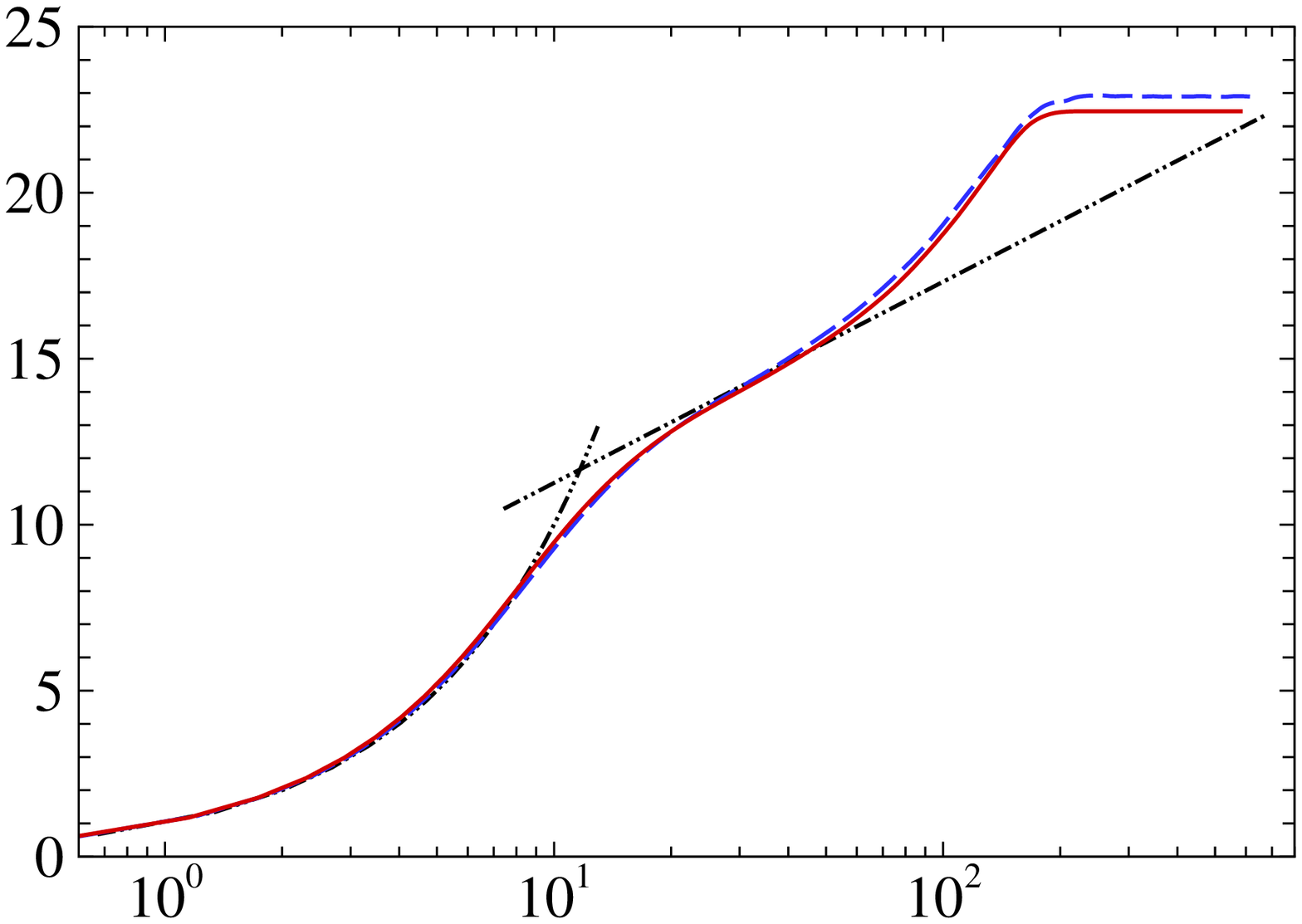}};
      \begin{scope}[x={(a.south east)},y={(a.north west)}]
        \node [align=center] at (0.04,0.95) {(a)};
        \node [align=center] at (0.55,0.03) {\large $y^+$};
        \node [align=center] at (0.25,0.84) {$\boxed{u_\text{VD}^+}$};
        \node [align=center] at (0.28,0.27) {\small{$u^+{=}y^+$}};
        \node [align=center,rotate=28] at (0.72,0.61) {\small{$u^+{=}\frac{1}{0.38}\ln(y^+)+5.2$}};
      \end{scope}
\end{tikzpicture}
\hspace{-0.3cm}
\begin{tikzpicture}
      \node[anchor=south west,inner sep=0] (a) at (0,0) {\includegraphics[width=0.49\columnwidth]
      {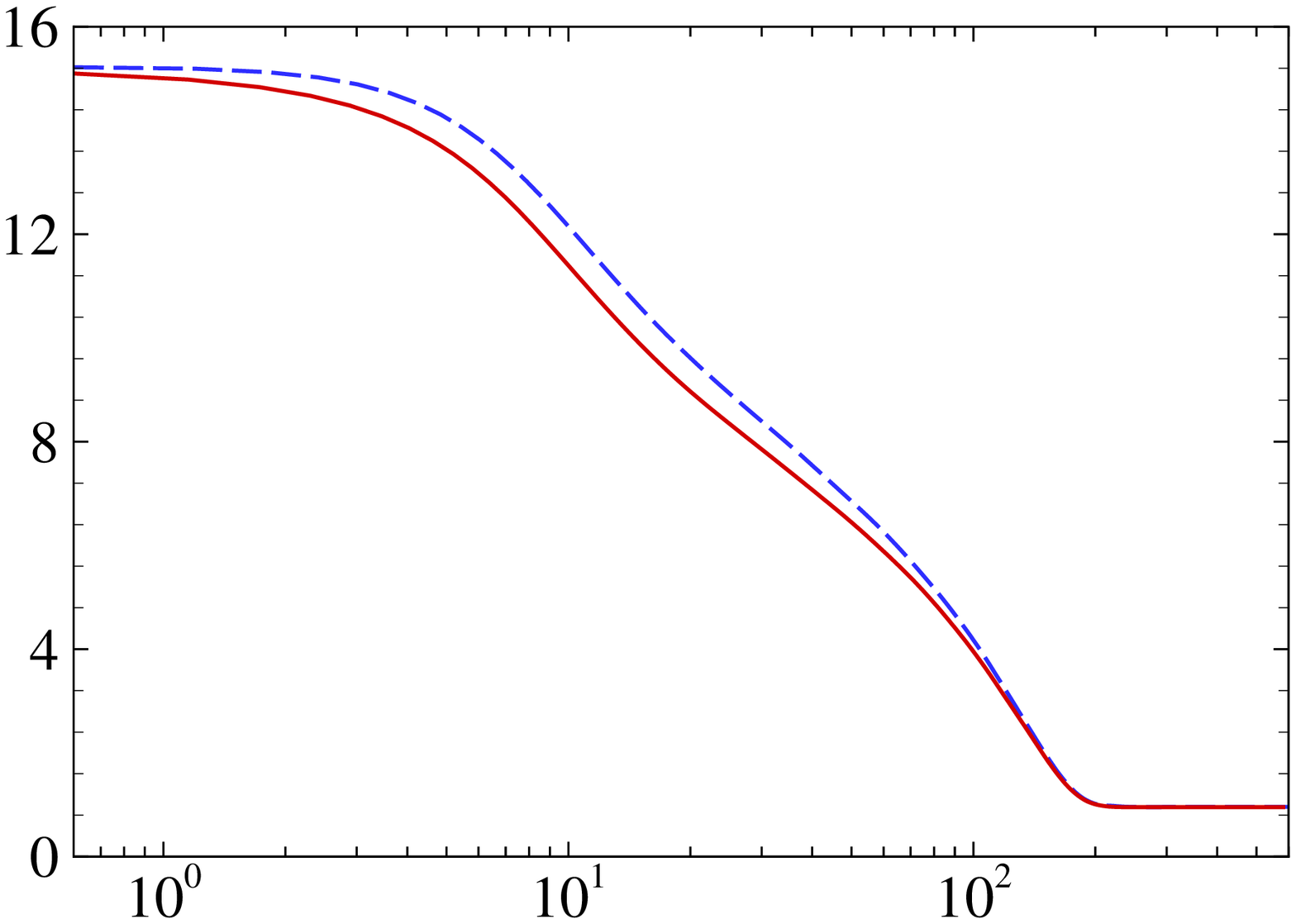}};
      \begin{scope}[x={(a.south east)},y={(a.north west)}]
        \node [align=center] at (0.04,0.95) {(b)};
        \node [align=center] at (0.55,0.03) {\large $y^+$};
        \node [align=center] at (0.81,0.84) {{$\boxed{\overline{T}/T_\infty}$}};
      \end{scope}
\end{tikzpicture}
\vspace{-0.5cm}
\caption{Wall-normal profiles of the van Driest-transformed streamwise velocity (a) and of the normalized mean temperature (b) at $Re_\tau=185$. (\protect\lsolid{red}), CN case; (\protect\ldash{blue}), FR case.}
\label{fig:wall_law}
\end{figure}
Figure~\ref{fig:wall_law}(a) displays the longitudinal velocity profiles rescaled according to the classical Van Driest transformation
\begin{equation}
u_\text{VD}^+ = \frac{1}{u_\tau} \int_0^{\overline{u}} {\sqrt{\frac{\overline{\rho}}{\overline{\rho}_w}}} \, \text{d}u,
\end{equation}
as a function of the inner wall coordinate. The van Driest scaling collapses well velocity profiles for both CN and FR cases in the linear and logarithmic zone; in the outer region the profiles are not perfectly superposed due to the different behavior of $\overline{\tau}_w$ and $\overline{\rho}_w$ which results in slightly different values of $u_\tau$. Of note, the value of the von Karman constant used in the log law is slightly smaller than the classical one ($\kappa \approx 0.38$, as suggested by Nagib \& Chauhanb\cite{nagib2008variations} and Monkewitz \cite{monkewitz2017revisiting}), resulting in a more accurate prediction of the slope of the velocity profile in the logarithmic region.
The wall-normal mean temperature profiles reported in figure~\ref{fig:wall_law}(b) show that values for the CN boundary layer are below the FR ones by approximately 5\% in the buffer region, as an expected consequence of the preponderant endothermic behavior of chemical processes. On the contrary, the mean density profiles (not shown) are not particularly affected by chemical activity.
%
\begin{figure}
\centering
\begin{tikzpicture}
      \node[anchor=south west,inner sep=0] (a) at (0,0) {\includegraphics[width=0.47\columnwidth]
      {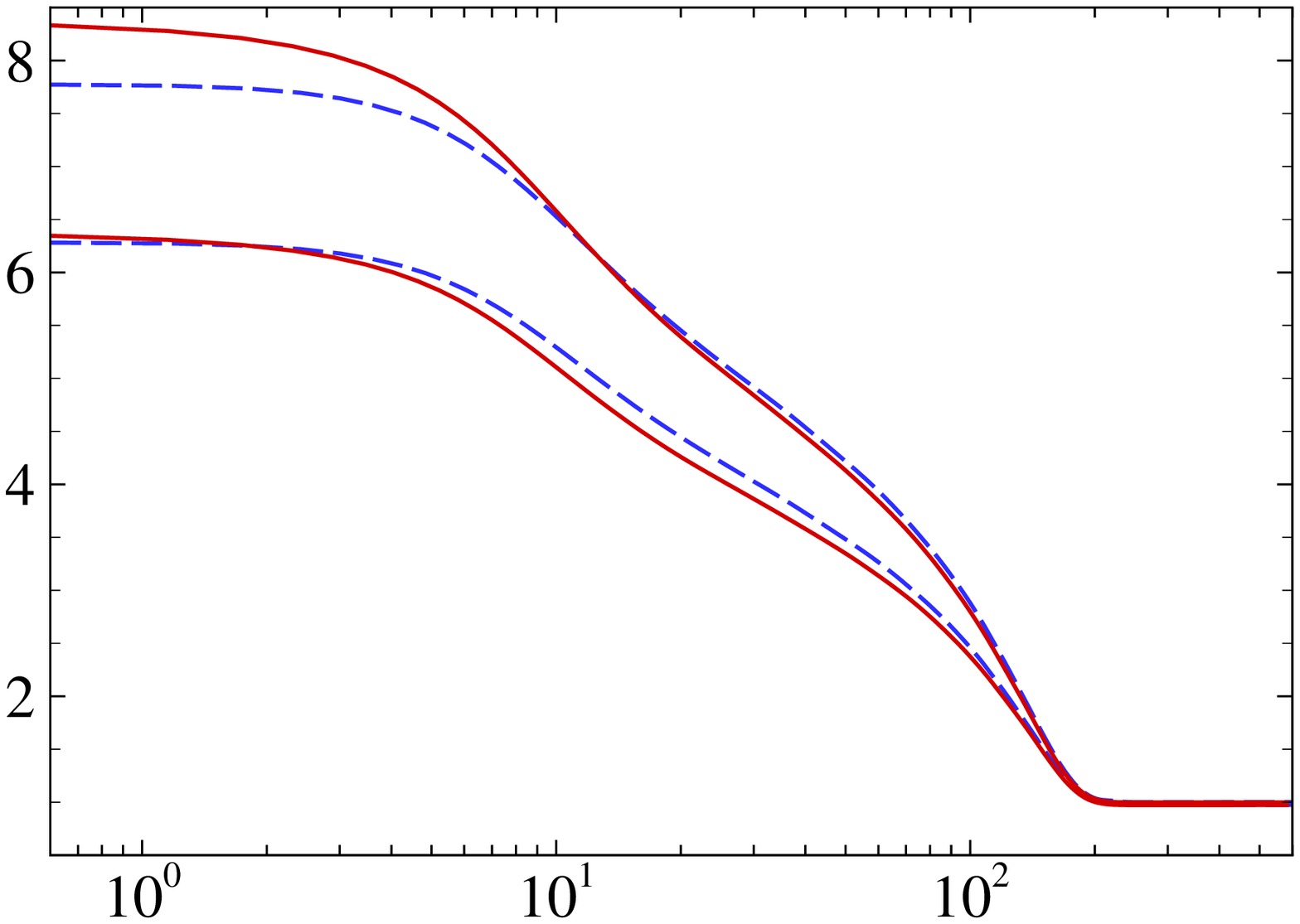}};
      \begin{scope}[x={(a.south east)},y={(a.north west)}]
        \node [align=center] at (0.03,0.94) {(a)};
        \node [align=center] at (0.55,0.03) {\large $y^+$};
        \node [align=center] at (0.56,0.85) {{$\boxed{\overline{\lambda}/\lambda_\infty}$}};
        \node [align=center] at (0.28,0.60) {{$\boxed{\overline{\mu}/\mu_\infty}$}};
      \end{scope}
\end{tikzpicture}
\hspace{-0.5cm}
\begin{tikzpicture}
      \node[anchor=south west,inner sep=0] (a) at (0,0) {\includegraphics[width=0.47\columnwidth]
      {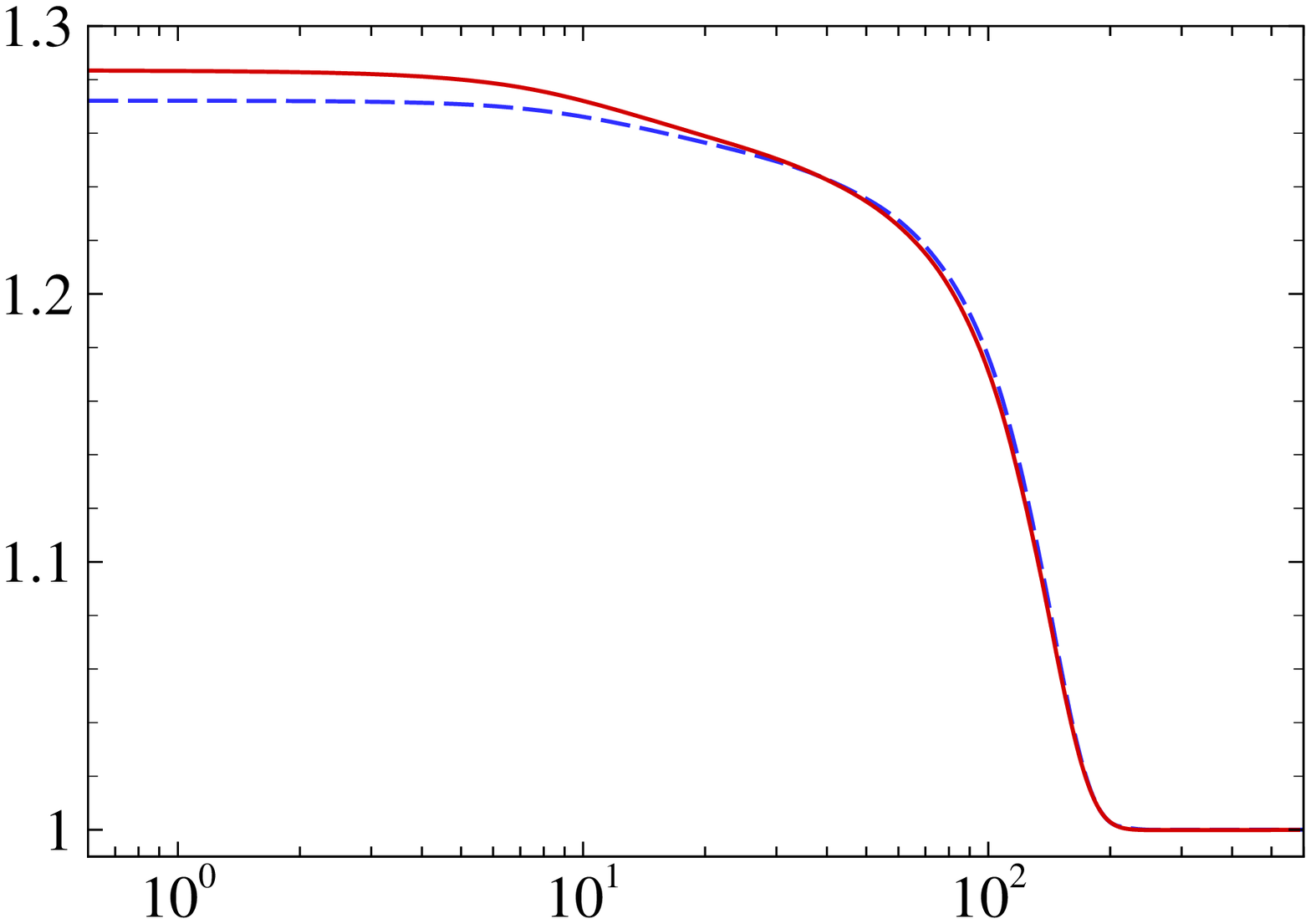}};
      \begin{scope}[x={(a.south east)},y={(a.north west)}]
        \node [align=center] at (0.03,0.94) {(b)};
        \node [align=center] at (0.55,0.03) {\large $y^+$};
        \node [align=center] at (0.30,0.25) {{$\boxed{\overline{c_p}/c_{p,\infty}}$}};
      \end{scope}
\end{tikzpicture}
\hspace{-0.5cm}
\begin{tikzpicture}
      \node[anchor=south west,inner sep=0] (a) at (0,0) {\includegraphics[width=0.47\columnwidth]
      {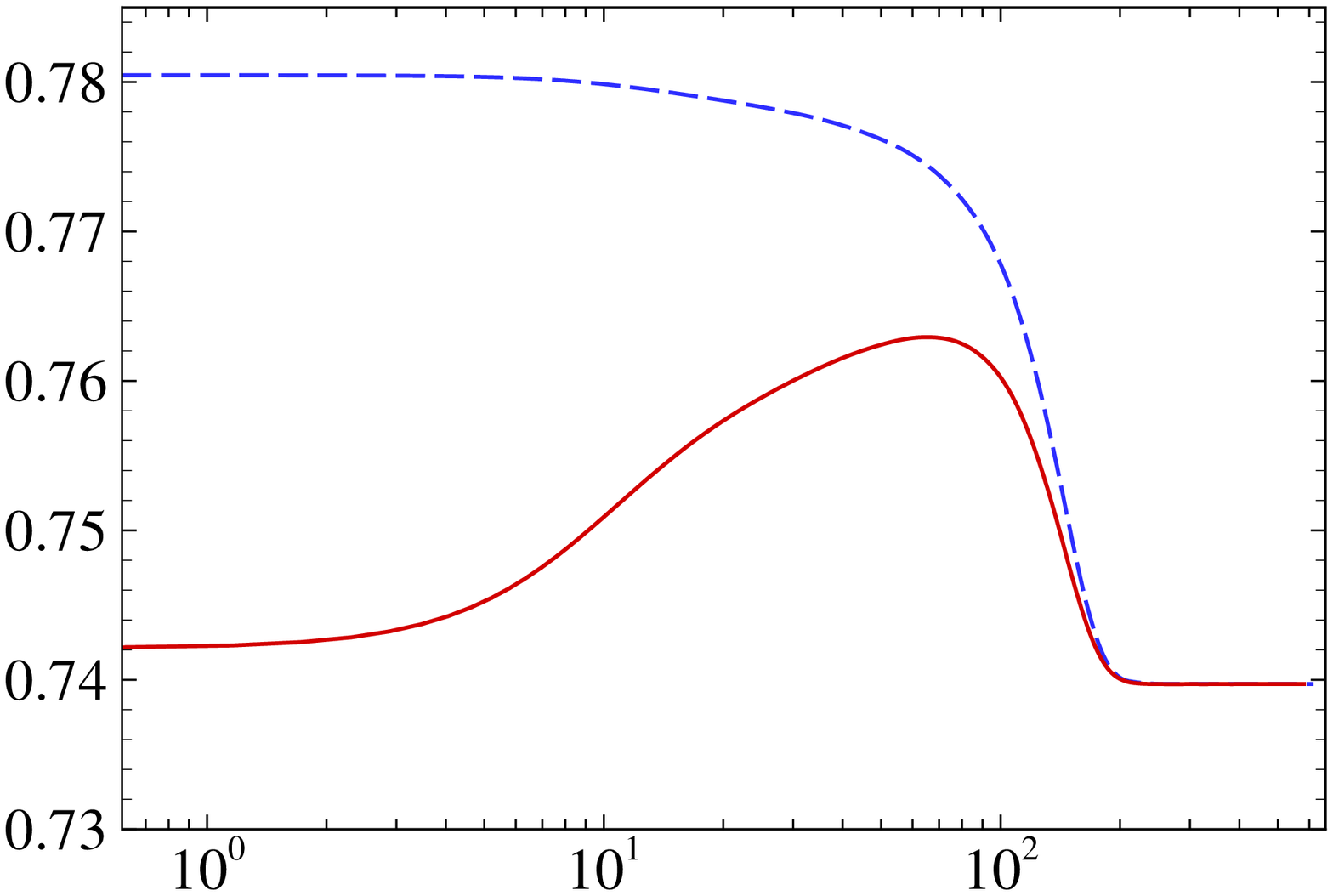}};
      \begin{scope}[x={(a.south east)},y={(a.north west)}]
        \node [align=center] at (0.03,0.94) {(c)};
        \node [align=center] at (0.55,0.03) {\large $y^+$};
        \node [align=center] at (0.85,0.84) {{$\boxed{\overline{Pr}}$}};
      \end{scope}
\end{tikzpicture}
\hspace{-0.5cm}
\begin{tikzpicture}
      \node[anchor=south west,inner sep=0] (a) at (0,0) {\includegraphics[width=0.47\columnwidth]
      {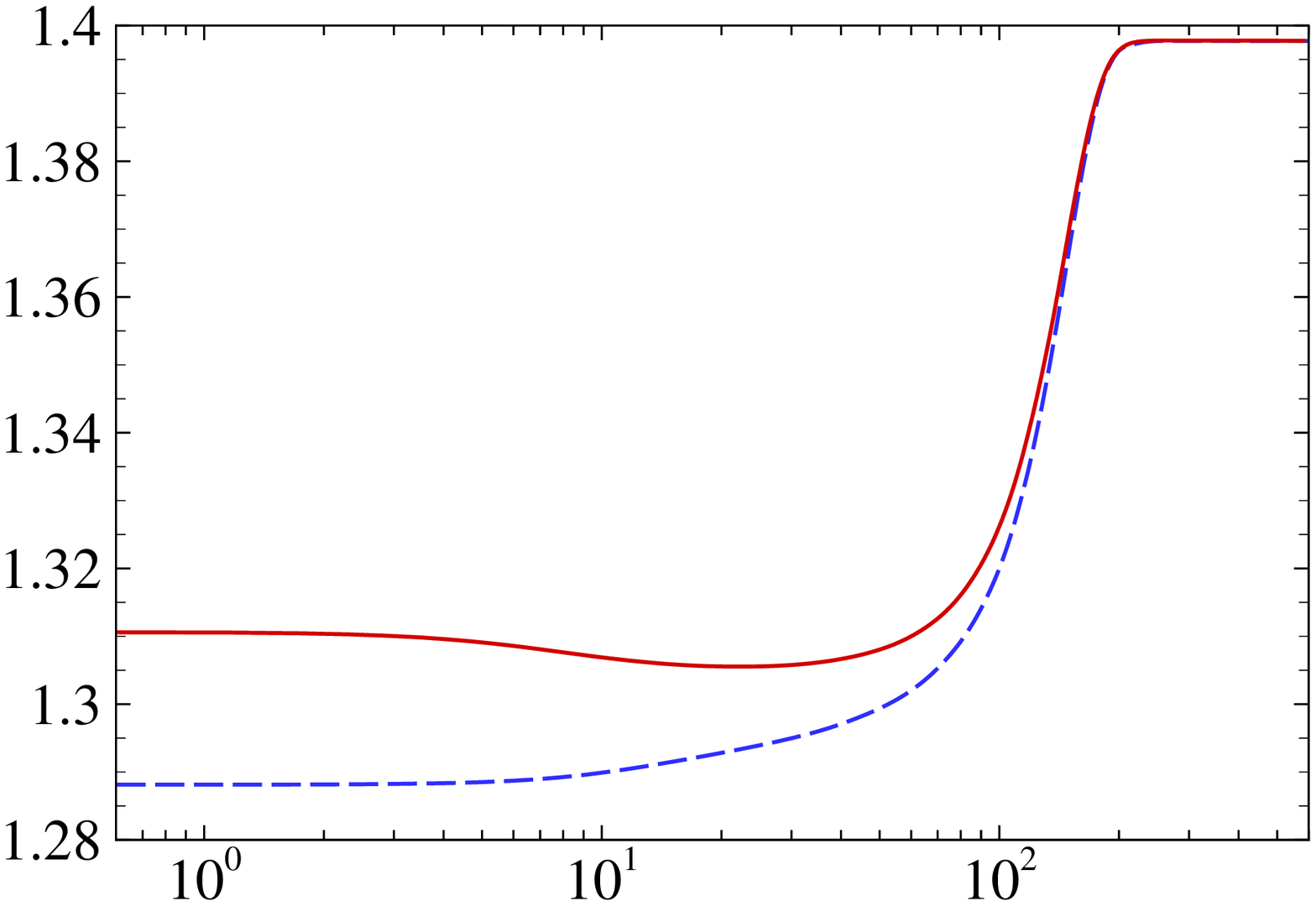}};
      \begin{scope}[x={(a.south east)},y={(a.north west)}]
        \node [align=center] at (0.03,0.94) {(d)};
        \node [align=center] at (0.55,0.03) {\large $y^+$};
        \node [align=center] at (0.25,0.84) {{$\boxed{\overline{\gamma}}$}};
      \end{scope}
\end{tikzpicture}
\vspace{-0.5cm}
\caption{Wall-normal mean profiles of viscosity and thermal conductivity (a), specific heat capacity at constant pressure (b), Prandtl number (c) and specific heat ratio $\gamma$ (d) at Re$_\tau=185$.  (\protect\lsolid{red}), CN case; (\protect\ldash{blue}), FR case.}
\label{fig:tranprop}
\end{figure}

Wall-normal average distributions of transport properties, Prandtl number $\overline{Pr}= \overline{\mu c_p/\lambda}$, specific heat capacity $\overline{c_p}$ and specific heat ratio $\overline{\gamma} = \overline{c_p/c_v}$ are shown in figure~\ref{fig:tranprop}. While chemical activity weakly affects the viscosity profile (as discussed before), larger thermal conductivity and isobaric specific heat values (by approximately 6\% and 15\%, respectively) are observed in the near wall region for the chemically reacting flow compared to the frozen one.
These deviations alter the wall-normal profile of $\overline{Pr}$ (albeit the absolute values never differ by more than 10\%), which exhibits a non-monotonic behavior, with a minimum at the wall and a peak in the logarithmic region. No major differences are observed by comparing $\overline{Pr}$ and $\overline{\mu} \, \overline{c_p}/ \overline{\lambda}$, indicating that that the peculiar behavior of  $\overline{Pr}$ is an effect of the modified mean-flow flow properties and not of the turbulent activity. Finally, changes in chemical composition also lead to minor modifications of the mean specific heat ratio $\overline{\gamma}$ in the inner region of the boundary layer. The latter varies with the temperature in both simulations, deviating from the classical value of 1.4. A larger near-wall value is observed in the reacting flow, also leading to a slightly higher average speed of sound $\overline{c} = \overline{\sqrt{\gamma R T}}$ (not shown).

\begin{figure}
\centering
\begin{tikzpicture}
      \node[anchor=south west,inner sep=0] (a) at (0,0) {\includegraphics[width=0.5\columnwidth]
      {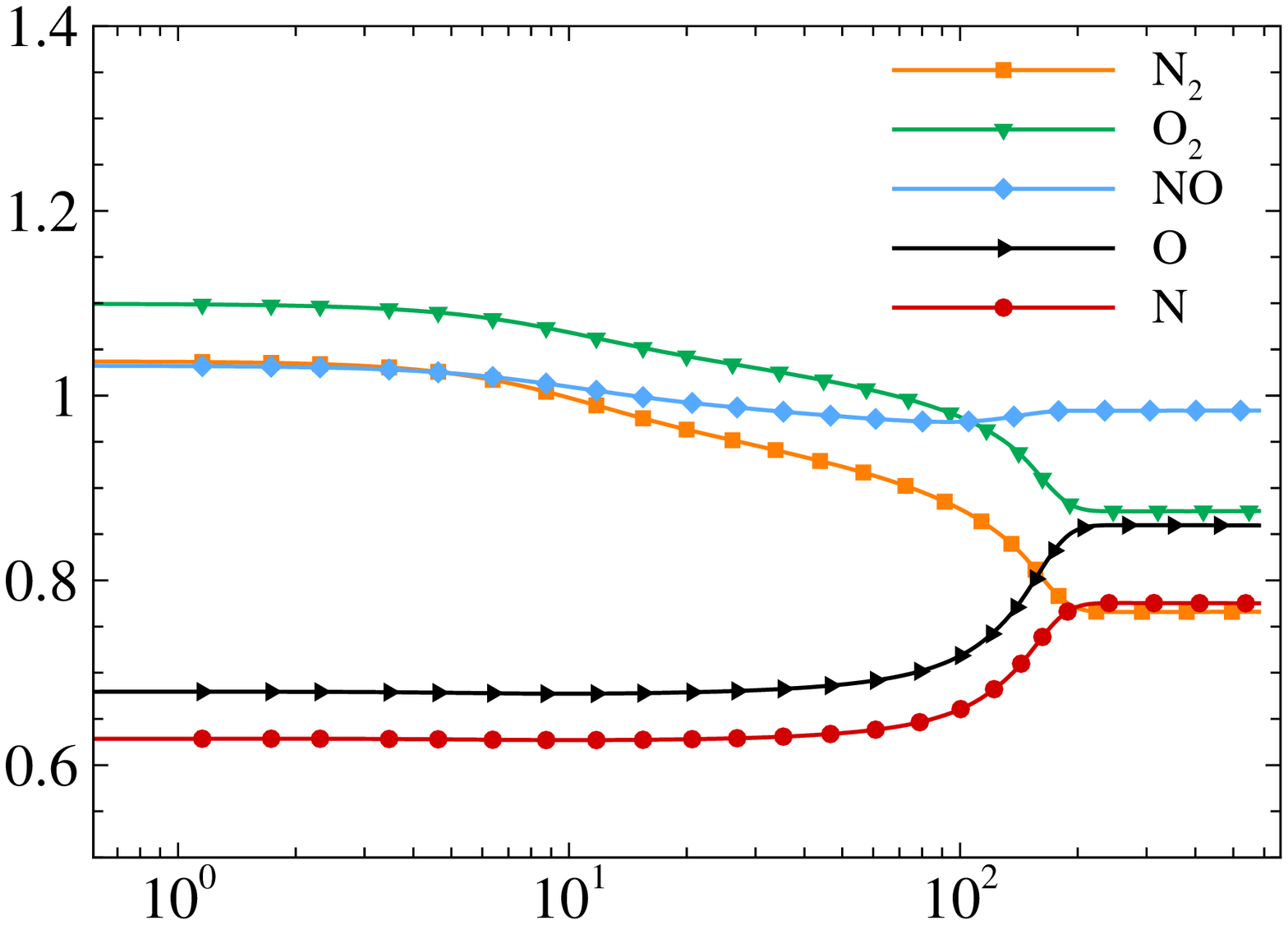}};
      \begin{scope}[x={(a.south east)},y={(a.north west)}]
        \node [align=center] at (0.05,0.94) {(a)};
        \node [align=center] at (0.55,0.03) {\large $y^+$};
        \node [align=center] at (0.84,0.20) {{$\boxed{\overline{Le_n}}$}};
      \end{scope}
\end{tikzpicture}
\hspace{-0.4cm}
\begin{tikzpicture}
      \node[anchor=south west,inner sep=0] (a) at (0,0) {\includegraphics[width=0.5\columnwidth]
      {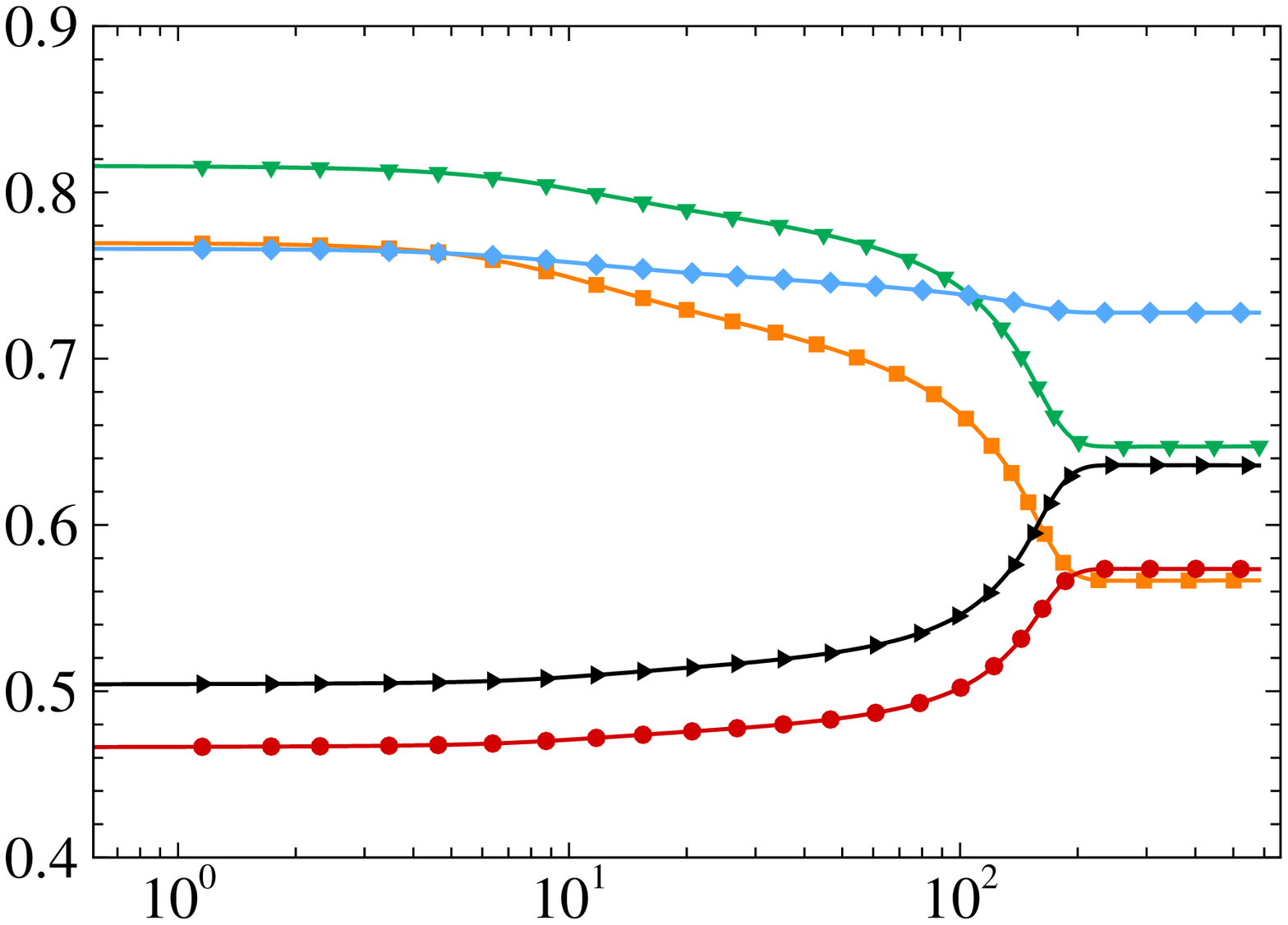}};
      \begin{scope}[x={(a.south east)},y={(a.north west)}]
        \node [align=center] at (0.05,0.94) {(b)};
        \node [align=center] at (0.55,0.03) {\large $y^+$};
        \node [align=center] at (0.83,0.21) {{$\boxed{\overline{Sc_n}}$}};
      \end{scope}
\end{tikzpicture}
\vspace{-0.25cm}
\caption{Wall-normal evolution of mean Lewis number (a) and Schmidt number (b) at $Re_\tau=185$, for the chemically-reacting simulation, in inner scaling.}
\label{fig:lewis_schmidt}
\end{figure}

The next series of figures focuses on the behavior of the chemically reacting mixture.
Figure~\ref{fig:lewis_schmidt} shows the wall-normal evolution of the average Lewis and Schmidt numbers for each species, $\overline{Le_n} =\overline{\lambda/\rho c_p D_n}$ and $\overline{Sc_n} = \overline{\mu/\rho D_n}$. Larger Lewis numbers indicate that thermal diffusivity effects tend to dominate mass diffusivity; similarly, for higher Schmidt numbers diffusion of momentum dominates mass diffusion. N and O, characterized by higher diffusion coefficients than other species, also exhibit smaller $\overline{Le}$ and $\overline{Sc}$ numbers in the reacting near-wall layer, meaning that the two atoms diffuse faster into the rest of the mixture. Moving towards the edge of the boundary layer, both non-dimensional coefficients decrease for N$_2$ and O$_2$, and increase for the atoms.
Overall, variations of $\overline{Le}$ and $\overline{Sc}$ across the boundary layer are of the order of 20\% for all species except NO, for which the profiles are almost constant; moreover, the local fluctuations amount to less than 1\% of the corresponding mean values. We conclude therefore that the use of simplified transport models based on the assumption of constant $Le$ and $Sc$ numbers constitute an acceptable first approximation, at least for thermodynamic conditions similar to those currently under investigation.

\begin{figure}
\centering
\begin{tikzpicture}
      \node[anchor=south west,inner sep=0] (a) at (0,0) {\includegraphics[trim={0 2 0 2}, clip, width=0.48\columnwidth]
      {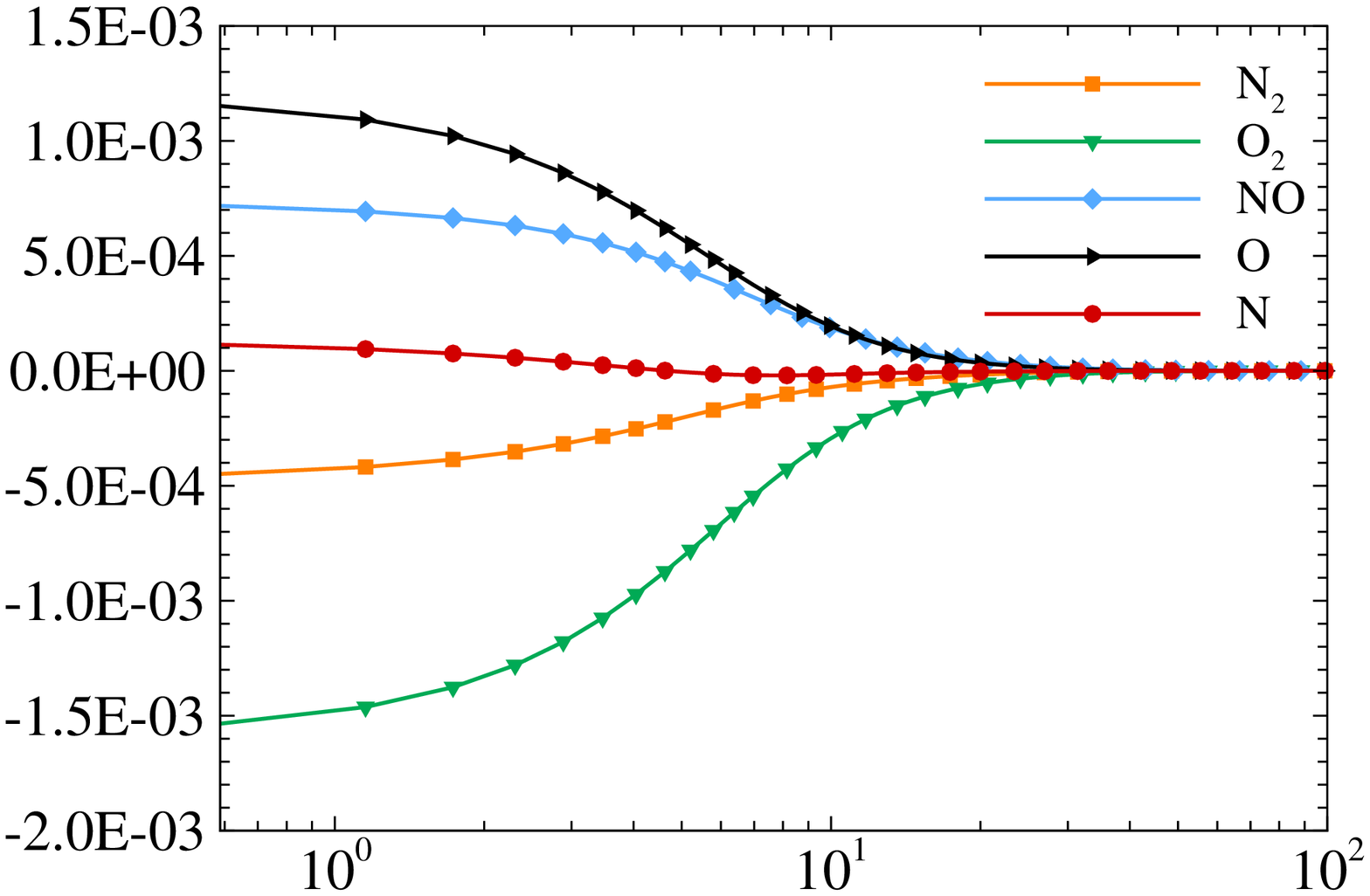}};
      \begin{scope}[x={(a.south east)},y={(a.north west)}]
        \node [align=center] at (0.01,0.95) {(a)};
        \node [align=center] at (0.55,0.00) {\large $y^+$};
        \node [align=center] at (0.87,0.24) {{$\boxed{ Da_n}$}};
      \end{scope}
\end{tikzpicture}
\hspace{-0.30cm}
\begin{tikzpicture}
      \node[anchor=south west,inner sep=0] (a) at (0,0) {\includegraphics[trim={0 2 2 2}, clip,width=0.48\columnwidth]
      {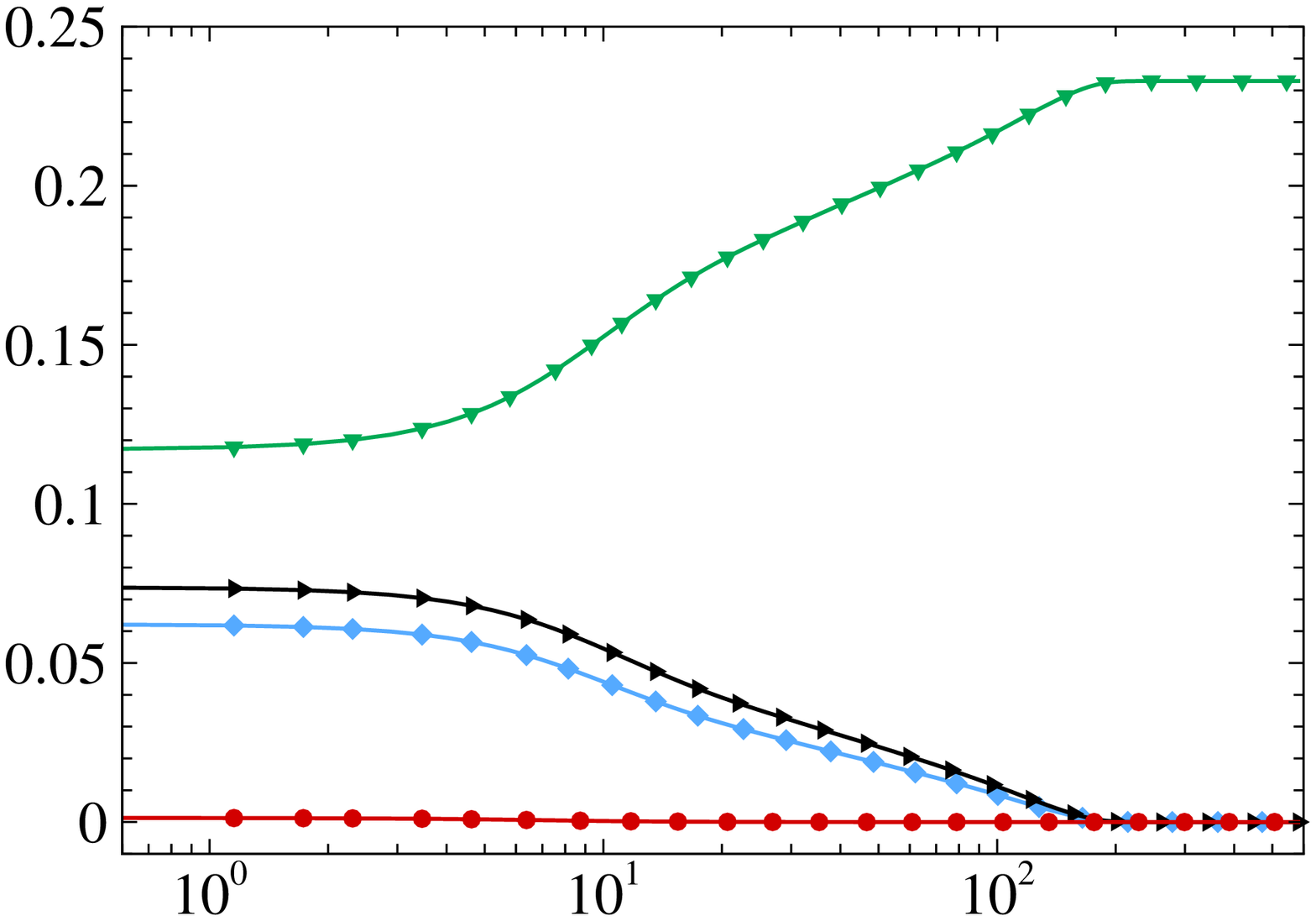}};
      \begin{scope}[x={(a.south east)},y={(a.north west)}]
        \node [align=center] at (0.03,0.95) {(b)};
        \node [align=center] at (0.55,0.00) {\large $y^+$};
        \node [align=center] at (0.84,0.25) {{$\boxed{\overline{Y_n}}$}};
      \end{scope}
\end{tikzpicture}
\hspace{-0.30cm}
\begin{tikzpicture}
      \node[anchor=south west,inner sep=0] (a) at (0,0) {\includegraphics[width=0.50\columnwidth]
      {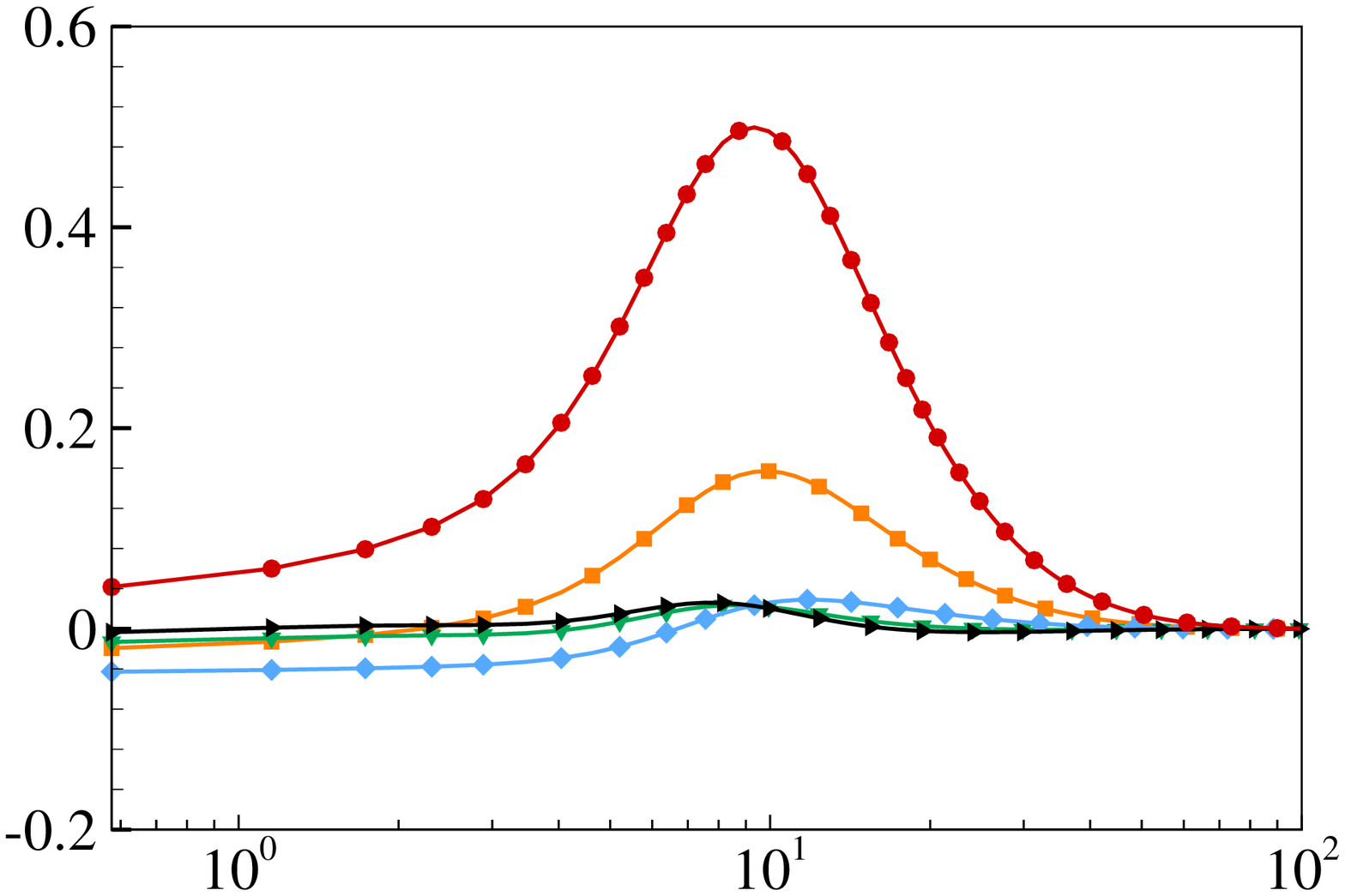}};
      \begin{scope}[x={(a.south east)},y={(a.north west)}]
        \node [align=center] at (0.05,0.94) {(c)};
        \node [align=center] at (0.57,0.02) {\large $y^+$};
        \node [align=center] at (0.30,0.84) {$\boxed{Da_n^I}$};
      \end{scope}
\end{tikzpicture}
\caption{Wall-normal profiles of the average Damk\"{o}hler number (a), species mass fractions (b), and species Damk\"{o}hler number  $Da_n^I$ at $Re_\tau=185$. In panel (b), N$_2$ is not shown, being $Y_{\text{N}_2}>0.25$.}
\label{fig:species_mean}
\end{figure}
The average profiles of the $n$-th species Damk\"{o}hler number $Da_n=\overline{\dot{\omega}_n/ \rho} \, \overline{\mu}_w /\overline{\tau}_w $ and mass fraction $\overline{Y_n}$ are reported in figure~\ref{fig:species_mean}(a) and \ref{fig:species_mean}(b), respectively. The magnitude of $Da_n$ represents the ratio of the flow characteristic time scale to the chemical time scale, while its positive or negative sign indicates production or depletion of a species, respectively. Note that the present definition of $Da_n$ is based on the characteristic time scale of the inner boundary layer region. The small values observed in figure~\ref{fig:species_mean}(a) imply that chemical reactions are characterized by much longer time scales than the residence time of the flow structures; in other terms, the flow is never too far from frozen-chemistry conditions, even in the near-wall region, justifying the relatively small differences registered between the FR and CN cases. Mass fraction profiles confirm that most of the chemical activity is localized in the viscous sublayer, where O$_2$ dissociates at a high rate. To quantify the strength of turbulence/chemistry interactions, i.e. the influence of temperature and species mass fractions fluctuations on species production rates, profiles of the species interaction Damk\"{o}hler number\cite{duan2011assessment} are reported in figure \ref{fig:species_mean}(c). This quantity, defined as
\begin{equation}
Da^I_n = \frac{\overline{\omega_n(T,\rho_n)} - \omega_n(\overline{T}, \overline{\rho_n})}{\overline{\omega_n(T,\rho_n)}_w},
\end{equation}
represents a measure of chemical production due to the turbulent fluctuations in Arrhenius' law, i.e. of the difference $\overline{\omega_n(T,\rho_n)} - \omega_n(\overline{T}, \overline{\rho_n})\neq 0$ due to the nonlinearity of $\omega_n$.
$Da_n^I$ takes rather small values across most of the boundary layer, except in the buffer region, where turbulent fluctuations are large enough to generate a significant contribution. Coherently with figure~\ref{fig:species_mean}(a), the species most affected are the ones with the smaller $Da_n$ (i.e., N and N$_2$), because of their larger sensitivity to temperature and density fluctuations deriving from turbulent motions. Conversely, the most chemically-active species are characterized by small values of $Da^I_n$, indicating that turbulence/chemistry interactions are somewhat contained in the flow under investigation. A direct consequence is that the evolution of the dynamic quantities is mostly decoupled from that of the chemical species, as discussed in the following by examination of turbulence intensities and energy spectra.
%
%

%

\subsection{Second-order statistics}
\begin{figure}
\begin{tikzpicture}
      \node[anchor=south west,inner sep=0] (a) at (0,0) {\includegraphics[width=0.49\columnwidth]
      {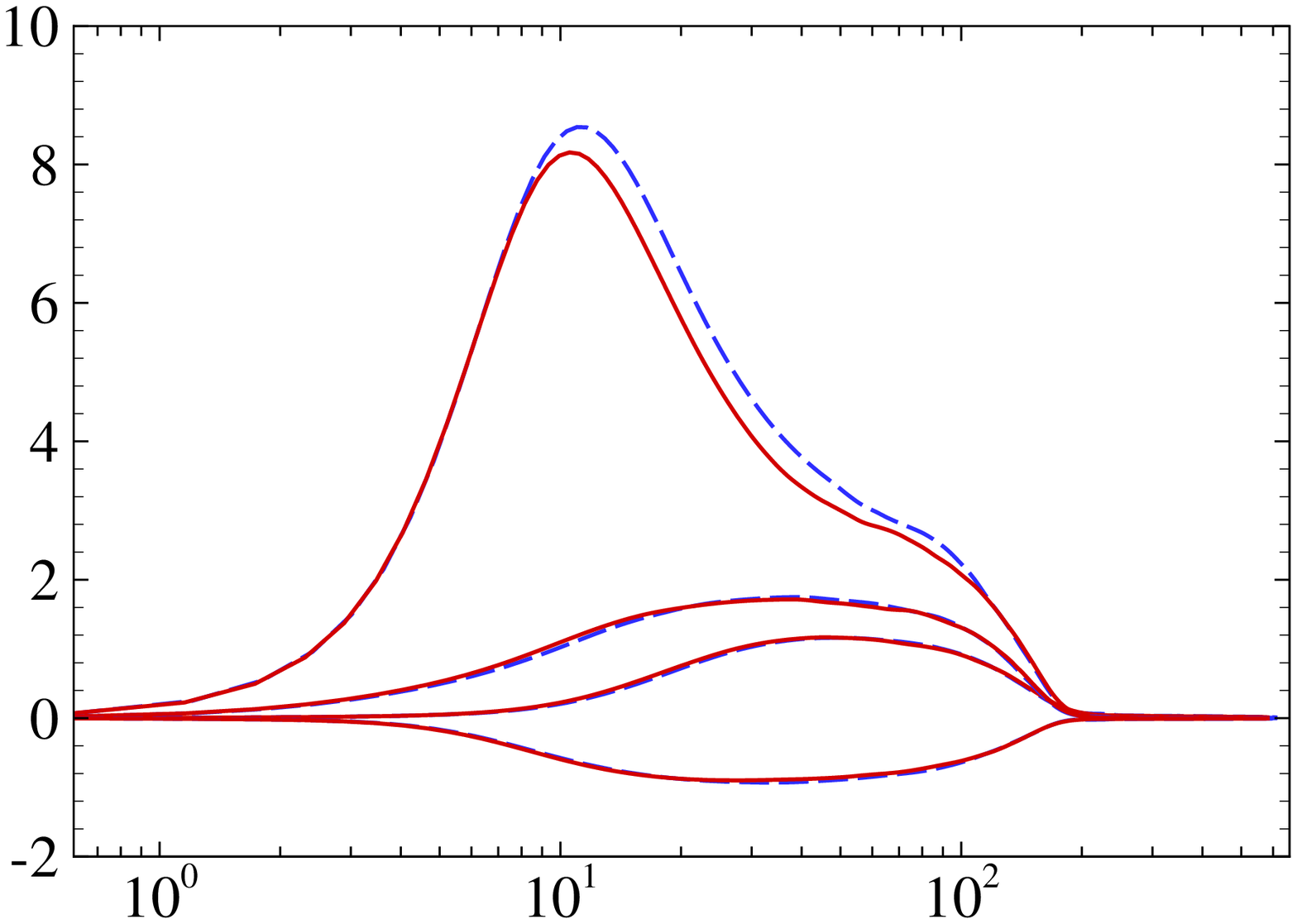}};
      \begin{scope}[x={(a.south east)},y={(a.north west)}]
        \node [align=center] at (0.03,0.95) {(a)};
        \node [align=center] at (0.55,0.03) {\large $y^+$};
        \node [align=center] at (0.56,0.72) {$u$};
        \node [align=center] at (0.48,0.38) {$w$};
        \node [align=center] at (0.59,0.28) {$v$};
        \node [align=center] at (0.80,0.21) {$-uv$};
        \node [align=center] at (0.83,0.83) {\scriptsize $\boxed{\frac{\overline{\rho u_i'' u_j''}}{\overline{\rho}_w u_{\tau}^2}}$};
      \end{scope}
\end{tikzpicture}
\hspace{-0.3cm}
\begin{tikzpicture}
      \node[anchor=south west,inner sep=0] (a) at (0,0) {\includegraphics[width=0.49\columnwidth]
      {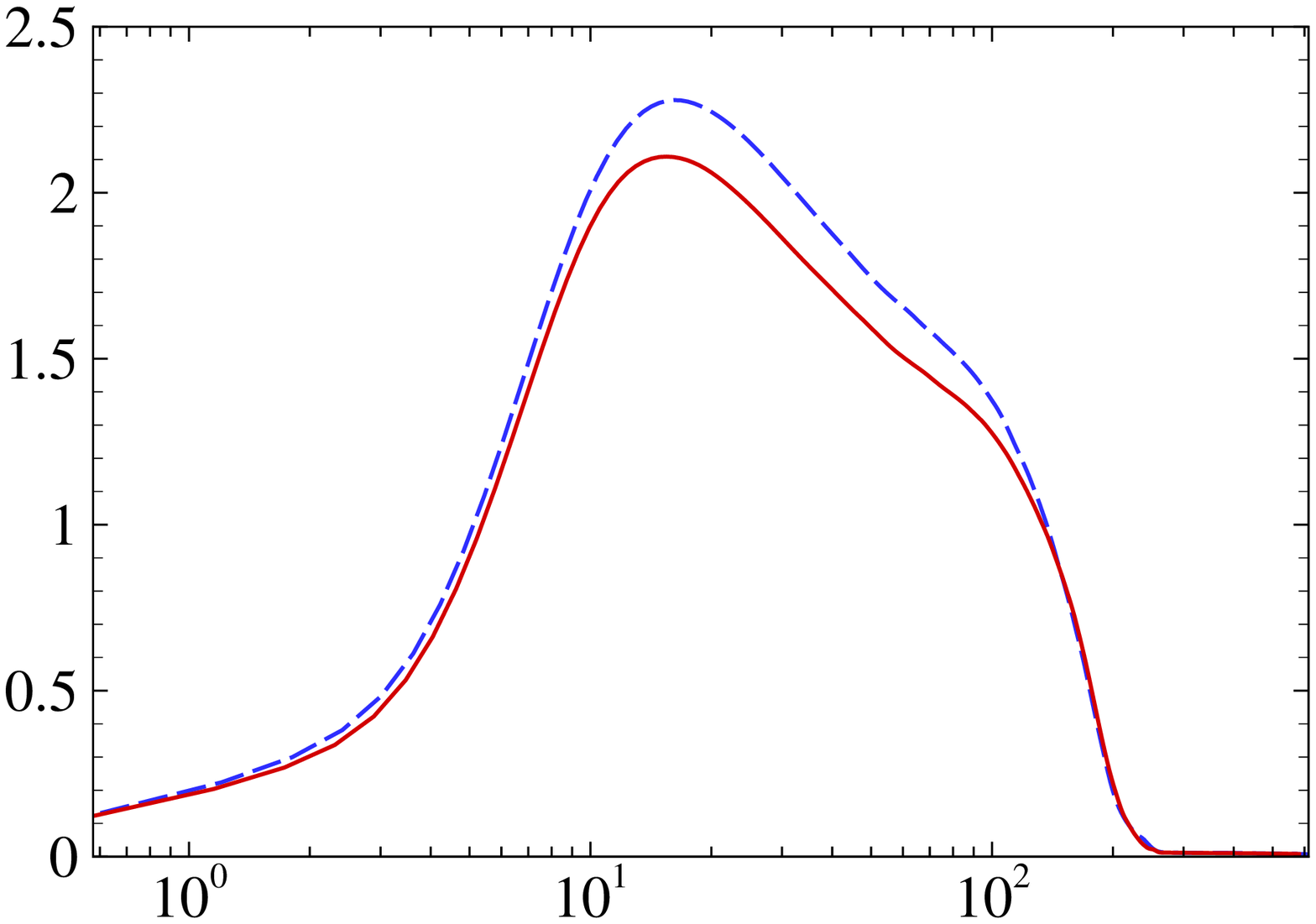}};
      \begin{scope}[x={(a.south east)},y={(a.north west)}]
        \node [align=center] at (0.03,0.95) {(b)};
        \node [align=center] at (0.55,0.03) {\large $y^+$};
        \node [align=center] at (0.80,0.85) {{$\boxed{ \sqrt{\overline{T'^2}}/ T_\infty}$}};
      \end{scope}
\end{tikzpicture}
\vspace{-0.5cm}
\caption{Wall-normal profiles of Reynolds stresses (a) and of the normalized r.m.s. temperature (b), at $Re_\tau=185$.
 (\protect\lsolid{red}), CN case; (\protect\ldash{blue}), FR case.}
\label{fig:reynolds_stresses}
\end{figure}

\begin{figure}
\centering
\begin{tikzpicture}
      \node[anchor=south west,inner sep=0] (a) at (0,0) {\includegraphics[width=0.49\columnwidth]
      {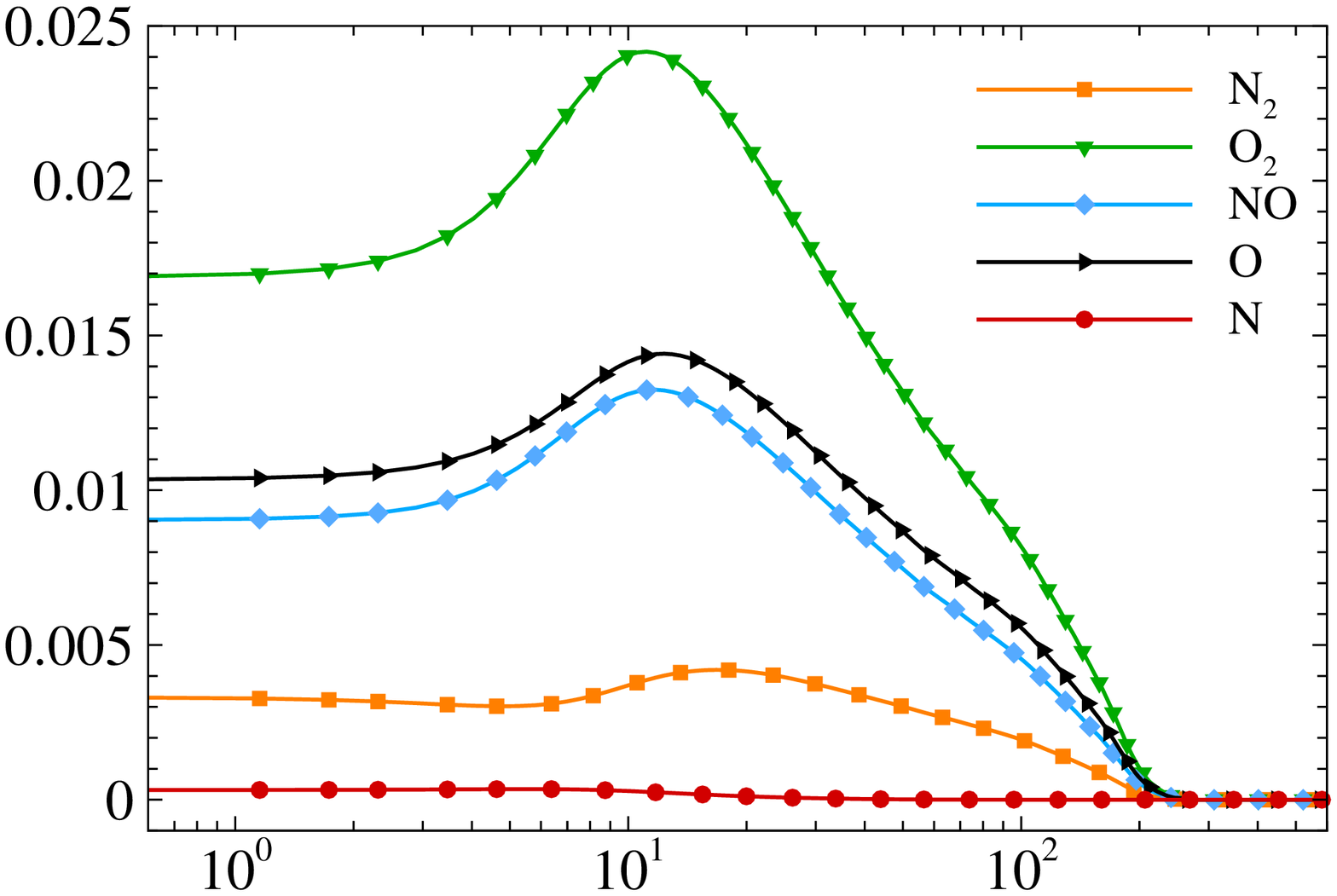}};
      \begin{scope}[x={(a.south east)},y={(a.north west)}]
        \node [align=center] at (0.55,0.03) {\large $y^+$};
        \node [align=center] at (0.24,0.83) {{$\boxed{\sqrt{\overline{Y'^2_n}}}$}};
      \end{scope}
\end{tikzpicture}
\vspace{-0.5cm}
\caption{Wall-normal profiles of r.m.s. mass fractions (b) at Re$_{\tau}=185$ for the CN case.}
\label{fig:mass_fraction_rms}
\end{figure}

To further analyze the effect of finite-rate chemistry on turbulent quantities, the wall-normal profiles of Favre-average-based Reynolds stresses are reported in figure~\ref{fig:reynolds_stresses}(a). The frozen-flow solution is also reported for reference on the same figure. We observe that finite-rate chemical reactions in the near-wall region partly drain energy from the turbulent fluctuations, which reach a slightly lower peak value of $\displaystyle\overline{\rho u_i'' u_j''}/\overline{\rho}_w u_{\tau}$ in the region of maximum turbulent production. No significant effects are observed for the other Reynolds stress components; a similar behavior has been also registered by Duan \textit{et al.}\cite{duan2011direct4}.
Contrary-wise, chemical activity does have an effect on the root mean square (r.m.s.) temperature fluctuations $\sqrt{\overline{T'^2}}/ T_\infty$ (reported in figure~\ref{fig:reynolds_stresses}b), which exhibit a peak value reduced by approximately 10\% with respect to the non-reacting case. Interestingly, the location of the largest temperature fluctuations corresponds to a peak in the species fluctuating mass fractions, also located in the buffer layer (figure~\ref{fig:mass_fraction_rms}), although the near-wall region is hotter. \\

\begin{figure}
\centering
\begin{tikzpicture}
      \node[anchor=south west,inner sep=0] (a) at (0,0) {\includegraphics[width=0.48\columnwidth]
      {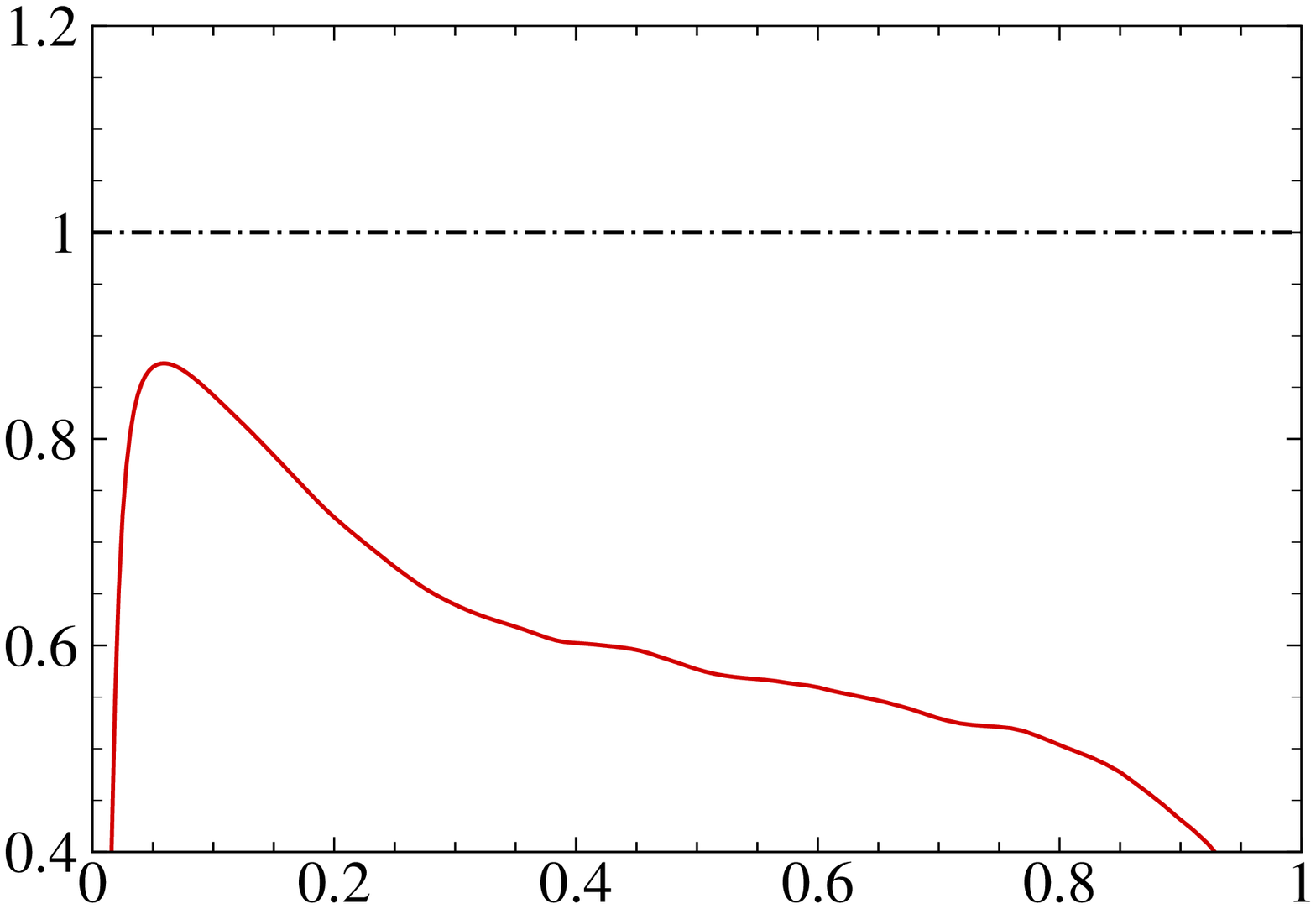}};
      \begin{scope}[x={(a.south east)},y={(a.north west)}]
        \node [align=center] at (0.02,0.94) {(a)};
        \node [align=center] at (0.55,0.01) {\large $y/\delta$};
        \node [align=center] at (0.83,0.84) {{$\boxed{-R_{u''T''}}$}};
      \end{scope}
\end{tikzpicture}
\hspace{-0.5cm}
\begin{tikzpicture}
      \node[anchor=south west,inner sep=0] (a) at (0,0) {\includegraphics[width=0.48\columnwidth]
      {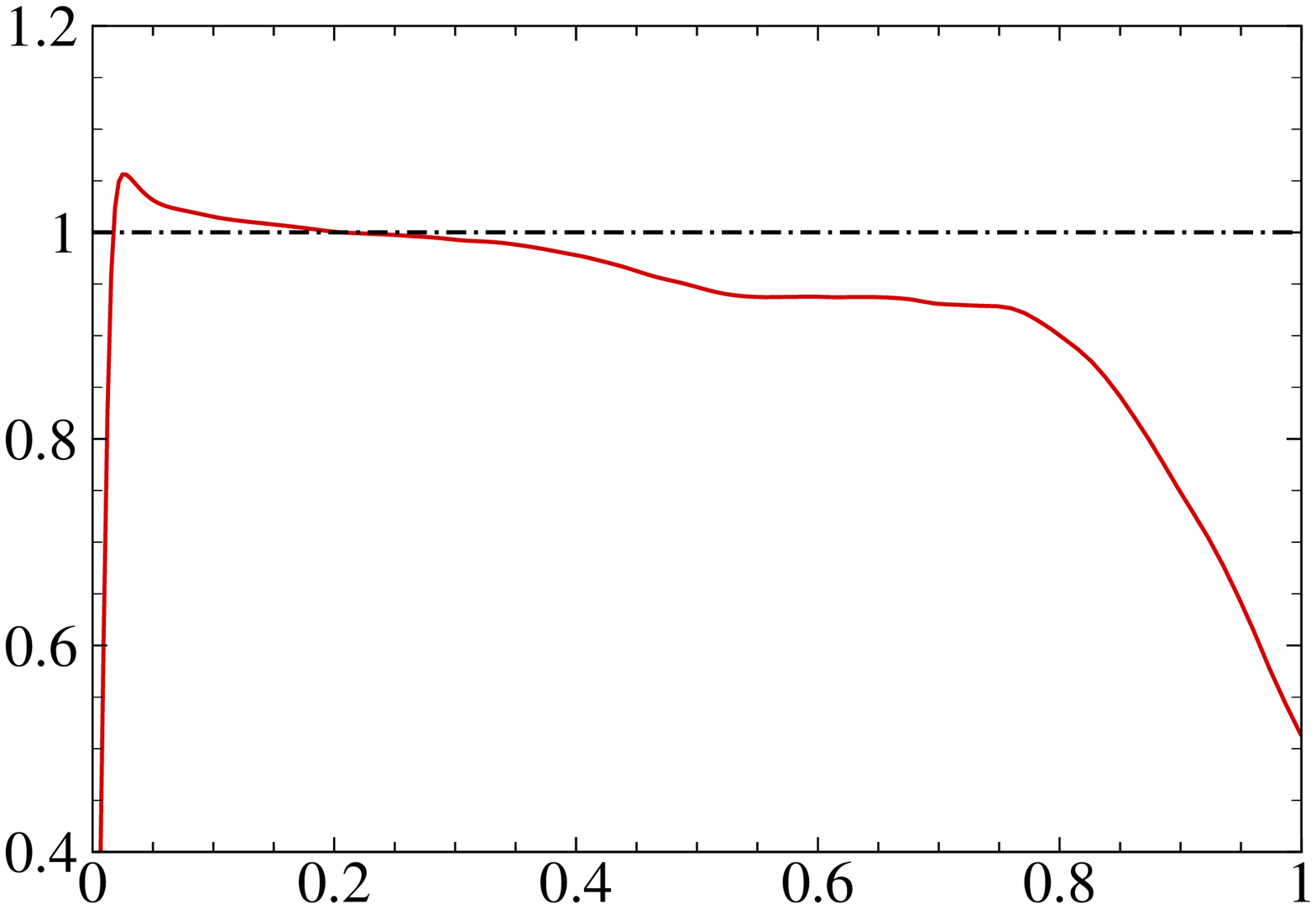}};
      \begin{scope}[x={(a.south east)},y={(a.north west)}]
        \node [align=center] at (0.02,0.94) {(b)};
        \node [align=center] at (0.55,0.01) {\large $y/\delta$};
        \node [align=center] at (0.36,0.28) {$\boxed{-R_{u''T''} + \frac{\overline{T_0''^2}}{2\overline{T''^2}}}$};
      \end{scope}
\end{tikzpicture}
\vspace{-0.25cm}
\caption{ Correlation coefficient between $u''$ and $T''$ without total temperature correction (a) and with total temperature correction (b), at Re$_{\tau}=185$. (\protect\lsolid{red}), CN case; (\protect\ldashdot{black}), SRA estimation.}
\label{fig:rey_analogy1}
\end{figure}

\begin{figure}
\centering
\begin{tikzpicture}
      \node[anchor=south west,inner sep=0] (a) at (0,0) {\includegraphics[width=0.48\columnwidth]{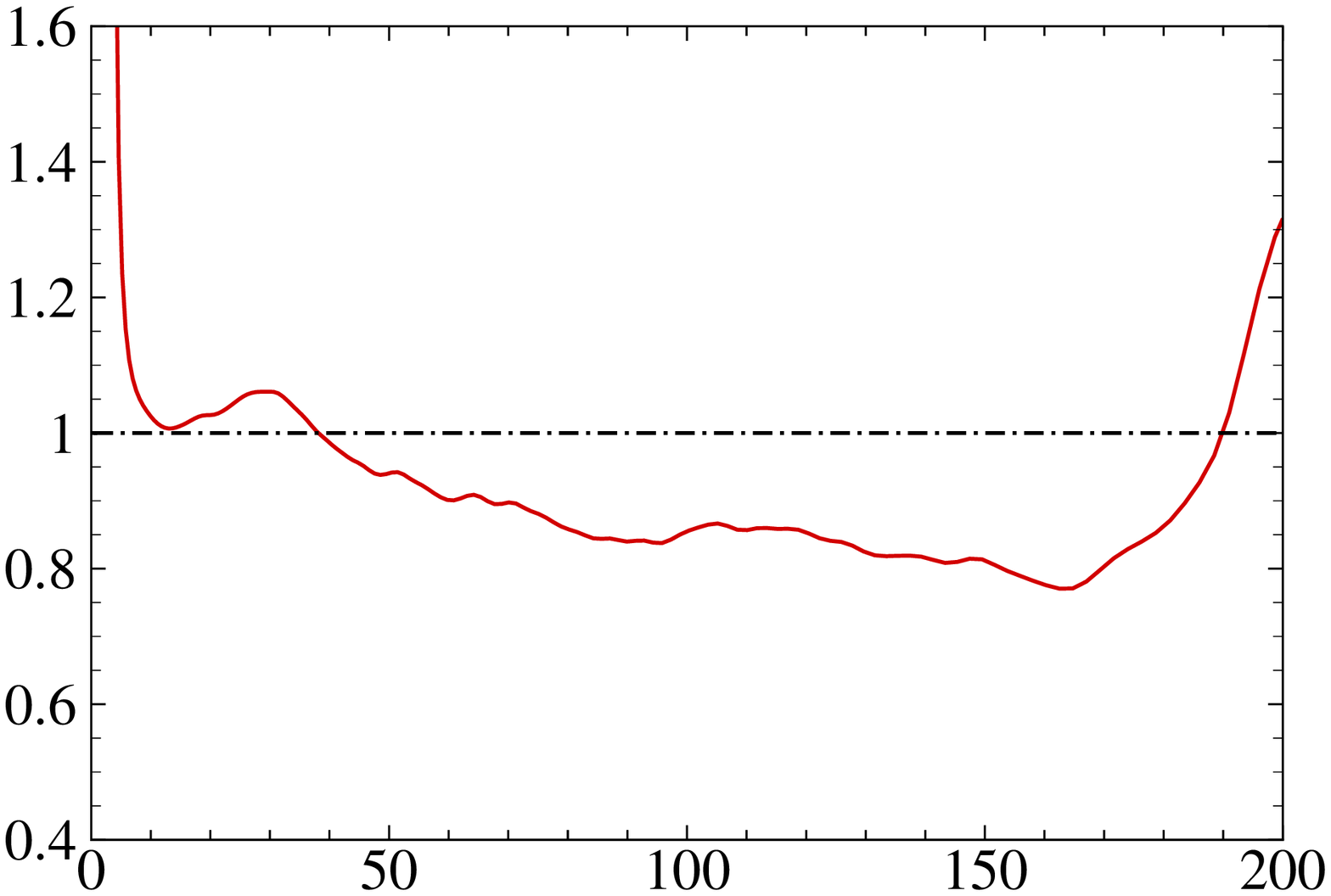}};
      \begin{scope}[x={(a.south east)},y={(a.north west)}]
        \node [align=center] at (0.55,0.01) {\large $y^+$};
        \node [align=center] at (0.85,0.84) {$\boxed{Pr_t}$};
      \end{scope}
\end{tikzpicture}
\vspace{-0.5cm}
\caption{Wall-normal profiles of the turbulent Prandtl number for CN case, at Re$_{\tau}=185$. The horizontal dashed-dotted lines denote the SRA estimation. }
\label{fig:prandtl_turb}
\end{figure}

Averages of unclosed convective and diffusive fluxes arising from Favre-averaging of the governing equations were also collected to verify the validity of some common modeling assumptions adopted in lower-fidelity simulations relying on the Reynolds-Averaged Navier-Stokes (RANS) equations. Special focus is put into closures of the turbulent heat transport terms arising in the averaged total energy and species transport equations. By applying a Favre averaging to the total energy equation \eqref{eq:energy} one obtains:
\begin{align}
\notag
\frac{\partial}{\partial t} \left(\overline{\rho} \widetilde{E}+ \frac{\overline{\rho u''_i u''_i}}{2} \right) & + \frac{\partial}{\partial x_j} \left[ \overline{\rho} \widetilde{u}_j \left(\widetilde{h} + \frac{\widetilde{u}_i\widetilde{u}_i}{2}\right) + \frac{\overline{\rho u''_i u''_j}}{2} \right] = \\
& \frac{\partial}{ \partial x_j}\left[ -\overline{q}_j -\overline{\rho u''_j h''} + \overline{\tau_{ij}u''_i} -\overline{\rho u''_j \frac{1}{2}u''_iu''_i} \right] +\frac{\partial}{\partial x_j} \left[ \widetilde{u}_i \left( \widetilde{\tau}_{ij} - \overline{\rho u''_iu''_j} \right) \right].
\end{align}
It is common practice to model the turbulent transport of a flow property $f$ as a linear function of its average gradient, e.g.:
\begin{align}
- \overline{\rho u'' f''} = \frac{\mu_t}{C_t} \frac{\partial \widetilde{f}}{\partial x}, \qquad
- \overline{\rho v'' f''} = \frac{\mu_t}{C_t} \frac{\partial \widetilde{f}}{\partial y}, \qquad
- \overline{\rho w'' f''} = \frac{\mu_t}{C_t} \frac{\partial \widetilde{f}}{\partial z},
\end{align}
the last term being zero for the present statistically 2D flow. Here, $\mu_t$ denotes the turbulent viscosity and $C_t$ an \textit{ad hoc} coefficient. The turbulent heat transport terms are usually modeled by introducing a ``turbulent'' Prandtl number defined as
\begin{equation}
\label{eq:prt}
Pr_t=\frac{\overline{\rho u''v''} \partial \widetilde{T}/\partial y}{\overline{\rho v''T''} \partial \widetilde{u}/\partial y},
\end{equation}
which is expected to be approximately equal to 1 throughout the flow, according to the classical so-called Strong Reynolds Analogy (SRA) first discussed by Morkovin\cite{morkovin1962mecanique}. The latter holds under the hypothesis of adiabatic wall, nearly constant total temperature and fully anticorrelated velocity and temperature fluctuations, that is
\begin{equation}
\label{eq:corr_uT}
-R_{u''T''} = -\frac{\overline{u'' T''}}{\sqrt{\overline{u''^2}} \sqrt{\overline{T''^2}}} \approx 1.
\end{equation}
For flows with large total temperature fluctuations (such as in the current case), equation~\eqref{eq:corr_uT} is corrected to account for them\cite{guarini2000direct}:
\begin{equation}
\label{eq:corr_uT2}
-R_{u''T''} + \frac{\overline{T_0''^2}}{2\overline{T''^2}}\approx 1.
\end{equation}
Figure~\ref{fig:rey_analogy1} presents the uncorrected \eqref{eq:corr_uT} and corrected \eqref{eq:corr_uT2} correlation distributions across the boundary layer, in outer scaling. Clearly, only the corrected correlation approaches reasonably well unity for the present high-enthalpy and high Mach number flow.  The assumption is however never satisfied in the outer part of the boundary layer, as also found in the work of Duan \textit{et al.}\cite{duan2011direct4}.
The turbulent Prandtl number is presented in figure~\ref{fig:prandtl_turb}. For air out of chemical equilibrium, $Pr_t$ follows essentially the trend registered by numerous authors in the literature \cite{huang1995compressible,duan2010direct2,duan2011direct3,zhang2018direct} and is not constant throughout the flow, in contrast with the SRA assumption which predicts $Pr_t\approx 1$. In the logarithmic and outer layers, $Pr_t$ approaches the value of 0.9, commonly used in turbulence models. A local maximum at about $y^+=30$ is observed; in the near-wall region, $Pr_t$ exceeds 1 and becomes singular at the wall due to the quasi-adiabatic boundary condition.

\begin{figure}
\centering
\begin{tikzpicture}
      \node[anchor=south west,inner sep=0] (a) at (0,0) {\includegraphics[width=0.47\columnwidth]
      {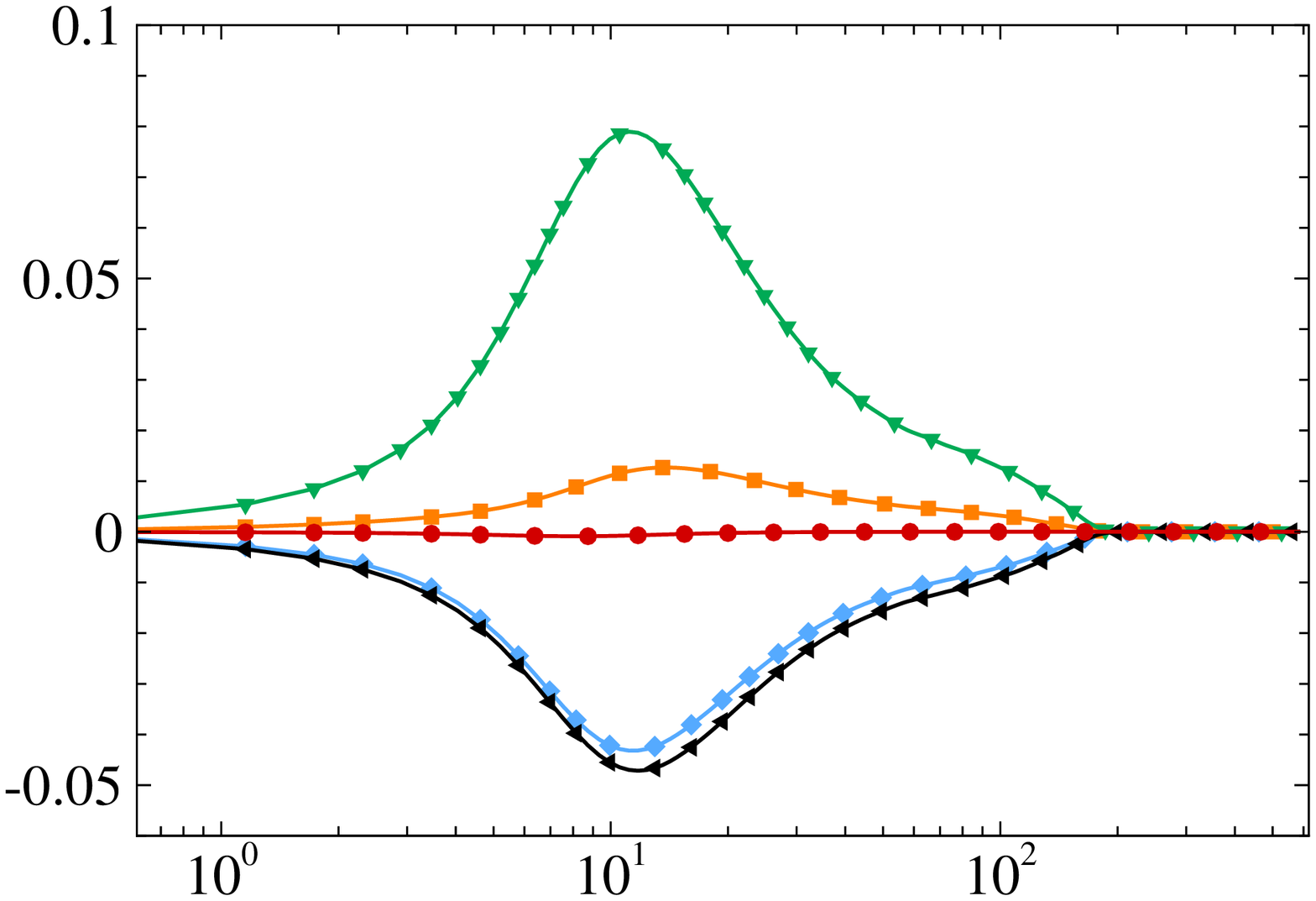}};
      \begin{scope}[x={(a.south east)},y={(a.north west)}]
        \node [align=center] at (0.0 ,0.95) {(a)};
        \node [align=center] at (0.55,0.03) {\large $y^+$};
        \node [align=center] at (0.80,0.81) {\scriptsize $\boxed{\frac{\overline{\rho u'' Y_{n}''}}{\overline{\rho}_w u_{\tau}}}$};
      \end{scope}
\end{tikzpicture}
\hspace{-0.1cm}
\begin{tikzpicture}
      \node[anchor=south west,inner sep=0] (a) at (0,0) {\includegraphics[width=0.47\columnwidth]
      {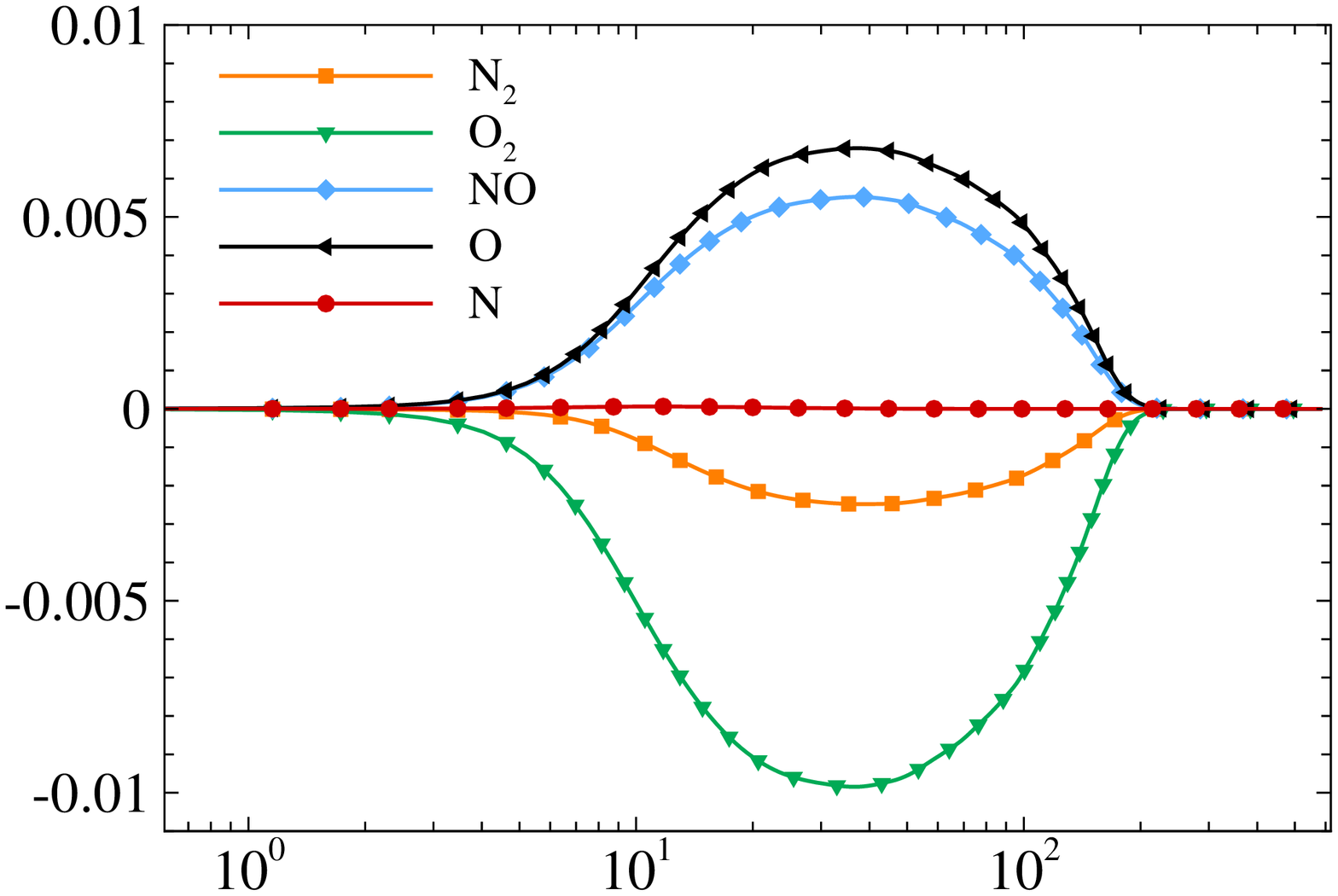}};
      \begin{scope}[x={(a.south east)},y={(a.north west)}]
        \node [align=center] at (0.0 ,0.95) {(b)};
        \node [align=center] at (0.55,0.03) {\large $y^+$};
        \node [align=center] at (0.25,0.25) {\scriptsize $\boxed{\frac{\overline{\rho v'' Y_{n}''}}{\overline{\rho}_w u_{\tau}}}$};
      \end{scope}
\end{tikzpicture}
\hspace{0cm}
\begin{tikzpicture}
      \node[anchor=south west,inner sep=0] (a) at (0,0) {\includegraphics[width=0.47\columnwidth]
      {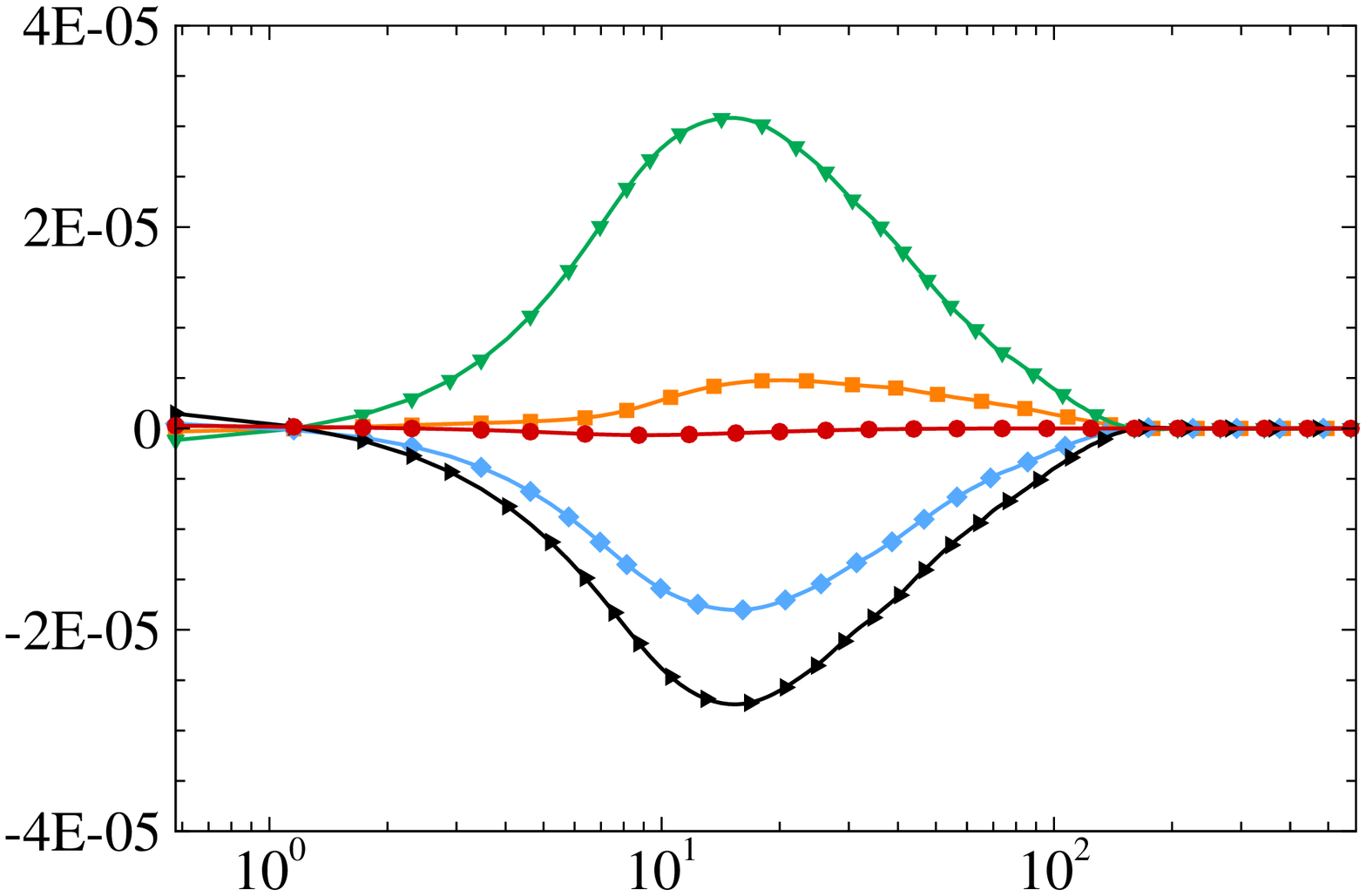}};
      \begin{scope}[x={(a.south east)},y={(a.north west)}]
        \node [align=center] at (0.0 ,0.95) {(c)};
        \node [align=center] at (0.55,0.03) {\large $y^+$};
        \node [align=center] at (0.80,0.81) {\scriptsize $\boxed{\frac{\overline{\rho D_n'' \frac{\partial Y_n''}{\partial x}}}{\overline{\rho}_w u_\tau}}$};
      \end{scope}
\end{tikzpicture}
\hspace{-0.1cm}
\begin{tikzpicture}
      \node[anchor=south west,inner sep=0] (a) at (0,0) {\includegraphics[width=0.47\columnwidth]
      {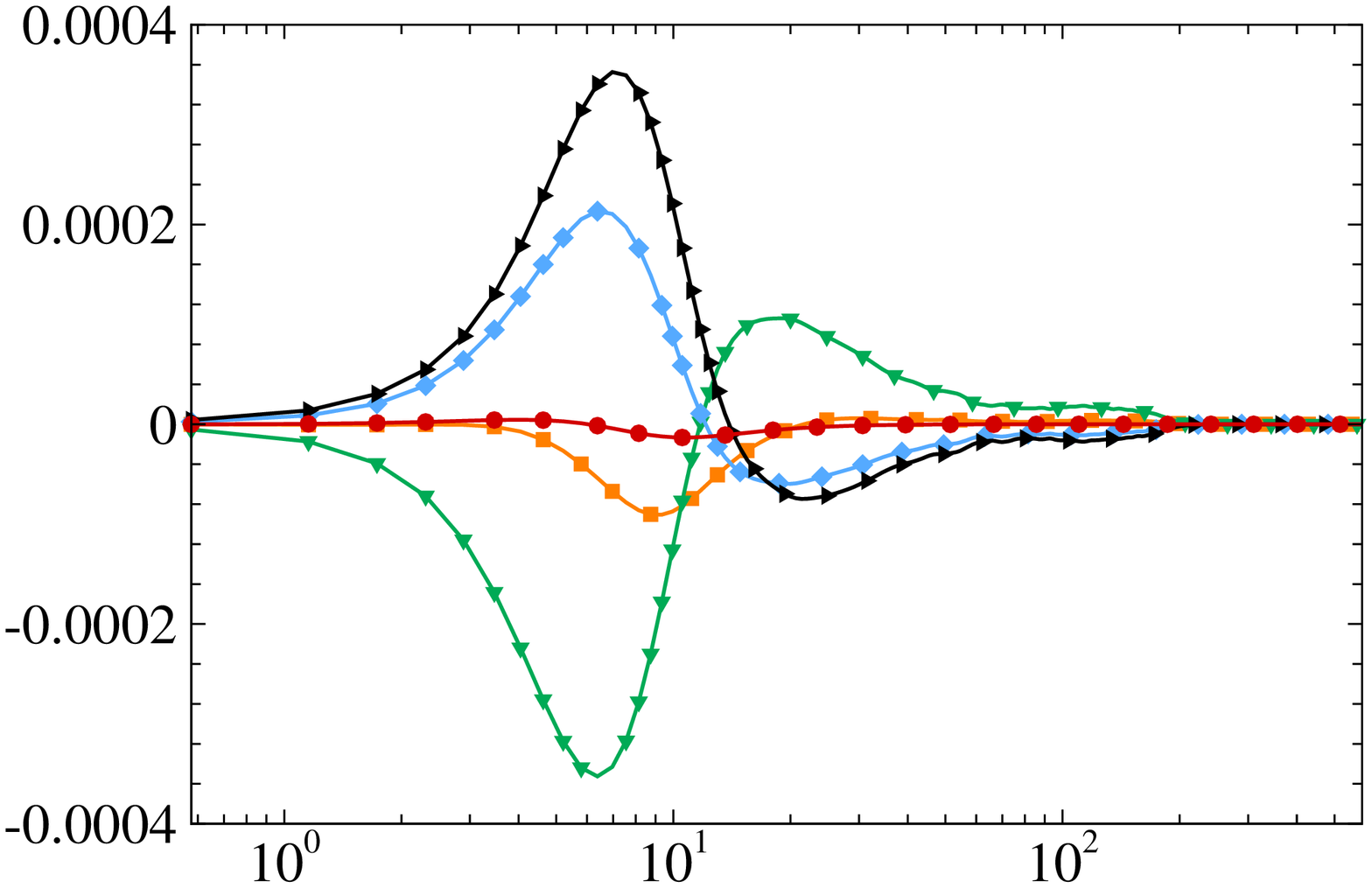}};
      \begin{scope}[x={(a.south east)},y={(a.north west)}]
        \node [align=center] at (0.0 ,0.95) {(d)};
        \node [align=center] at (0.55,0.03) {\large $y^+$};
        \node [align=center] at (0.80,0.81) {\scriptsize $\boxed{\frac{\overline{\rho D_n'' \frac{\partial Y_n''}{\partial y}}}{\overline{\rho}_w u_\tau}}$};
      \end{scope}
\end{tikzpicture}
\caption{Normalized turbulent transport of species mass fractions in the streamwise (a) and wall-normal (b) directions; normalized turbulent diffusion fluxes  in the streamwise (c) and wall-normal (d) directions, at Re$_{\tau}=185$.}
\label{fig:mass_diffusion_flux}
\end{figure}
%

Similarly to the procedure followed for the total energy equation, the Favre averaging of the species transport equation~\eqref{eq:species} leads to
\begin{equation}
\frac{\partial \overline{\rho} \widetilde{Y}_n}{\partial t} + \frac{\partial \left(\overline{\rho} \widetilde{Y}_n \widetilde{u}_j \right)}{\partial x_j} = \frac{\partial}{\partial x_j} \left( \overline{\rho} \widetilde{D}_n \frac{\partial \widetilde{Y}_n}{\partial x_j}\right) + \overline{\rho} \widetilde{\omega}_n \\
 + \frac{\partial}{ \partial x_j}  \overline{\rho u''_j Y''_n} + \frac{\partial}{\partial x_j} \overline{\rho D''_n \frac{\partial Y''_n}{\partial x_j}} + \overline{\rho} V_c,
\end{equation}
where the unclosed terms are the turbulent transport of chemical species, $\displaystyle\frac{\partial}{\partial x_j} \overline{\rho u''_j Y_n''}$, and the turbulent species diffusion, $\displaystyle\frac{\partial}{\partial x_j} \overline{\rho D''_n \frac{\partial Y_n''}{\partial x _j}}$. The last term on the r.h.s., $\overline{\rho} V_c$, resulting from Favre-averaging of the mass diffusion term in equation \eqref{eq:fick}, was found to be negligibly small throughout the flow and is not discussed further. The exact turbulent transport and diffusion terms computed from the DNS data are reported in figure~\ref{fig:mass_diffusion_flux}.
The turbulent transport of species in the streamwise and wall-normal directions (panels a and b) are preponderant in the buffer and logarithmic zone. The turbulent diffusion terms along $x$ and $y$ directions (panels c and d) are 1 or 2 orders of magnitude smaller; in particular, $\overline{\rho D_n'' \frac{\partial Y_n''}{\partial x}}$ is active in the region where $\overline{\rho u'' Y_{n}''}$ peaks, and its contribution can certainly be neglected.
However, $\overline{\rho D_n'' \frac{\partial Y_n''}{\partial y}}$ is not completely negligible compared to the other terms in the viscous sublayer and the buffer region, even though, in the RANS approach, the turbulent fluxes deriving from the diffusive terms are not accounted for explicitly.
\begin{figure}
\centering
\begin{tikzpicture}
      \node[anchor=south west,inner sep=0] (a) at (0,0) {\includegraphics[width=0.48\columnwidth]{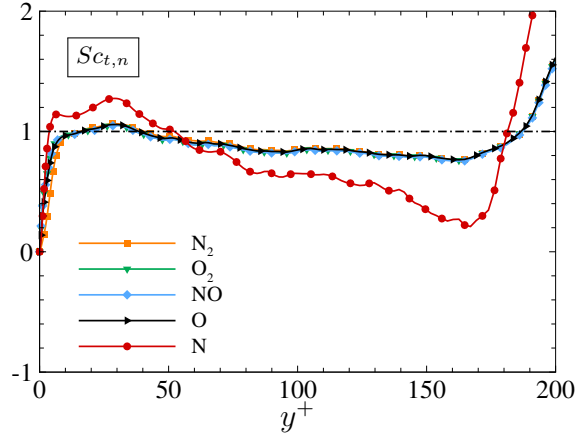}};
      \begin{scope}[x={(a.south east)},y={(a.north west)}]
        \node [align=center] at (0.55,0.01) {\large $y^+$};
        \node [align=center] at (0.25,0.84) {$\boxed{Sc_{t,n}}$};
      \end{scope}
\end{tikzpicture}
\vspace{-0.5cm}
\caption{Wall-normal profiles of the turbulent Schmidt number for CN case, at Re$_{\tau}=185$. The horizontal dashed-dotted lines denote the SRA estimation. }
\label{fig:schmidt_turb}
\end{figure}

Likewise the turbulent heat transport fluxes, the species transport fluxes are modeled by introducing a turbulent Schmidt number, e.g.:
\begin{equation}
\label{eq:schmidt_turb}
Sc_{t,n}=\frac{\overline{\rho u''v''} \partial \widetilde{Y}_n/\partial y}{\overline{\rho v''Y_n''} \partial \widetilde{u}/\partial y},
\end{equation}
such that
\begin{equation}
- \overline{v'' Y_n''} = \frac{\mu_t}{Sc_{t,n}} \frac{\partial \widetilde{Y}_n}{\partial y}
\end{equation}
in the wall normal direction, and similarly for the other directions. Figure~\ref{fig:schmidt_turb} reports the $Sc_{t,n}$ profiles for the five species. In all cases, a value reasonably close to unity (corresponding to the common modeling practice) is observed for all species (except $N$, characterized by a very low species mass flux), in the logarithmic and outer regions of the boundary layer. Such approximation fails in the near-wall region, where $\partial \widetilde{Y_n}/\partial y \approx 0$ due to the non-catalytic boundary condition.

\subsection{Spectral content}
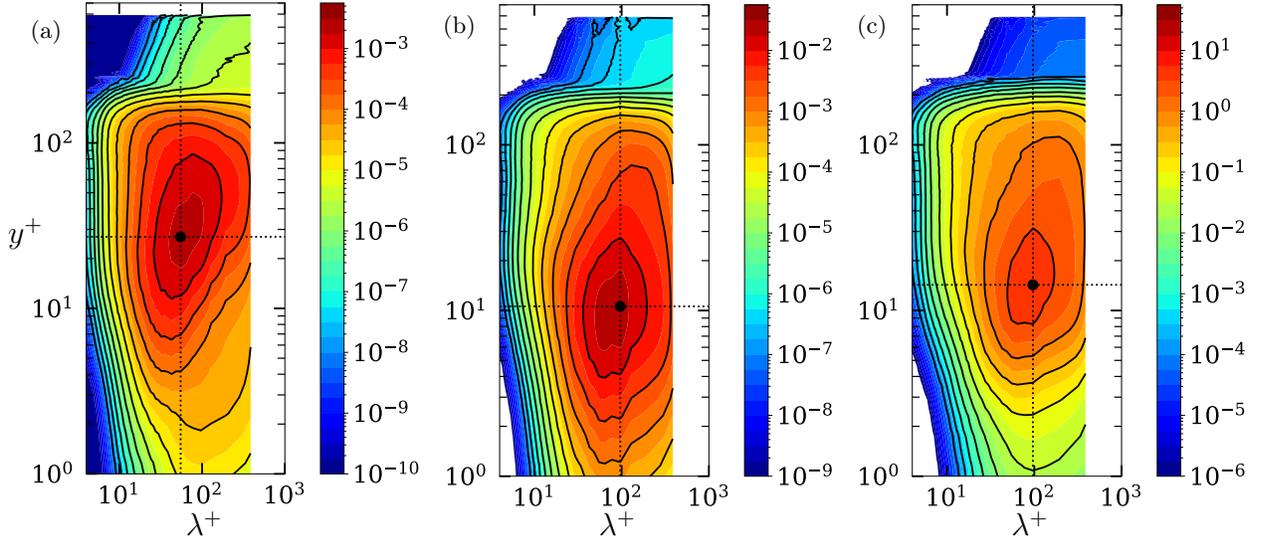
\begin{figure}
\centering
\begin{tikzpicture}
      \node[anchor=south west,inner sep=0] (a) at (0,0) {\includegraphics[width=0.3\columnwidth, trim={0 0 2 6}, clip]
      {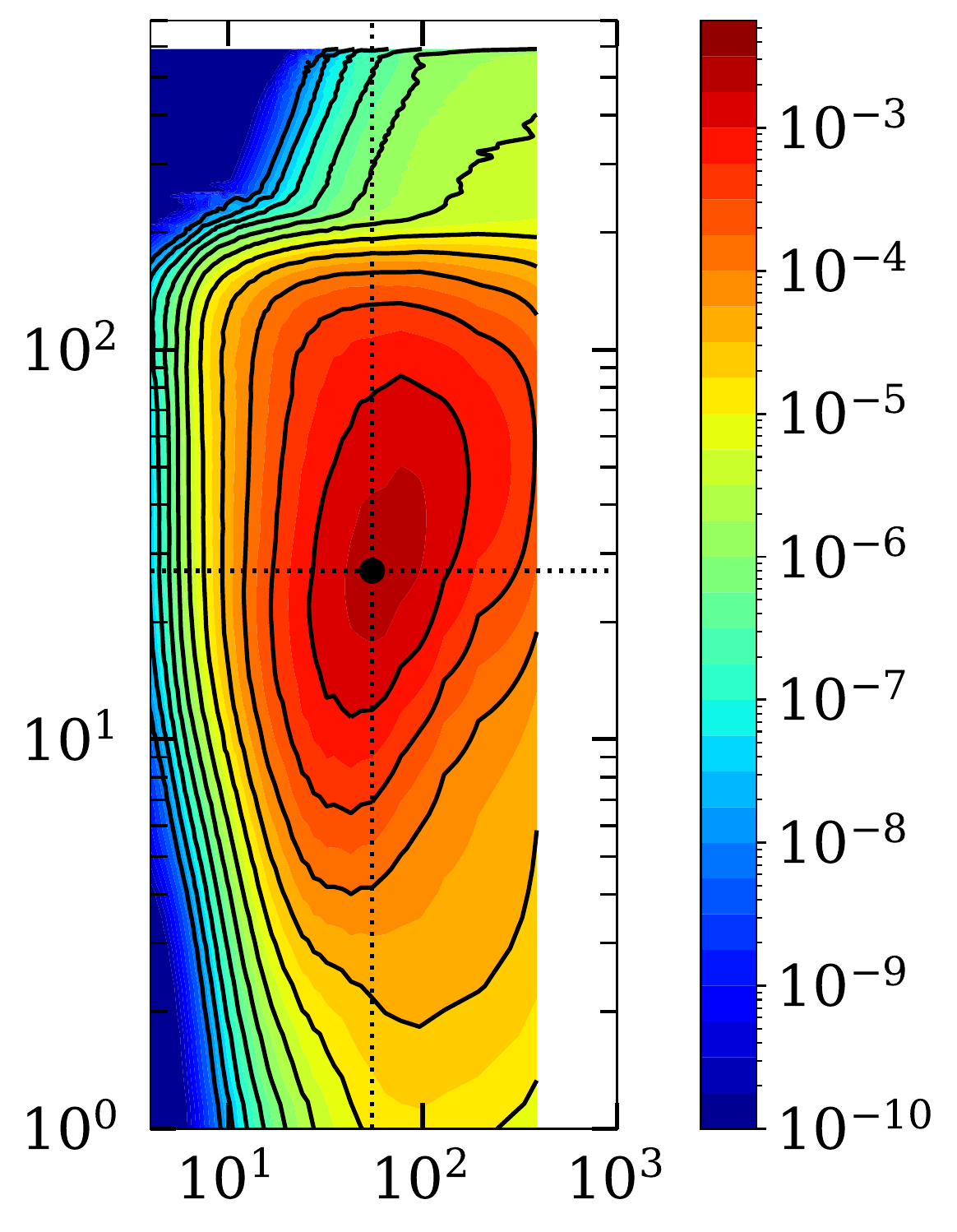}};
      \begin{scope}[x={(a.south east)},y={(a.north west)}]
        \node [align=center] at (0.06,0.94)  {(a)};
        \node [align=center] at (0.45,-0.01) {\large $\lambda^+$};
        \node [align=center] at (0.01,0.55)  {\large $y^+$};
      \end{scope}
\end{tikzpicture}
\begin{tikzpicture}
      \node[anchor=south west,inner sep=0] (a) at (0,0) {\includegraphics[width=0.3\columnwidth, trim={0 2 2 0}, clip]
      {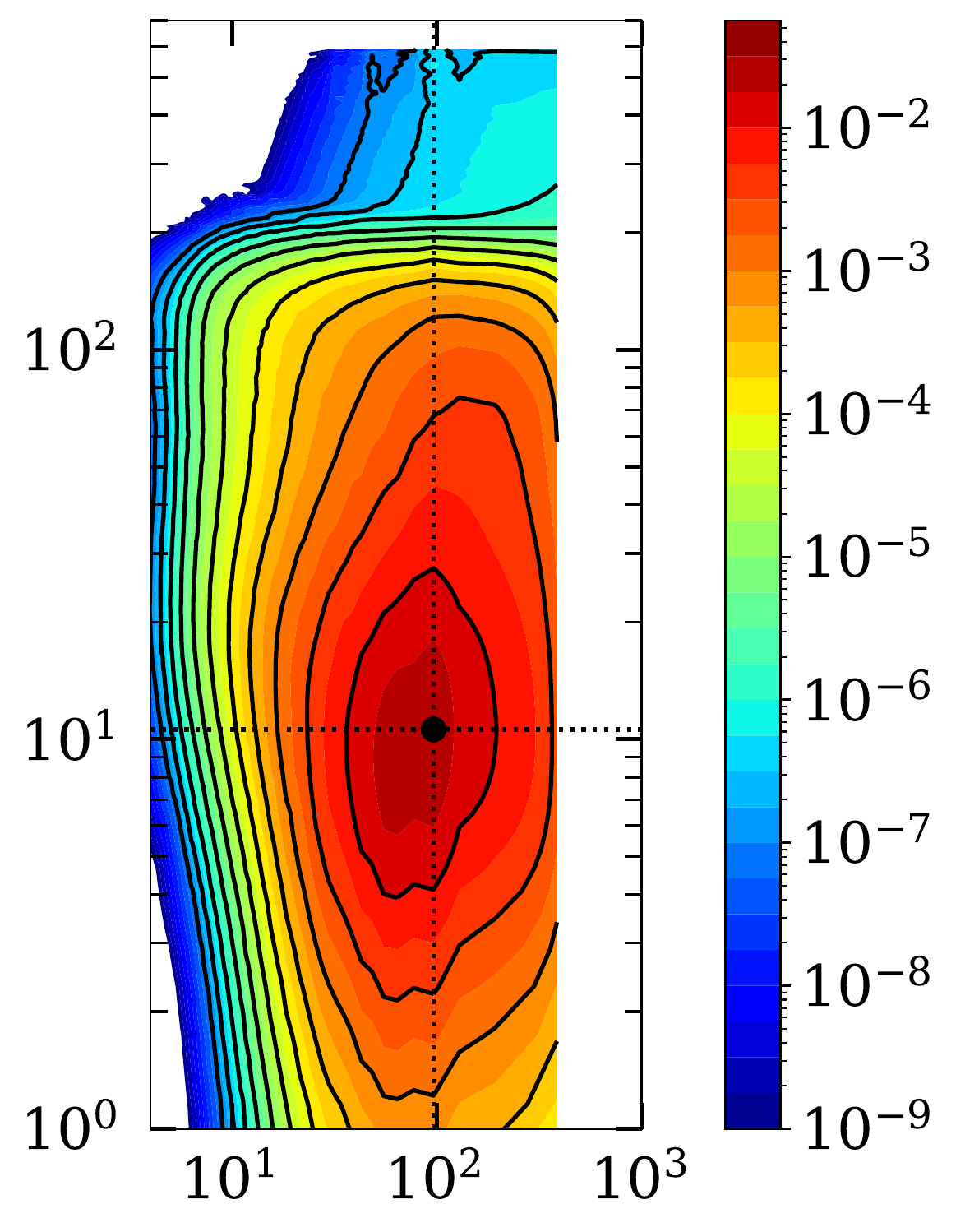}};
      \begin{scope}[x={(a.south east)},y={(a.north west)}]
        \node [align=center] at (0.06,0.94)   {(b)};
        \node [align=center] at (0.45,-0.01) {\large $\lambda^+$};
      \end{scope}
\end{tikzpicture}
\begin{tikzpicture}
      \node[anchor=south west,inner sep=0] (a) at (0,0) {\includegraphics[width=0.3\columnwidth, trim={0 2 2 0}, clip]
      {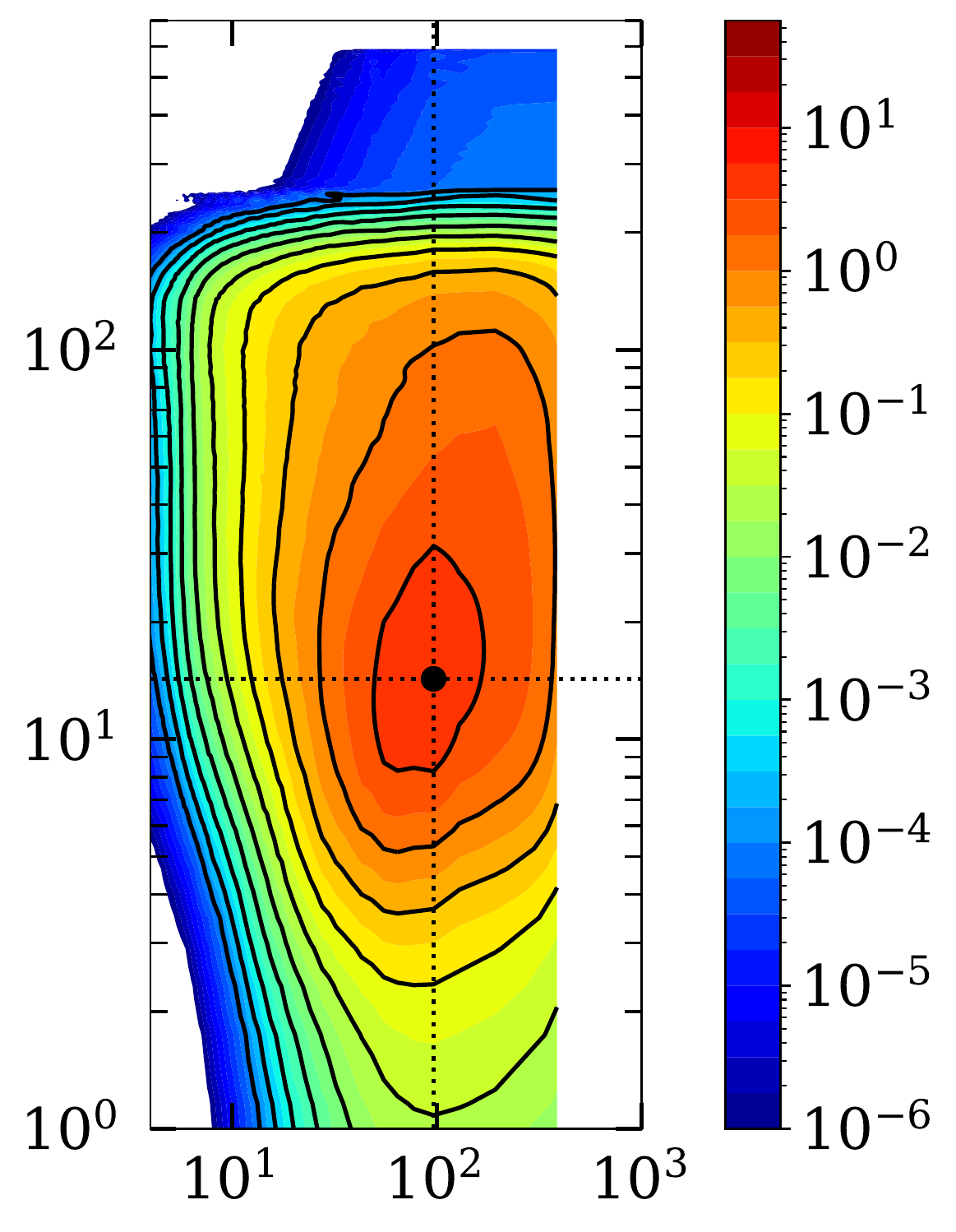}};
      \begin{scope}[x={(a.south east)},y={(a.north west)}]
        \node [align=center] at (0.06,0.94)   {(c)};
        \node [align=center] at (0.45,-0.01) {\large $\lambda^+$};
      \end{scope}
\end{tikzpicture}
\caption{Premultiplied spanwise spectra $k_z E_{v v}/u_\infty^2$ (a), $k_z E_{u u}/u_\infty^2$ (b) and $k_z E_{T T}/T_\infty^2$ (c) for the chemical nonequilibrium simulation, at Re$_\tau=185$.}
\label{fig:spectra2d}
\end{figure}

To characterize the near-wall turbulent structures, premultiplied spectra of the fluctuating wall-normal and streamwise velocities, as well as of the temperature, are reported in fig.~\ref{fig:spectra2d} as a function of the normalized spanwise wavenumber $\lambda ^+$.
The well developed spectra indicate that a fully turbulent state has been reached at this position.
The spectral content is not significantly altered by chemical effects, and the overall trend is similar to perfect-gas simulations \cite{sciacovelli2020numerical}.
All of the spectra exhibit a peak in the buffer layer, at the same location were the Reynolds stresses and the temperature fluctuations peak (see figure~\ref{fig:reynolds_stresses}).

\subsection{Skin friction analysis}
\begin{figure}
\centering
\begin{tikzpicture}
      \node[anchor=south west,inner sep=0] (a) at (0,0) {\includegraphics[width=0.47\columnwidth]
      {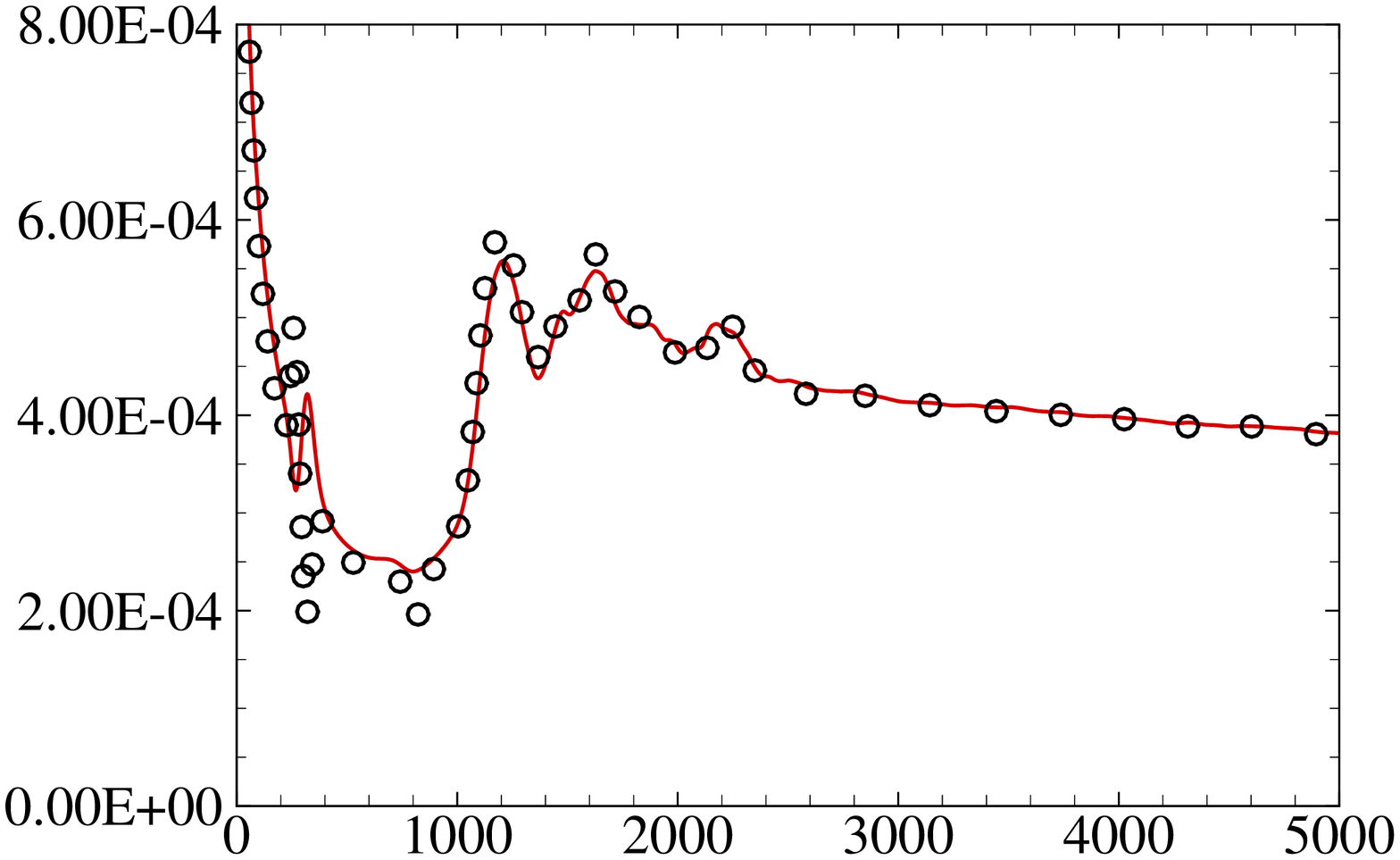}};
      \begin{scope}[x={(a.south east)},y={(a.north west)}]
        \node [align=center] at (-0.01,0.94) {(a)};
        \node [align=center] at (0.56 ,0.02) {\large{$\hat{x}$}};
        \node [align=center] at (0.85 ,0.85) {{$\boxed{C_f}$}};
      \end{scope}
\end{tikzpicture}
\hspace{-0.5cm}
\begin{tikzpicture}
      \node[anchor=south west,inner sep=0] (a) at (0,0) {\includegraphics[width=0.47\columnwidth]
      {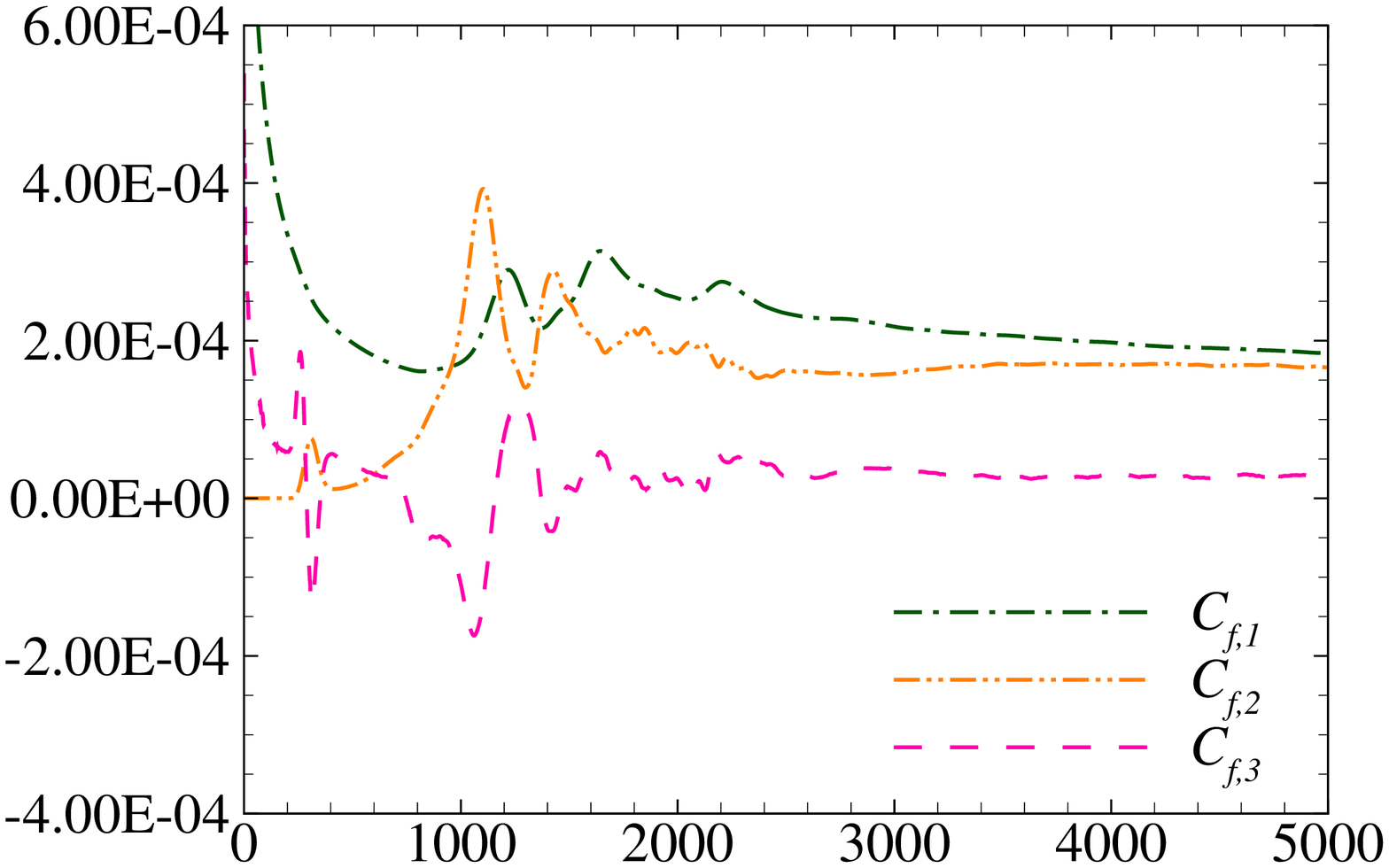}};
      \begin{scope}[x={(a.south east)},y={(a.north west)}]
        \node [align=center] at (-0.01,0.94) {(b)};
        \node [align=center] at (0.56 ,0.02) {\large{$\hat{x}$}};
      \end{scope}
\end{tikzpicture}
\vspace{-0.5cm}
\caption{Streamwise evolution of skin friction coefficient (\protect\lsolid{red}) superposed with Renard-Deck decomposition (symbols) (a) and contribution of each terms of equation~\eqref{eq:renard_deck} (b) for the chemical nonequilibrium case. }
\label{fig:RD}
\end{figure}

We complete the analysis by studying the contributions of the mean field and turbulent quantities to the generation of skin friction at the plate wall. To this aim, we consider the Renard and Deck\cite{renard2016theoretical} decomposition of the mean skin friction in an incompressible boundary layer, extended to compressible boundary layers by Li \textit{et al.}\cite{li2019decomposition}.
Based on the kinetic energy transport equation, and under the assumptions of: i) no-slip condition at the wall, ii) homogeneity in the spanwise direction and iii) no body force,
the mean skin friction drag coefficient can be expressed as:
\begin{align}
\notag
C_f = & \underbrace{\frac{2}{\rho_\infty u^3_\infty} \int_0^\delta \overline{\tau_{xy}} \frac{\partial \widetilde{u}}{\partial y} \text{d}y}_{C_{f,1}}+ \underbrace{\frac{2}{\rho_\infty u^3_\infty} \int_0^\delta -\overline{\rho u''v''} \frac{\partial \widetilde{u}}{\partial y} \text{d}y}_{C_{f,2}} \\
&+ \underbrace{\frac{2}{\rho_\infty u^3_\infty} \int_0^\delta ( \widetilde{u} - u_\infty) \left[ \overline{\rho} \left( \widetilde{u}\frac{\partial \widetilde{u} }{\partial x} +\widetilde{v}\frac{\partial \widetilde{u}}{\partial y} \right) - \frac{\partial}{\partial x} \left( \overline{\tau_{xx}} - \overline{\rho} \widetilde{u''u''} -\overline{p} \right) \right]  \text{d}y } _{C_{f,3}}
\label{eq:renard_deck}
\end{align}
The terms denoted as $C_{f,1}$ and $C_{f,2}$ represent the contributions of the mean-field molecular dissipation and the dissipation due to the Reynolds stresses, respectively; $C_{f,3}$ accounts for the boundary layer spatial growth and includes the effects of streamwise heterogeneity. Figure~\ref{fig:RD}(a) shows that the sum of the preceding terms computed from DNS data is in excellent agreement with the total skin friction coefficient, with minor discrepancies next to the suction-and-blowing forcing location. The presence of chemical reactions does not affect the validity of the decomposition, which is derived under rather general hypotheses. In the fully turbulent region, $C_{f,1}$ and $C_{f,2}$ are preponderant with respect to the third term (figure~\ref{fig:RD}(b)).
We also observe that $C_{f,1}$ contributes more than $C_{f,2}$, consistently with the relatively low Reynolds numbers reached at the end of the plate\cite{fan2019decomposition}.

\section{Conclusions}
\label{sec:conclusions}
A high-enthalpy hypersonic turbulent boundary layer at Mach 10, spatially developing  along a quasi-adiabatic flat plate, is investigated by means of Direct Numerical Simulation (DNS). The fluid is air, modeled by using the five-species model of Park. The free-stream thermodynamic conditions are such that gas dissociation phenomena occur, giving rise to a non-equilibrium chemical state. At the considered wall temperature, oxygen and a small amount of nitrogen dissociate. The influence of such high-temperature effects on turbulence dynamics is then investigated by comparing first- and second-order flow statistics to those obtained for a turbulent boundary layer of an oxygen/nitrogen mixture with frozen chemistry.

The main effect of gas dissociation is to change the mixture composition near the wall, in the region comprised between the viscous and buffer sublayers, modifying the fluid thermo-physical properties. Additionally, the endothermic chemical reactions drain energy from the flow in the turbulence production region, leading to slightly smaller streamwise turbulent intensity and to a reduction of temperature fluctuations by approximately 10\%. The small values observed for the mean Damk\"{o}hler numbers of the species  point out that chemical dissociation is much slower than the characteristic time scale of the flow, i.e. the flow is not far from frozen flow conditions; turbulence/chemistry interactions are thus found to be small.
In this situation, the main differences with respect to a frozen flow are observed for the thermodynamic quantities; on the contrary, the presence of chemical reactions does not alter substantially velocity fluctuations or turbulent spectra.
The DNS data have been used to assess some classical modeling assumptions derived from the Strong Reynolds Analogy theory. The latter is shown to remain valid, provided that corrections accounting for total temperature fluctuations are applied.
The well known assumptions of a constant turbulent Prandtl number and constant turbulent Schmidt number, classically used to model turbulent transport of heat and species mass fractions, respectively, is also investigated.
The results show that, while the constant Prandtl number is a rather crude one, in agreement with previous results in the literature, the constant Schmidt number is better respected, except in the immediate neighborhood of the wall were a non-catalytic condition is applied. However, the Schmidt number does not take a nearly unit value for all of the present species, and species-dependent coefficients should be adopted for better accuracy.
Finally, the Renard-Deck decomposition for the skin friction coefficient has been found to remain valid also at the present severe hypersonic conditions. \\
The present DNS was carried out under the hypotheses of quasi-adiabatic, non-catalytic wall.
Further work is required for elucidating the influence of wall cooling. In that case, it is expected that the region of maximum chemical activity moves away from the wall approaching that of maximum turbulence production, thus enhancing the turbulence-chemistry interactions. Thermal non-equilibrium conditions will be considered as well, with the aim of quantifying the influence of thermal relaxation phenomena on wall turbulence dynamics.

\acknowledgements
This work was partially supported by the Italian Ministry of Education, University and Research under the Programme Department of Excellence Legge 232/2016 (Grant No. CUP - D94I18000260001). The present project was granted access to the HPC resources of IDRIS and TGCC under the allocation A0072B10947 made by GENCI (Grand Equipement National de Calcul Intensif). We also acknowledge CINECA for awarding access to the Galileo supercomputer under the allocation HP10CLMXP0.\\

\revappendix

\section{Assessment of mesh adequacy}
\label{app:grid_study}
\begin{figure}
\centering
\begin{tikzpicture}
      \node[anchor=south west,inner sep=0] (a) at (0,0) {\includegraphics[width=0.46\columnwidth]
      {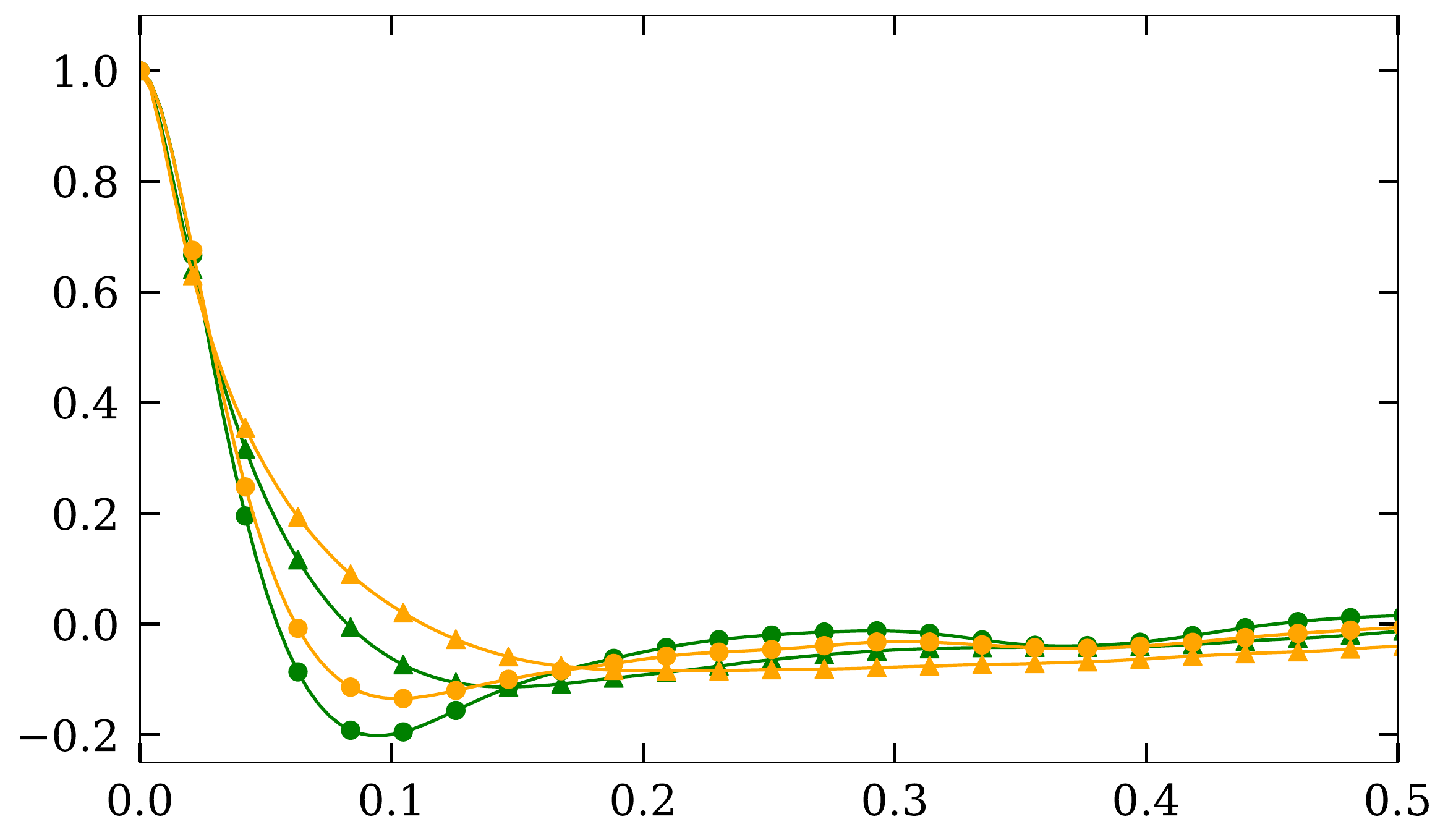}};
      \begin{scope}[x={(a.south east)},y={(a.north west)}]
        \node [align=center] at (0.00,0.96) {(a)};
        \node [align=center] at (0.55,-0.02) {$\Delta_z/L_z$};
        \node [align=center] at (0.82,0.78) {$\boxed{\frac{R_{\alpha\alpha}}{R_{\alpha \alpha (0)}}}$};
      \end{scope}
\end{tikzpicture}
\begin{tikzpicture}
      \node[anchor=south west,inner sep=0] (a) at (0,0) {\includegraphics[width=0.46\columnwidth]
      {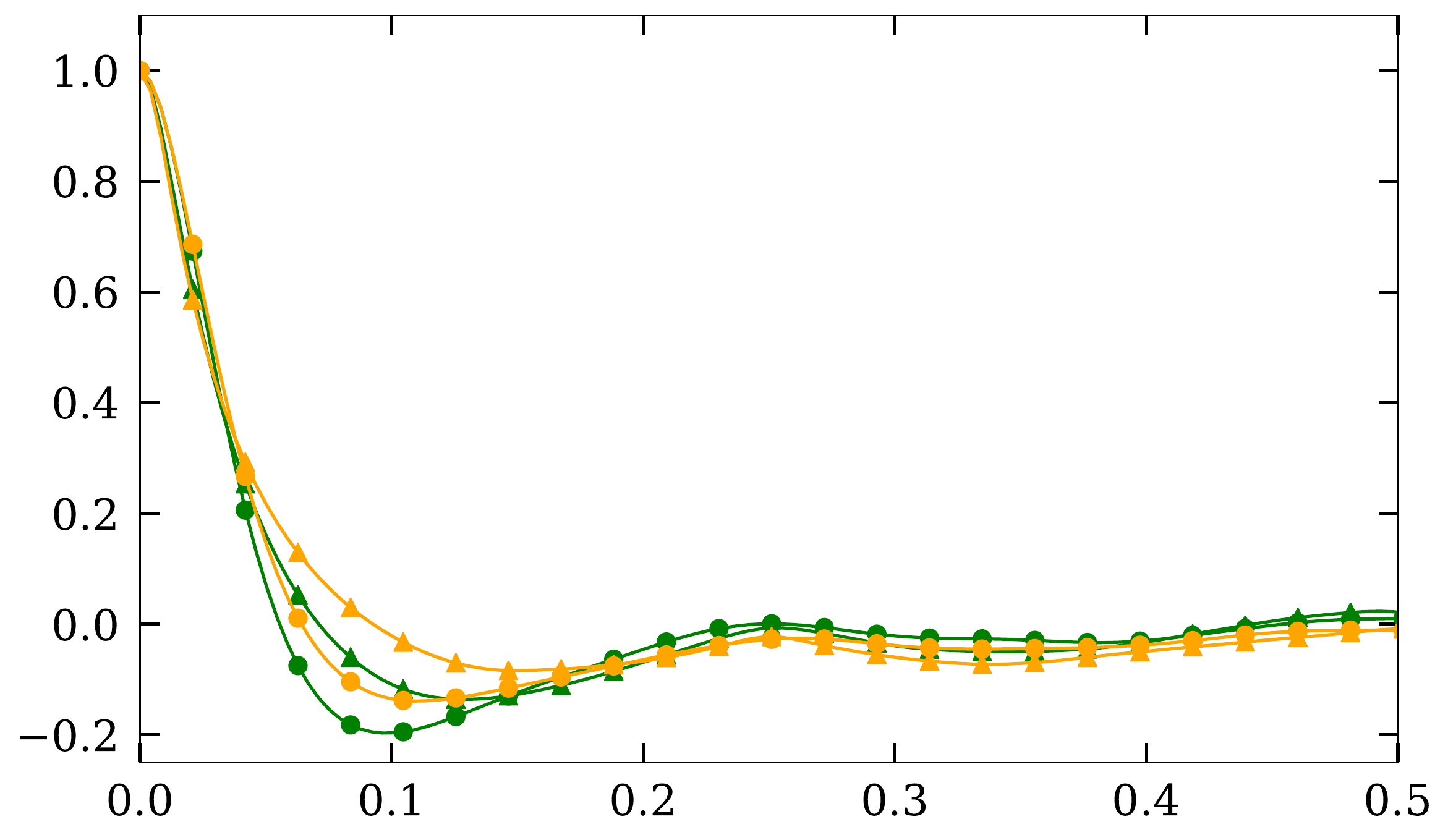}};
      \begin{scope}[x={(a.south east)},y={(a.north west)}]
        \node [align=center] at (0.00,0.96) {(b)};
        \node [align=center] at (0.55,-0.02) {$\Delta_z/L_z$};
      \end{scope}
\end{tikzpicture}
\vspace{-0.5cm}
\caption{ Distribution of two-point correlations in the spanwise direction, at $Re_\tau=185$: $\alpha=u$, at $y^+=10$ (\protect\linecircle{darkgreen}) and $y^+=150$ (\protect\linetriangle{darkgreen}); $\alpha=\rho$, at $y^+=10$ (\protect\linecircle{oorange}) and $y^+=150$ (\protect\linetriangle{oorange}). Panel (a), CN case; panel (b), FR case.}
\label{fig:two_point_correlation}
\end{figure}
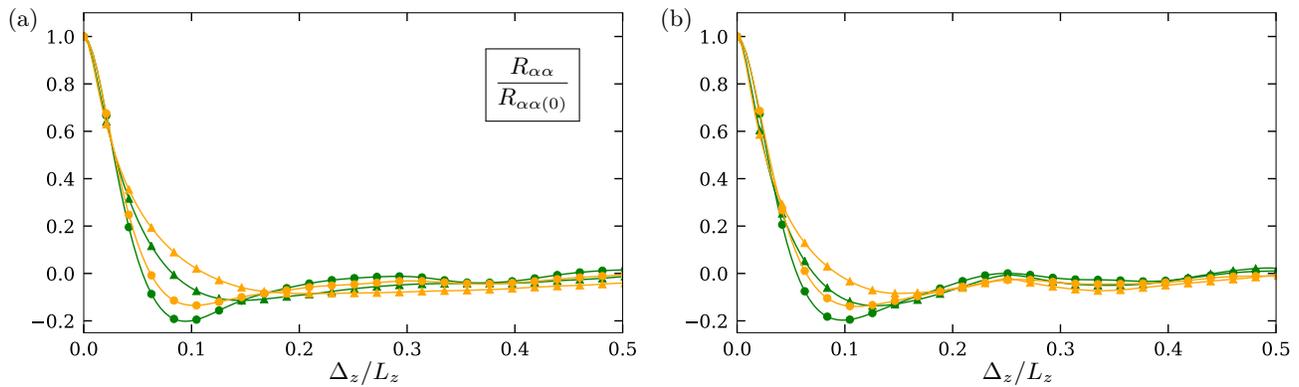

\begin{figure}
\centering
\begin{tikzpicture}
      \node[anchor=south west,inner sep=0] (a) at (0,0) {\includegraphics[width=0.30\columnwidth]{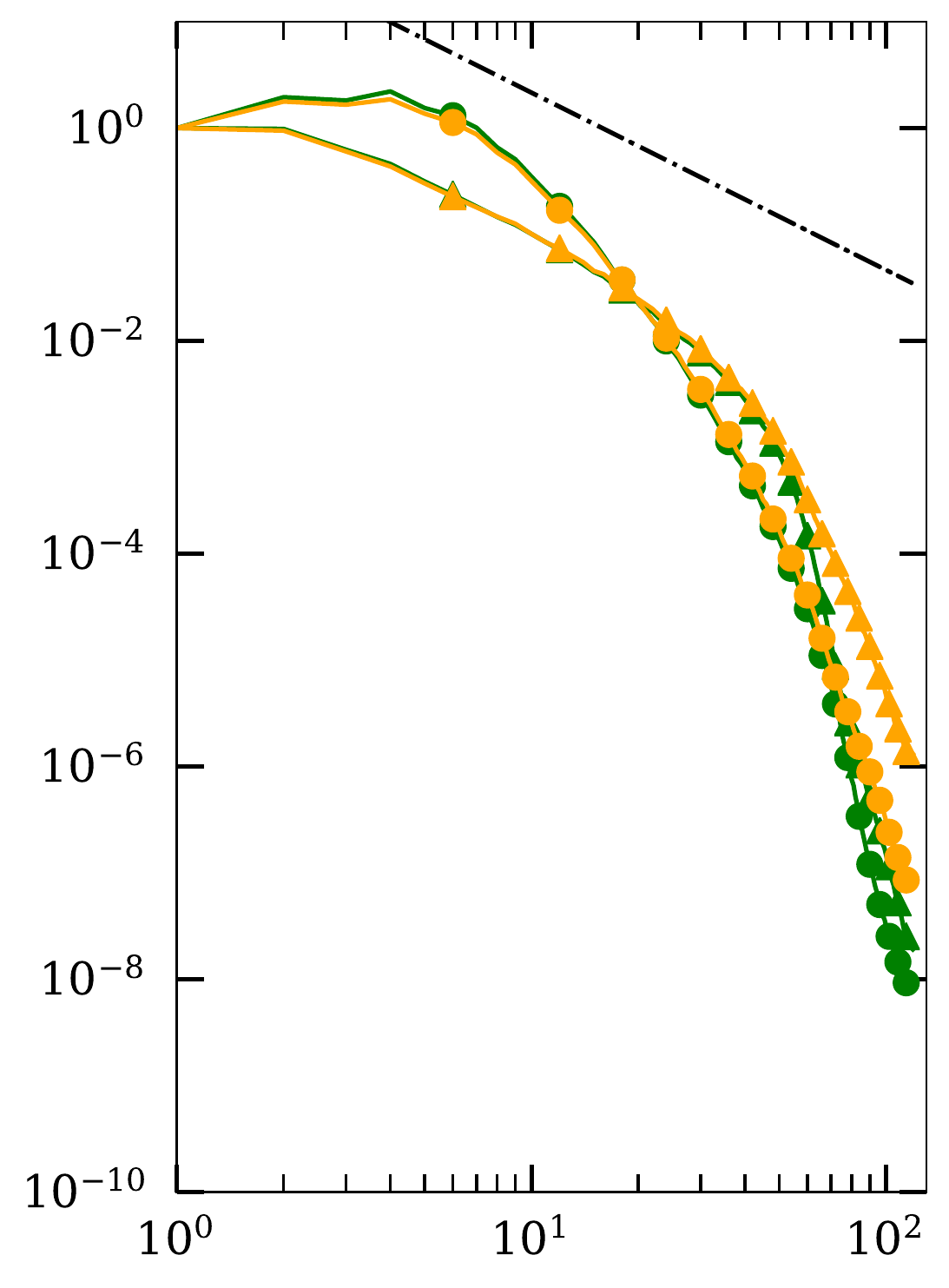}};
      \begin{scope}[x={(a.south east)},y={(a.north west)}]
        \node [align=center] at (0.00,0.96) {(a)};
        \node [align=center] at (0.32,0.17) {$\boxed{\frac{E_\alpha}{E_\alpha(0)}}$};
        \node [align=center] at (0.55,0.0) {\large $k_z$};
        \node [align=center] at (0.84,0.90) {$k_z^{-5/3}$};
      \end{scope}
\end{tikzpicture}
\hspace{0.8cm}
\begin{tikzpicture}
      \node[anchor=south west,inner sep=0] (a) at (0,0) {\includegraphics[width=0.30\columnwidth]{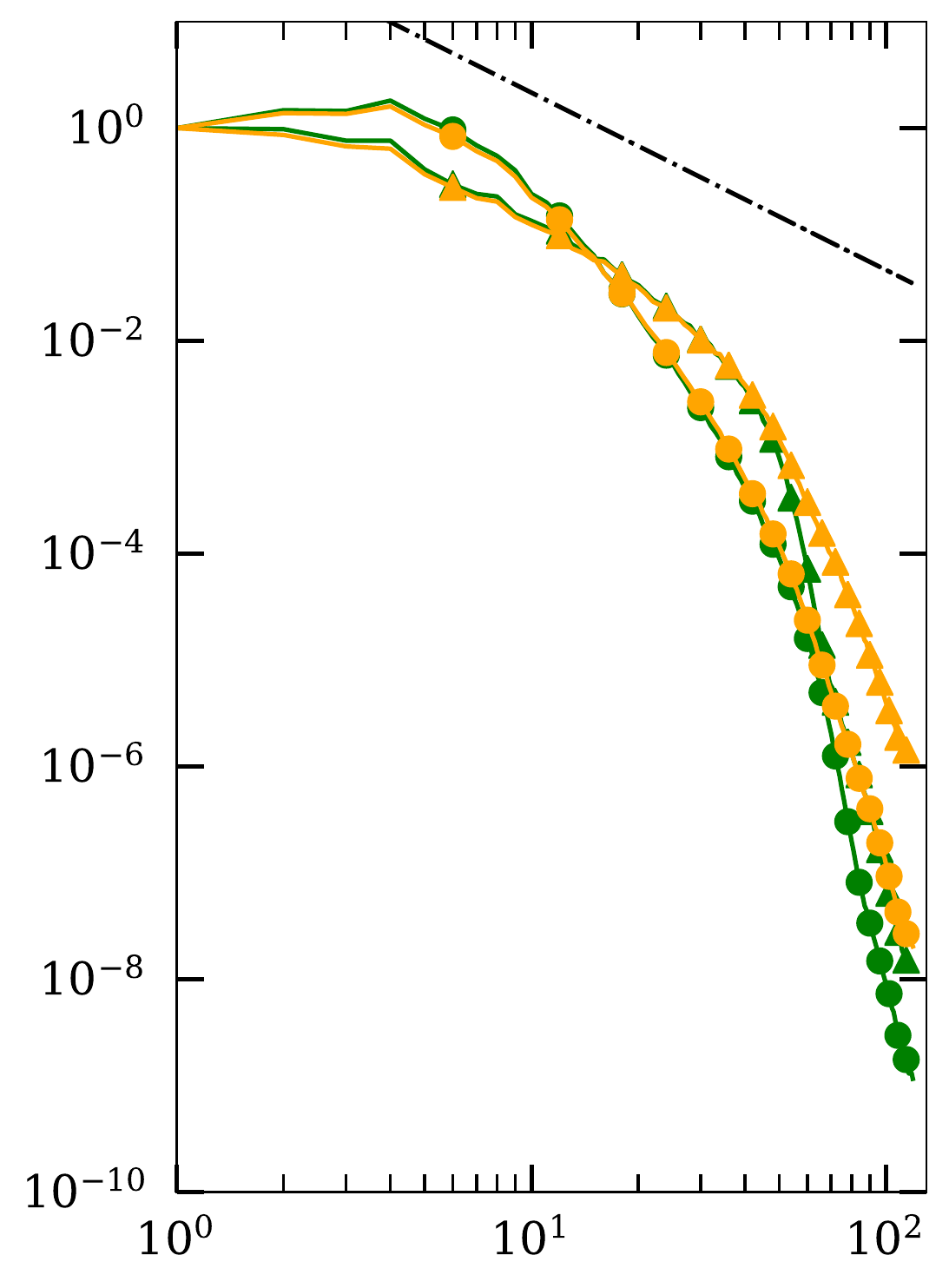}};
      \begin{scope}[x={(a.south east)},y={(a.north west)}]
        \node [align=center] at (0.00,0.96) {(b)};
        \node [align=center] at (0.55,0.0) {\large $k_z$};
        \node [align=center] at (0.84,0.90) {$k_z^{-5/3}$};
      \end{scope}
\end{tikzpicture}
\vspace{-0.5cm}
\caption{One-dimensional energy spectra in the spanwise direction, at $Re_\tau=185$ : $\alpha=\rho u^2$, at $y^+=10$ (\protect\linecircle{darkgreen}) and $y^+=150$ (\protect\linetriangle{darkgreen});  $\alpha=T$, at $y^+=10$ (\protect\linecircle{oorange}) and $y^+=150$ (\protect\linetriangle{oorange}) . Panel (a), CN case; panel (b), FR case.}
\label{fig:spectra1d}
\end{figure}

In this appendix, additional material is provided in order to verify the adequacy of the selected computational mesh. According to Poggie \textit{et al.} \cite{poggie2015resolution}, the spanwise extent of the computational domain should be at least twice the local boundary layer thickness, $\delta_\text{99}$, in order to ensure proper decorrelation in the spanwise direction. By considering the $\delta_\text{99}$ value at the end of the flat plate, we obtain for the present simulation $(L_x \times L_y \times L_z) / \delta_{99,\text{end}} = 65 \times 3 \times 2$, which meets the suggested recommendation. Direct confirmation is obtained by inspection of the two-points spanwise correlations, shown in figure~\ref{fig:two_point_correlation} for density and streamwise velocity at two different wall-normal locations (one near the turbulent production inner peak, $y^+\approx10$, and another in the logarithmic zone). With regard to the resolution, the authors indicate that using $\Delta x^+ < 10$, $\Delta y^+ < 1$ and $\Delta z^+ < 5$ is sufficient to obtain well-converged first- and second-order statistics; resolutions shown in table~\ref{tab:dns_param} are well below the suggested limits and ensure very accurate representation of the small details of the flow. Another important check consists in verifying the absence of energy pileup at high wavenumbers; to this aim, we report the one-dimensional kinetic energy spectra in figure~\ref{fig:spectra1d}, at the same two wall positions. The energy distribution cascades down smoothly for approximately nine orders of magnitude, with an inertial range extending for more than a decade in the logarithmic region. The cutoff wavenumber indicates a good grid resolution and no energy accumulation is observed at the smallest scales.

\bibliography{biblio}

\end{document}